\documentclass[11pt, oneside]{article}   	

\usepackage{jheppub}

\usepackage{amsmath}
\usepackage{amssymb}
\usepackage{amsfonts}
\usepackage{amsthm}
\usepackage{amsbsy}
\usepackage{array}

\usepackage{mathtools}
\usepackage[usenames,dvipsnames]{xcolor}

\usepackage{subcaption}

\usepackage{dsdshorthand}

\usepackage[vcentermath]{youngtab}

\usepackage{tikz}
\usetikzlibrary{trees}
\usetikzlibrary{decorations.pathmorphing}
\usetikzlibrary{decorations.markings}
\usetikzlibrary{shapes.misc}

\newcommand\wL{\mathbf{L}}

\renewcommand\vol{\mathop{\mathrm{vol}}}

\newcommand{\tsym}{\cT}
\newcommand{\la}{\langle}
\newcommand{\ra}{\rangle}
\newcommand{\pa}{\partial}

\newcommand\Disc{\mathrm{Disc}}

\newcommand{\thalf}{\tfrac{1}{2}}
\newcommand*\link[1]{\hspace*{0em plus 1fill}\makebox{#1}}

\newtheorem* {conjecture} {Conjecture}

\makeatletter
\def\@fpheader{\ }
\makeatother

\title{  Shocks, Superconvergence, and a Stringy Equivalence Principle}
\author{Murat Kolo\u{g}lu$^{1}$, Petr Kravchuk$^{2}$, David Simmons-Duffin$^{1}$, and Alexander Zhiboedov$^{3}$}
\affiliation{${}^1$Walter Burke Institute for Theoretical Physics, Caltech, Pasadena, California 91125, USA \\
${}^2$School of Natural Sciences, Institute for Advanced Study, Princeton, New Jersey 08540, USA \\
${}^3$CERN, Theoretical Physics Department, 1211 Geneva 23, Switzerland
}
\date{}
\abstract{
We study propagation of a probe particle through a series of closely situated gravitational shocks. We argue that in any UV-complete theory of gravity the result does not depend on the shock ordering --- in other words, coincident gravitational shocks commute. Shock commutativity leads to nontrivial constraints on low-energy effective theories. In particular, it excludes non-minimal gravitational couplings unless extra degrees of freedom are judiciously added.  In flat space, these constraints are encoded in the vanishing of a certain ``superconvergence sum rule." In AdS, shock commutativity becomes the statement that average null energy (ANEC) operators commute in the dual CFT. We prove commutativity of ANEC operators in any unitary CFT and establish sufficient conditions for commutativity of more general light-ray operators. Superconvergence sum rules on CFT data can be obtained by inserting complete sets of states between light-ray operators. In a planar 4d CFT, these sum rules express $\frac{a-c}{c}$ in terms of the OPE data of single-trace operators.
}

\preprint{CALT-TH 2019-012 \\
\link{CERN-TH-2019-040}}

\begin{document}

\maketitle

\newpage

\section{Introduction}

In General Relativity (GR), particles follow geodesics regardless of their polarizations or internal composition. This is sometimes called the ``strong equivalence principle" \cite{Will:2014kxa}. However, in the presence of non-minimal (higher-derivative) couplings, this principle is no longer true --- the path of a particle can depend on its polarization and is not given by a geodesic. Such modifications of GR are known to be in tension with causality and unitarity.\footnote{In quantum field theory, causality is a statement about commutativity of local operators at spacelike-separated points. In gravitational theories, we do not have local operators but the asymptotic structure of the gravitational field is weakly coupled and relatively simple. We can therefore introduce gravitational field operators at the asymptotic boundary of spacetime and impose their commutativity at spacelike separations. This leads to a notion of asymptotic causality \cite{Gao:2000ga}. In AdS/CFT, this becomes a familiar statement about commutativity of local CFT operators at spacelike-separated points on the boundary of AdS. In flat space, it is related to the rate of growth of the amplitude in the forward limit \cite{Camanho:2014apa}.} 

A simple example is the propagation of a probe particle through a gravitational shock (gravitational field of a highly-boosted particle). In GR, propagation through a shock leads to a velocity kick and a Shapiro time delay. By contrast, in theories with non-minimal gravitational couplings, there can be gravitational birefringence: depending on the polarization of the probe particle, the effect of the shock can be different. Moreover, for certain polarizations, the probe particle can experience a time advance \cite{Camanho:2014apa}. By arranging many shocks one after the other, one can accumulate the time advances and produce macroscopic violations of asymptotic causality. The restoration of causality requires an infinite set of massive higher-spin particles. It was argued in \cite{Camanho:2014apa} that the masses of these higher-spin particles must be related to the scale that enters the modified gravitational coupling\footnote{The subscript GB stands for the common Gauss-Bonnet modification of GR, but we mean it more generally as a statement about any non-minimal gravitational coupling.}
\be
\label{eq:cartoon}
\alpha_{{\rm GB}} \lesssim {1 \over m_{{\rm gap}}^2}.
\ee 
Indeed, this is what happens in string theory, where the higher-spin particles are string excitations. Similar bounds on three-point couplings were derived in \cite{Hofman:2009ug,Camanho:2014apa,Meltzer:2017rtf,Afkhami-Jeddi:2016ntf,Kulaxizi:2017ixa,Costa:2017twz}. A common feature of these arguments is the lack of a sharp equality relating the non-minimal couplings to the extra degrees of freedom required for causality.

In this work, we provide such an equality between $\alpha_{{\rm GB}}$ and contributions of massive states in a general gravitational theory. We note that non-minimal gravitational couplings introduce another feature that is absent in GR, namely non-commutativity of coincident gravitational shocks.\footnote{A shock is a region of curvature localized on a null surface. We say that two shocks become coincident in the limit that their null surfaces coincide.} This is another violation of the strong equivalence principle. Indeed, as we review below, geodesics are insensitive to the ordering of gravitational shocks. On the other hand, for theories with non-minimal gravitational couplings, the effect of propagation through multiple shocks depends on the ordering of the shocks. What is less trivial is that this effect can be traced to pathological behavior of the scattering amplitude in the Regge limit. We find that the converse is also true: in any UV-complete gravitational theory, soft Regge behavior guarantees that coincident gravitational shocks must commute. This can be readily checked in tree-level string theory. 

Therefore, we suggest that a weaker ``stringy" equivalence principle does hold in general UV-complete gravitational theories: coincident gravitational shocks commute. In contrast to the causality discussion above, commutativity of coincident shocks leads to quantitative sum rules that equate the size of non-minimal couplings to the extra degrees of freedom that are present in the theory. 

In section~\ref{sec:shocksinflatspace}, we explain how commutativity of coincident shocks is equivalent to boundedness of amplitudes $\cA(s,t)$ in the Regge limit ($t\to \oo$ with fixed $s$). Specifically, shocks with spins $J_1$ and $J_2$ commute if and only if the Regge intercept of the theory $J_0$ satisfies\footnote{The Regge intercept depends on the value of the Mandelstam variable $s$, but we suppress that dependence here for simplicity.}
\be
\label{eq:reggecondition}
J_1+J_2 > J_0+1.
\ee For example, gravitational (spin-2) shocks commute if $J_0<3$. It has been argued that consistent weakly-coupled gravity theories in flat space actually obey $J_0\leq 2$ \cite{Camanho:2014apa}, so gravitational shocks certainly commute in this case (as do higher-spin shocks).

In section~\ref{sec:shocksandreggeflat}, we show that commutativity of coincident shocks is equivalent to certain dispersion relations called ``superconvergence" sum rules \cite{DEALFARO1966576,russiansuperconvergence}. When applied to gravitational amplitudes, these sum rules express (squares of) non-minimal couplings in terms of three-point couplings of massive states. This shows that non-minimal couplings cannot exist without additional massive states, recovering a result from \cite{Camanho:2014apa} in a different way. In subsequent sections, we study superconvergence sum rules in several examples, showing explicitly how they are obeyed in GR (section~\ref{sec:shocksineinstein}), disobeyed in higher-derivative gravity theories (section~\ref{sec:nonminimalflat}), but obeyed in string theories (sections~\ref{sec:gravitonamplitudestring} and \ref{sec:shocksmatrix}). Indeed, the failure of superconvergence sum rules in higher-derivative theories like Gauss-Bonnet gravity gives an efficient way to show that they violate the Regge boundedness condition $J_0<3$ without computing full amplitudes in those theories.

In AdS, commutativity of coincident shocks translates into a statement that can be proven nonperturbatively using CFT techniques. As we review in section~\ref{sec:shocksinads}, we can create shocks by integrating local operators along null lines on the boundary of AdS. It is most natural to study propagation through AdS shocks using observables called ``event shapes" \cite{Basham:1977iq,Basham:1978zq,Basham:1978bw,Hofman:2008ar,Belitsky:2013xxa}, which we review in sections~\ref{sec:eventshapreview} and~\ref{sec:computinginbulk}. Commutativity of coincident shocks becomes the statement that two null-integrated operators commute when placed on the same null plane:
\be
\label{eq:nullintegratedoperatorcommutativity}
\left[
\int_{-\oo}^\oo dv_1\,\cO_{1;v\cdots v}(u=0,v_1,\vec y_1),
\int_{-\oo}^\oo dv_2\,\cO_{2;v\cdots v}(u=0,v_2,\vec y_2)
\right]
\stackrel{?}{=}
0.
\ee
Here, we use lightcone coordinates $ds^2 = -du\,dv + d\vec y^2$. The CFT operators lie on the same plane $u=0$ but at different transverse positions $\vec y_1,\vec y_2\in \R^{d-2}$. Furthermore, their vector indices are aligned with the direction of integration (the $v$-direction). For example, when $\cO_1$ and $\cO_2$ are both the stress-tensor $T_{\mu\nu}$, (\ref{eq:nullintegratedoperatorcommutativity}) becomes a commutator of average null energy operators. An average null energy operator on the boundary creates a gravitational shock in the bulk.

The commutativity statement (\ref{eq:nullintegratedoperatorcommutativity}) might seem obvious, since $\cO_1$ and $\cO_2$ are spacelike-separated everywhere along their integration contours. However, the spacelike separation argument is too quick, and is actually {\it wrong\/} in some examples (see appendix~\ref{app:lighttransformscalarsubtleties}). The problem is that the positions of $\cO_1$ and $\cO_2$ coincide at the endpoints of their integration contours (in an appropriate conformal frame), and one must be careful to analyze what happens there.

We perform a careful analysis of the commutator (\ref{eq:nullintegratedoperatorcommutativity}) in section~\ref{sec:commutativityone}, explaining the circumstances when it is well-defined (but not necessarily zero), and the additional conditions required for it to vanish. A necessary condition for vanishing is
\be
J_1 + J_2 &> J_0 + 1,
\label{eq:conditionforvanishingofcommutator}
\ee
where $J_1$ and $J_2$ are the spins of $\cO_1$ and $\cO_2$, and $J_0$ is the Regge intercept of the CFT \cite{Brower:2006ea,Cornalba:2007fs,Costa:2012cb,Caron-Huot:2017vep,Kravchuk:2018htv}. In upcoming work \cite{AnecOPE}, we also show that a non-vanishing commutator necessarily leads to a Regge pole at $J=J_1+J_2-1$.

In section~\ref{sec:reggelimit}, we prove that $J_0\leq 1$ in nonperturbative CFTs (generalizing arguments of \cite{Hartman:2015lfa,Caron-Huot:2017vep} to spinning correlators). This establishes commutativity of average null energy operators in nonperturbative theories\footnote{Commutativity of average null energy (ANEC) operators is important for understanding information-theoretic aspects of CFTs \cite{Faulkner:2016mzt,Casini:2017roe}, and plays a central role in the recently proposed BMS symmetry in CFT \cite{Cordova:2018ygx}. As far as we are aware, it is usually argued for using the fact that the stress tensors are spacelike separated. Our analysis closes a loophole in this argument.} 
\be
\label{eq:stresscommutatorintro}
\left[
\int_{-\oo}^\oo dv_1\,T_{vv}(u=0,v_1,\vec y_1),
\int_{-\oo}^\oo dv_2\, T_{vv}(u=0,v_2,\vec y_2)
\right]= 0  ,
\ee
for $\vec y_1 \neq \vec y_2$. For large-$N$ theories in the planar limit, the bound on chaos \cite{Maldacena:2015waa} implies that $J_0 \leq 2$. Thus, average null energy operators commute in planar theories as well. However, commutativity can be lost at higher orders in large-$N$ perturbation theory (and only recovered nonperturbatively).

The condition (\ref{eq:conditionforvanishingofcommutator}) is in direct analogy to the condition  (\ref{eq:reggecondition}) in flat space. When it holds, one can derive analogous superconvergence sum rules for CFTs by evaluating event shapes of (\ref{eq:nullintegratedoperatorcommutativity}). In section~\ref{sec:computingeventshapes}, we show how to compute these event shapes using the conformal block decomposition, expressing them as sums over intermediate CFT states.\footnote{The conformal blocks we study in this work are for the ``lightray-local $\to$ lightray-local" channel. This is the conventional OPE. By contrast, in \cite{AnecOPE} we develop a new type of OPE that allows one to describe the ``lightray-lightray $\to$ local-local" channel.} The relevant conformal blocks can be computed explicitly in any spacetime dimension. The blocks for stress tensors agree perfectly with our bulk calculations from section~\ref{sec:shocksinads}. The result is an infinite set of superconvergence sum rules for CFT data.

Of course, the usual crossing symmetry equations \cite{Ferrara:1973yt,Polyakov:1974gs} are also an infinite set of sum rules for CFT data. However, CFT superconvergence sum rules have some nice properties.
In large-$N$ theories in the planar limit, they get contributions only from single-trace operators and non-minimal three-point structures (i.e.\ three-point structures that do not arise from GR in AdS). Thus, one obtains expressions for non-minimal three-point coefficients in terms of massive ``stringy" states, analogous to superconvergence sum rules in flat space.

As an example, in 4d CFTs, we find the superconvergence sum rules
\be
\label{eq:superconvergenceexamples}
(t_4+2t_2)^2 &= \sum_\f |\l_{TT\f}|^2 \frac{15\.2^4\pi^4 \Gamma(\De_\f-1)\Gamma(\De_\f)}{C_T \Gamma(4-\frac{\De_\f}{2})^2 \G(2+\frac{\De_\f}{2})^6} + \textrm{non-scalar}, \nn\\
(t_4+2t_2)^2 &= -\sum_\f |\l_{TT\f}|^2 \frac{360^2\pi^4 \Gamma(\De_\f-1)\Gamma(\De_\f)}{7C_T \Gamma(4-\frac{\De_\f}{2})^2 \G(2+\frac{\De_\f}{2})^6} + \textrm{non-scalar},
\ee
along with an infinite number of others.
Here, $t_2$ and $t_4$ are coefficients of non-Einstein three-point structures in the correlator $\<TTT\>$, see \cite{Hofman:2008ar}. For example, in 4d $\cN=1$ theories, we have $t_2=\frac{6(c-a)}{c}$ and $t_4=0$. The sums in (\ref{eq:superconvergenceexamples}) run over scalar operators $\phi$ with dimensions $\De_\f$ and OPE coefficients $\l_{TT\phi}$ in the $T\x T$ OPE. The term ``non-scalar" refers to contributions of operators with spin $J\geq 2$, not including the stress-tensor (whose contribution is on the left-hand side of (\ref{eq:superconvergenceexamples})). For simplicity, we have written only the sum rules that get contributions from scalar operators. Other sum rules give expressions for other combinations of $t_2$ and $t_4$, but involve exclusively non-scalars. The factors $\G(4-\frac{\De_\f}{2})^{-2}$ in (\ref{eq:superconvergenceexamples}) ensure that contributions of double-trace operators are suppressed by $O(1/N^4)$ in the large-$N$ limit. This is a generic feature of superconvergence sum rules and it stems from the fact that they can be written in terms of a double-discontinuity \cite{Caron-Huot:2017vep}. In particular, one can see explicitly that if no single-trace operators are present other than the stress tensor, then $t_2$ and $t_4$ must vanish in the planar limit.\footnote{Strictly speaking the above equations only fix $t_4+2t_2$, but there are also linearly independent constraints from other components of the sum rule.}${}^{,}$\footnote{Our methods for computing event shapes may be also useful for investigating positivity conditions. In \cite{CamanhoUN} it was shown that positivity of multi-point energy correlators also leads to vanishing non-minimal couplings, e.g. $t_2,t_4=0$ for $\<TTT\>$, in theories with only gravitons and photons in the bulk.}

In \cite{Heemskerk:2009pn}, it was conjectured that for any CFT with a large gap $\De_\mathrm{gap}$ in the spectrum of spin $J\geq 3$ single-trace operators (``stringy states"), non-minimal couplings in the effective bulk Lagrangian should be suppressed by powers of $1/\De_\mathrm{gap}$.
 In section~\ref{sec:bounds}, we argue (non-rigorously) that the contributions of stringy states to superconvergence sum rules are suppressed by powers of $1/\De_\mathrm{gap}$, and this establishes the conjecture of \cite{Heemskerk:2009pn} (in the case of three-point couplings) in a way different from the arguments of \cite{Hofman:2009ug,Camanho:2014apa,Meltzer:2017rtf,Afkhami-Jeddi:2016ntf,Kulaxizi:2017ixa,Costa:2017twz}. We conclude in section~\ref{sec:conclusions}.

In appendix~\ref{app:moresuperflat}, we give more details about superconvergence sum rules in flat space. In appendix~\ref{app:lighttransformscalarsubtleties}, we give an example of the phenomenon of ``detector cross-talk," where na\"ively spacelike light-ray operators can fail to commute. In the remaining appendices we provide details about sum rules in CFT.

\paragraph{Note added:} After this work had been largely completed, we became aware of \cite{BelinEE} which has some overlap with this paper.

\section{Shocks and superconvergence in flat space}
\label{sec:shocksinflatspace}

In General Relativity (GR), test bodies follow geodesics, and it follows that the effects of coincident shocks are commutative.
To see this, consider a shockwave in flat space \cite{Aichelburg:1970dh, tHooft:1987vrq}
\be
\label{eq:flatshock}
ds^2 = - d u\, d v + {4 \Gamma({D-4 \over 2}) \over \pi^{{D-4 \over 2}}}{G p^v \over |\vec y|^{D-4}} \delta(u) d u^2 + d \vec y^2,
\ee
and let us study null geodesics in this geometry.\footnote{Obtaining such solutions from a smooth Cauchy data (or limits thereof) in gravitational theories can be subtle and was discussed for example in \cite{Papallo:2015rna}. For us, the solution (\ref{eq:flatshock}) is a convenient way to think about the high energy limit of  gravitational scattering and per se does not play any role.}

A shockwave is a gravitational field created by a relativistic source. The Aichelburg-Sexl geometry (\ref{eq:flatshock}) is an exact solution of Einstein's equations with a stress-energy source $T_{uu}(u , \vec y) = p^v \delta(u) \delta^{(D-2)}(\vec y)$ localized on a null geodesic. The only non-trivial metric component $h_{uu}(u, \vec y) = {4 \Gamma({D-4 \over 2}) \over \pi^{{D-4 \over 2}}}{G p^v \over |\vec y|^{D-4}} \delta(u) $ is a solution of the Laplace equation in the transverse plane parametrized by $\vec y$ with a non-trivial source
\be
\Box_{\vec y} h_{uu} (u, \vec y) = -16 \pi G T_{uu}(u, \vec y) .
\ee
Famously, (\ref{eq:flatshock}) continues to be an exact solution in any higher derivative theory of gravity as well \cite{Horowitz:1989bv}. Higher derivative interactions, however, lead to nontrivial corrections to the propagation of probe particles on the shockwave backgrounds. 

Consider a probe particle on the shockwave background that follows a null geodesic. We can parameterize the null geodesic by $u$. Suppose the geodesic approaches the shock at impact parameter $\vec y(u=0) = \vec b$. Crossing the shock causes both a Shapiro time delay
\be
v_\mathrm{after} - v_\mathrm{before} &= {4 \Gamma({D-4 \over 2}) \over \pi^{{D-4 \over 2}}}{G p^v \over |\vec b|^{D-4}} ,
\ee
and a velocity kick in the transverse plane due to gravitational attraction
\be
\label{eq:geodesiccalculation}
\dot{\vec y}_\mathrm{after} - \dot{ \vec y}_\mathrm{before} &= - {4 \Gamma({D-2 \over 2}) \over \pi^{{D-4 \over 2}}} G p^v {\vec b \over |\vec b|^{D-2}} .
\ee
The same result can be obtained by analyzing wave equations on the shockwave backgrounds or using scattering amplitudes (as we review in detail below). In General Relativity the effect of a shockwave on a probe particle does not depend on the the polarization of the particle. This is no longer true in higher-derivative theories of gravity. This can lead to Shapiro time advances and violations of asymptotic causality \cite{Camanho:2014apa}. 

We can also consider a more complicated geometry constructed by a superposition of relativistic sources localized at different retarded times $u_i$ and transverse positions $\vec b_i$. The exact gravitational field created by such a superposition is simply a sum of the shockwaves (\ref{eq:flatshock}).

\begin{figure}[t]
	\centering
		\includegraphics[scale=1.2]{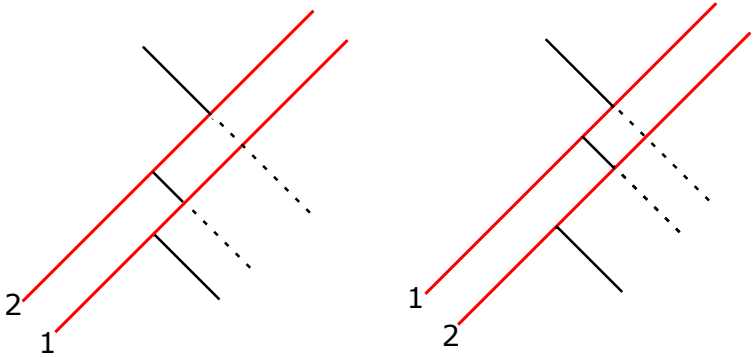}
		\caption{The probe geodesic is denoted by a black line and shock waves by red lines. Dashed lines mark time delay associated to each shock. In General Relativity propagation through a pair of closely situated shockwaves is commutative, namely the overall effect does not depend on the order of the shocks. This is no longer true in theories with higher derivative corrections. We argue that commutativity must hold in any UV complete theory.}
		\label{fig:commutativity}
\end{figure}

 If a probe particle follows a geodesic, then propagation through a series of closely situated shocks leads to an additive effect. In particular, the result does not depend on the ordering of shocks and is commutative (figure~\ref{fig:commutativity}). This is no longer true in theories with higher derivative corrections. In this case, the result of the propagation through a pair of closely situated shocks will generically depend on their ordering. 

In this section we show that commutativity of shockwaves is directly related to the Regge limit. In particular, we argue that in any UV complete theory (gravitational or not) the shock waves must commute. Therefore, any non-commutativity of shocks present in the low energy effective theory should be exactly canceled by the extra degrees of freedom. The mathematical expression of this fact is encapsulated in the superconvergence sum rules which we describe in detail below. 

\subsection{Shockwave amplitudes}
\label{sec:shockwaveamplitudes}

It will instructive for our purposes to restate the discussion of the previous section in terms of scattering amplitudes. This has been done in \cite{Camanho:2014apa}, whose setup we review momentarily. In the simplest case of a propagation through a single shock, we consider an absorption of a virtual graviton by a probe particle
\be
\label{eq:singleshock}
g^* X \to X' ,
\ee
where $X$ and $X'$ describe a particle in an initial and final state (these could be different) and $g^*$ stands for a virtual graviton that is emitted from some extra source that we do not write down explicitly.\footnote{In other words, the full description of (\ref{eq:singleshock}) is in terms of a four-point amplitude.} We denote the corresponding scattering amplitude ${\cal A}_{g^* X \to X'}$.

To discuss causality, we consider the high-energy behavior of the scattering amplitude in the forward direction, see \cite{Camanho:2014apa}.  A convenient choice of momenta and polarization for the process (\ref{eq:singleshock}) is
\be
\label{eq:singleshockkin}
p_X &= \p{p^u, {\vec q^2 \over 4 p^u} , -{\vec q \over 2}} , ~~~ p_{X'} = \p{-p^u, -{\vec q^2 \over 4 p^u} , -{\vec q \over 2}} , \nn \\
p_{g^*} &= (0, 0 , \vec q) , ~~~ \e_{g^*} = (0 , - 2 , 0) .
\ee
Here, we use lightcone coordinates $(u,v,\vec y)$ with metric
\be
\label{eq:lightconeycoordinates}
ds^2 = -du\, dv + d\vec y^2,\qquad \vec y\in \R^{D-2}.
\ee
The virtuality of the graviton is $\vec q^2$.

An interesting phenomenon occurs when studying this process in  impact parameter space. In this case, the virtual graviton emitted by a source comes with the following wavefunction
\be
 \int d^{d-2} \vec q {e^{i \vec b \cdot \vec q} \over \vec q^2} {\cal A}_{g^* X \to X'}(p^u, \vec q) .
\ee
The remarkable property of this integral is that it can be evaluated by taking the residue at $\vec q^2 = 0$, so that the virtual graviton becomes on-shell! The on-shell condition $\vec q^2 =0$ requires that $\vec q$ becomes complex. The result is that the physical phase shift $\delta(p^v, \vec b)$ is computed by a scattering amplitude in spacetime with mixed signature (making $\vec q$ complex corresponds to a second Wick rotation). More precisely, we get 
\be
\label{eq:phaseshift}
\delta(p^v, \vec b) \propto {\cal A}_{g^* X \to X'}(p^u, - i \pa_{\vec b}) {1 \over | \vec b|^{D-4}} ,
\ee
where ${\cal A}_{g X \to X'}$ is now a usual on-shell amplitude, albeit evaluated in slightly unusual kinematics. Note that the on-shell condition $\vec q^2 =0$ is reflected in (\ref{eq:phaseshift}) by the fact that ${1 \over | \vec b|^{D-4}}$ is a harmonic function, so it is killed by $(-i\ptl_{\vec b})^2$. The causality discussion of \cite{Camanho:2014apa} then focuses on the properties of the on-shell amplitude ${\cal A}_{g^* X \to X'}$ and shows that in gravitational theories with higher derivative corrections it can lead to causality violations unless new degrees of freedom are added. 

We would like to emphasize that the whole discussion can be formulated in terms of on-shell amplitudes with shockwave gravitons $g^*$ and their properties. This is particularly useful in a gravitational theory, where there is no clear definition of off-shell observables. Therefore it is desirable to formulate the problem purely in terms of on-shell observables. This is the path we follow below.

\begin{figure}[t!]
	\centering
		\includegraphics[scale=1.2]{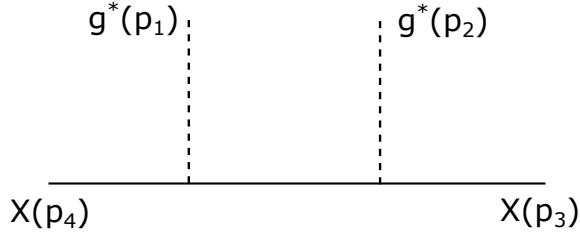}
		\caption{An elastic scattering of a probe particle and an on-shell shock state graviton. We will argue that the Regge behavior of this amplitude in consistent theories of gravity is such that gravitational shockwaves always commute. We adopt a CFT correlator-like prescription where the time in the diagram goes from right to left. }
		\label{fig:ampltwoshocks}
\end{figure}

In this paper we will be interested in elastic scattering of a probe $X$ and two shockwave gravitons $g^*$
\be
\label{eq:process}
g^*(p_2) X(p_3) \to g^*(p_1) X(p_4) ,
\ee
see figure \ref{fig:ampltwoshocks}. We choose the final state of the probe to be $X$ for simplicity, but the whole discussion goes through intact for any other one-particle state $X'$.

A convenient choice of momenta is
\be
\label{eq:momentumthreefour}
p_1 &= (0,-p^v,\vec q_1), ~~~p_2 = (0,p^v,\vec q_2), \nn\\
p_3 &= ( p^u, 0, -\vec q_2), ~~~ p_4 = (- p^u, 0, -\vec q_1),
\ee
and again the shockwave gravitons have polarizations $\e_{g^*} = (0 , - 2 , 0)$ as in (\ref{eq:flatshock}). As above, we can think of each shockwave graviton as originating from a pair of particles that we have not written explicitly. In this way, ${\cal A}_{g^* X \to g^* X}$ can be thought of as an economical description of the relevant part of the six-point amplitude. An on-shell condition is $\vec q_1^2 = \vec q_2^2 = 0$. As before this naturally arises in the impact parameter transform
\be
\int d^{d-2} \vec q_1 \int d^{d-2} \vec q_2 e^{i \vec q_1 \cdot \vec b_1} e^{i \vec q_2 \cdot \vec b_2} {1 \over \vec q_1^2} {1 \over \vec q_2^2} {\cal A}_{g^* X \to g^* X} .
\ee
Below, we simply study the on-shell amplitude ${\cal A}_{g^* X \to g^* X}$ as an object on its own, without referring to impact parameter space.\footnote{In a gravitational theory when trying to separate a probe from the rest of the system we should check that the joint system does not form a black hole. In particular, we would like the impact parameter $b$ in the discussion above to be larger than the Schwarzschild radius of the system $r_S^{D-3} = \sqrt{p^u p^v} G_N$. In momentum space this implies that ${r_S^2 \over b^2} < 1$, where $\vec q_1 , \vec q_2 \sim {1 \over b}$. This implies that for given $\vec q_1 , \vec q_2$ we can only consider ${p^u p^v  \over m_{Pl}^2} < \left( m_{Pl}^2 b^2 \right)^{D-3}  $. For energies ${p^u p^v  \over m_{Pl}^2} > \left( m_{Pl}^2 b^2\right)^{D-3}  $ the picture of a probe propagating through a shockwave is not the correct description of physics. This does not present a problem in a tree-level gravitational theory when we work to leading order in $G_N$ and thus can make it arbitrarily small (or, similarly, if we consider gravitational deep inelastic scatering in a gapped QFT, see section \ref{sec:gravitationalDIS}). At finite $G_{N}$, we can still formally define the amplitude ${\cal A}_{g^* X \to g^* X}$ in kinematics (\ref{eq:momentumthreefour}) with complex null transverse momenta, but its physical interpretation at arbitrarily high energies is less clear. We ignore this subtlety below and only consider tree-level examples in flat space, though we believe everything we say holds at finite $G_N$ as well. This problem does not appear in our CFT discussion where $g^*$ corresponds to a light-ray operator insertion in a boundary CFT and has, thus, a clear definition in a finite $N$ CFT.}

One new feature of (\ref{eq:momentumthreefour}) compared to (\ref{eq:singleshockkin}) is that the shockwave graviton transfers to the probe a large longitudinal momentum $p^v$.  In \cite{Camanho:2014apa} the shockwaves were carefully separated in the $u$ direction such that $p^v$ is effectively set to $0$ and the amplitude ${\cal A}_{g^* X \to g^* X}$ reduces  to a product of a one-shock interactions ${\cal A}_{g* X \to X'} {\cal A}_{g^* X' \to X}$. Our regime of interest is the opposite, namely we would like to put the shockwaves on top of each other. This effectively leads to studying ${\cal A}_{g^* X \to g^* X}$ at arbitrarily large values of $p^v$.

\subsection{Shock commutativity and the Regge limit}
\label{sec:shocksandreggeflat}

An important class of shockwave amplitudes arises when the shockwaves are localized on null planes. Such objects provide a natural translation into the language of on-shell amplitudes of propagation through classical shock backgrounds like (\ref{eq:flatshock}). For simplicity, we consider $2\to 2$ scattering amplitudes of massless scalars in this section, where particles $1$ and $2$ play the role of shocks. We generalize to the case of gravitational (or spinning) shocks in the next section.

A shock wavefunction localized at $u=u_0$ is given by
\be
\label{eq:shockwave}
4\pi \de(u-u_0) e^{i\vec q\. \vec y} &=\int_{-\oo}^\oo  dp^v e^{i (-\frac 1 2 p^v (u-u_0) + \vec q\.\vec y)}
= \int_{-\oo}^\oo dp^v  e^{\frac i 2 p^v u_0} e^{ip\.y},\nn\\
p^\mu &= (0,p^v, \vec q),
\ee
where $\vec q\in \C^{D-2}$ is null.  
In order for $p^\mu$ to be on-shell, $\vec q$ must be complex. Note also that $p^v$ is integrated over both positive and negative values. As explained in the previous section, such wavefunctions do not represent physical incoming or outgoing particles, but rather can be thought of as arising from poles of higher-point amplitudes.

Let shocks 1 and 2 be localized at $u_1$ and $u_2$. We take particles $3$ and $4$ to be momentum eigenstates. Overall, the momenta in ``all-incoming" conventions are given by (\ref{eq:momentumthreefour}), 
where we must integrate over $p^v$ to create the delta-function localized shocks. Let us define the Mandelstam variables $s=-(p_1+p_2)^2=-(\vec q_1+\vec q_2)^2$ and $t=-(p_2+p_3)^2=p^u p^v$.\footnote{This labeling is chosen for consistency with our CFT conventions in section~\ref{sec:computingeventshapes}. Note that the roles of $t$ and $s$ are swapped relative to usual discussions of high energy scattering.}

\begin{figure}[t!]
	\centering
		\begin{tikzpicture}
		
		\draw (-3,-2) -- (-3,2) -- (3,2) -- (3,-2) -- cycle;
		\draw[dashed] (-3,0) -- (-0.4,0);
		\draw[dashed] (0,0) -- (3,0);
		\draw[->,thick] (-2.8,-0.2) -- (-0.4,-0.2) to[out=0,in=180] (0.4,0.2) -- (2.8,0.2);
			
		\draw (-2.5,1.9) -- (-2.5,1.5) -- (-2.9,1.5);		
		\node[right] at (-2.95,1.7) {$t$};
		
		\end{tikzpicture}
		\caption{Integration contour for computing the shock amplitude. The integral is along the real axis, rotated by a small positive angle. The dashed lines represent $t$- and $u$-channel cuts. We assume that $s<0$.}
		\label{fig:originalcontour}
\end{figure}
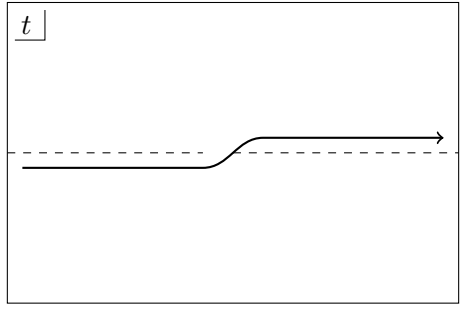

Denote the scattering amplitude for particles 1, 2, 3, and 4 in momentum space by $i\cA\p{s,t}$.
Plugging in shock-wavefunctions (\ref{eq:shockwave}) for particles $1$ and $2$, and applying momentum conservation $p_1^v = -p_2^v =-p^v$, we find that the amplitude for particles 3 and 4 scattering with shocks 1 and 2 is
\be
\label{eq:qdef}
\cQ(u_1-u_2)&\equiv i\int_{-\oo}^\oo dp^v\, \exp\p{-\frac i 2 p^{v}(u_1-u_2)} \cA\p{s = -(\vec q_1+\vec q_2)^2,t=p^v p^u} \nn\\
&= \frac i {p^u} \int_{-\oo}^\oo dt\, \exp\p{-\frac{i}{2}\frac{t}{p^u}(u_1-u_2)} \cA(s,t).
\ee
The amplitude $\cA(s,t)$ may have singularities on the real $t$-axis. As usual, the correct prescription is to approach these singularities from above for positive $t$ and below for negative $t$. The corresponding integration contour is shown in figure~\ref{fig:originalcontour}.

\begin{figure}[t!]
	\centering
	\begin{subfigure}[b]{0.45\textwidth}
		\begin{tikzpicture}
		
		\draw (-3,-2) -- (-3,2) -- (3,2) -- (3,-2) -- cycle;
		\draw[dashed] (-3,0) -- (-0.4,0);
		\draw[dashed] (0,0) -- (3,0);
		\draw[->,dotted,thick] (-1.5,-0.5) to[out=-90,in=180] (0,-1.7) to[out=0,in=-90] (1.5,-0.5);
		\draw[->,thick,opacity=0.3] (-2.8,-0.2) -- (-1.8,-0.2);
		\draw[thick,opacity=0.3] (-1.8,-0.2) -- (-0.4,-0.2) to[out=0,in=-90] (0,0);
		\draw[->,thick] (2.8,-0.2) -- (1.8,-0.2);
		\draw[->,thick] (1.8,-0.2) -- (0.4,-0.2) to[out=180,in=-90] (0,0) to[out=90,in=180] (0.4,0.2) -- (2.8,0.2);
			
		\draw (-2.5,1.9) -- (-2.5,1.5) -- (-2.9,1.5);		
		\node[right] at (-2.95,1.7) {$t$};
		
		\end{tikzpicture}
		\caption{When $\De u>0$, we can wrap the contour around the positive $t$-axis.}
		\label{fig:foldpositive}
		\end{subfigure}~
	\begin{subfigure}[b]{0.45\textwidth}
	\begin{tikzpicture}
		\draw (-3,-2) -- (-3,2) -- (3,2) -- (3,-2) -- cycle;
		\draw[dashed] (-3,0) -- (-0.4,0);
		\draw[dashed] (0,0) -- (3,0);

		\draw[->,dotted,thick] (1.5,0.5) to[out=90,in=0] (0,1.7) to[out=180,in=90] (-1.5,0.5);

		\draw[->,thick] (-2.8,-0.2) -- (-1.8,-0.2);
		\draw[->,thick] (-1.8,-0.2) -- (-0.4,-0.2) to[out=0,in=-90] (0,0) to[out=90,in=0] (-0.4,0.2) -- (-2.8,0.2);
		\draw[->,thick,opacity=0.3] (0,0) to[out=90,in=180] (0.4,0.2) -- (2.8,0.2);

		\draw (-2.5,1.9) -- (-2.5,1.5) -- (-2.9,1.5);		
		\node[right] at (-2.95,1.7) {$t$};
		
		\end{tikzpicture}
		\caption{When $\De u<0$, we can wrap the contour around the negative $t$-axis.}
		\label{fig:foldnegative}
	\end{subfigure}
	\caption{Depending on whether the integrand decays exponentially in the upper or lower half-plane, we can deform the contour in different ways. In both cases, the old contour is shown in gray and the new contour in black. The direction of deforming the contour is indicated with a dotted arrow.}
\end{figure}
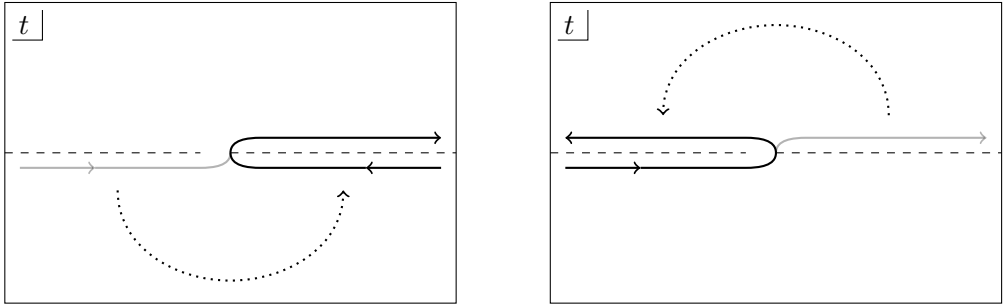

When the shocks are separated in the $u$ direction, the amplitude should factorize into a product of $S$-matrix elements describing successive interactions with each shock. This comes about as follows. Note that the factor
\be
\exp\p{-\frac{i}{2}\frac{t}{p^u}\De u}
\ee
 causes the integrand in (\ref{eq:qdef}) to be exponentially damped in $t$ in either the upper or lower half-plane, depending on the sign of $\De u$. For example, suppose $\De u>0$. The integrand is damped for $\Im t < 0$, so we can wrap the $t$-contour around the cut on the positive $t$-axis, giving a discontinuity (figure~\ref{fig:foldpositive})
\be
\label{eq:tchancutintegral}
\mathcal{Q}(\De u) &= \frac {2\pi} {p^u} \int_0^\oo dt\, \exp\p{-\frac{i}{2}\frac{t}{p^u}\De u} \Disc_t \cA(s,t) \qquad (\De u > 0),
\ee
where
\be
\Disc_t f(t) \equiv \frac{i}{2\pi} \p{f(t+i\e) - f(t-i\e)} ,
\ee
so that $\Disc_t {1 \over t - m_X^2} = \delta(t- m_X^2)$.

The formula (\ref{eq:tchancutintegral}) is true as long as $\cA(s,t)$ grows sub-exponentially in $t$.
The discontinuity factors into products of on-shell amplitudes in the $t$-channel
\be
\Disc_t \cA(s,t) &=- \sum_X \de(t-m_X^2) \cA(23,-X) \cA(X,14)\qquad (t>0).
\ee
Here, we use all-incoming notation, so $-X$ indicates the state $X$ with momentum and helicities flipped.
Plugging this into (\ref{eq:tchancutintegral}), we get an expression for the shock amplitude as a sum over intermediate states $X$. The physical interpretation is that particle $3$ propagates through shock $2$, creating an intermediate state $X$. The intermediate state $X$ propagates through shock $1$ to become particle $4$.

Next, suppose $\De u<0$, so that shock $1$ occurs before shock $2$. In this case, we can fold the $t$-contour to wrap around  the $u$-channel cut (figure~\ref{fig:foldnegative}),
\be
\mathcal{Q}(\De u) &= -\frac {2\pi} {p^u} \int_{-\oo}^{0} dt\, \exp\p{-\frac{i}{2}\frac{t}{p^u}\De u} \Disc_t \cA(s,t) \qquad (\De u < 0).
\ee
The discontinuity now factorizes into a product of on-shell amplitudes in the $u$-channel
\be
-\Disc_t \cA(s,t) &=- \sum_X \de(-t-s-m_X^2) \cA(13,-X)\cA(X,24)\qquad (t<0).
\ee

Let us take a limit where the shocks become coincident, $\De u \to 0^\pm$. We find
\be
\mathcal{Q}(0^+) &= \frac {2\pi} {p^u} \int_0^\oo dt\, \Disc_t \cA(s,t),\nn\\
\mathcal{Q}(0^-) &= -\frac {2\pi} {p^u} \int_{-\oo}^0 dt\, \Disc_t \cA(s,t).
\ee
For these quantities to be well-defined, the discontinuities should die faster than $|t|^{-1}$ along the real $t$-axis. Let us assume this is the case.
The commutator of coincident shocks is
\be
\label{eq:commutatorintegral}
\mathcal{Q}(0^+) - \mathcal{Q}(0^-) &= \frac {2\pi} {p^u} \int_{-\oo}^\oo dt\, \Disc_t \cA(s,t) \\
&=- \frac {2 \pi} {p^u} \sum_X \left[\cA(23,-X)\cA(X,14) - \cA(13,-X)\cA(X,24)\right].
\label{eq:superconvergencesumruleone}
\ee
It is not immediately obvious what the sum (\ref{eq:superconvergencesumruleone}) should be. However, if $\cA(s,t)$ decays faster than $t^{-1}$ in the limit of large complex $t$ with fixed $s$ (the Regge limit), then we can close the integration contour in (\ref{eq:commutatorintegral}) by including arcs at infinity and shrink it to zero.  Thus, coincident shocks commute if and only if the amplitude is sufficiently soft in the Regge limit.

When the amplitude decays faster than $t^{-1}$ for fixed $s$, then we obtain the condition
\be
\label{eq:superconvergencesumrule}
\int_{-\oo}^\oo dt\, \Disc_t \cA(s,t) &= 0
\ee
which is an example of a ``superconvergence" sum rule \cite{DEALFARO1966576,russiansuperconvergence}. In general, if the amplitude dies faster than $t^{-N-1}$ in the Regge limit, we can integrate its discontinuity against $t^n$ for $0\leq n<N$ to obtain additional superconvergence sum rules. Note that we obtain a different sum rule for each $s$. Superconvergence sum rules have been used, for example, to bootstrap the Veneziano amplitude \cite{SCHMID1968348}.

\subsubsection{Spinning shocks}

While scalar shock amplitudes must decay faster than $t^{-1}$ to have a superconvergence sum rule, this condition gets relaxed for spinning shocks. Consider a massless spin-$J$ particle described by a traceless symmetric tensor field $h_{\mu_1\cdots\mu_J}(x)$. It is convenient to define a symmetric $J$-differential $h_J(x,dx) = h_{\mu_1\cdots\mu_J}(x) dx^{\mu_1} \cdots dx^{\mu_J}$.

States are labeled by a momentum $p^\mu$ and a transverse traceless-symmetric polarization tensor $\e_{\mu_1\cdots\mu_J}$, modulo gauge redundancy. Let us parameterize the polarization tensor as a product of vectors $\e_{\mu_1\cdots\mu_J}=\e_{\mu_1}\cdots \e_{\mu_J}$, where $\e_\mu$ is transverse and null.\footnote{Monomials of this form span the space of transverse traceless symmetric tensors, so this parameterization is without loss of generality.} Thus, we can label momentum eigenstates by $|p,\e\>$ where $p^\mu$ and $\e^\mu$ are vectors, and the state is a homogeneous polynomial of degree-$J$  in $\e$.

A momentum eigenstate has wavefunction
\be
\label{eq:spinningmomentumeigenstate}
\<\Omega|h_J(x,dx)|p,\e\> &= (\e\. dx)^J e^{ip\.x}.
\ee
We are interested in shock-wavefunctions of the form
\be
\label{eq:spinningshockwave}
\<\Omega|h_J(x,dx)|\mathrm{shock}\> &= 4\pi \de(u) (du)^J e^{i\vec q\.\vec y}.
\ee
For example, when $J=2$, $h_J(x,dx)=h_{\mu\nu} dx^\mu dx^\nu$ has an interpretation as a metric perturbation. In this case, (\ref{eq:spinningshockwave}) is the perturbation in the Aichelburg-Sexl shockwave metric (Fourier-transformed in the transverse space) \cite{Aichelburg:1970dh}. Comparing to (\ref{eq:spinningmomentumeigenstate}), we see that
\be
|\mathrm{shock}\> &= \int_{-\oo}^\oo dp^v \big|p^\mu=(0,p^v,\vec q), \e^\mu=\eta^{\mu u}=(0,-2,0)\big\>.
\ee

The computation of the shock amplitude and its commutator goes through essentially unchanged from the scalar case. To understand how the spinning amplitude should behave at large $t$, it is useful to tie the Mandelstam variable $t$ to a symmetry generator. Consider a boost parameterized by $z \in \C$,
\be
\label{eq:ourboost}
\Lambda(z) : (u,v,\vec y) &\to \p{\frac{u}{z},z v, \vec y}.
\ee
The boost acts on the momenta $p_1,p_2$ by rescaling $p^v\to z p^v$, which also rescales $t\to z t$.\footnote{The action of the boost~(\ref{eq:ourboost}) is closely related to a BCFW deformation \cite{Fu:2013cza}.} Thus, the shock commutator can be written
\be
\cQ(0^+)-\cQ(0^-) &= \frac{2\pi t}{p^u}\int_{-\oo}^\oo dz\,\Disc_t \cA(\e_1,\e_2,\e_3,\e_4;s,z t),
\ee
where 
\be
\label{eq:polarizationvectors}
\e_1=\e_2=(0,-2,0)
\ee
are the polarizations of the shocks, and $\e_3,\e_4$ are the polarization vectors of the other particles.

Let us define a boosted amplitude $\cA(z)$ by acting with $\Lambda(z)$ on particles $1$ and $2$, keeping particles $3$ and $4$ fixed.
Note that the boost acts nontrivially on the polarization vectors (\ref{eq:polarizationvectors}):
\be
\Lambda(z):\e_{1,2} &\to z\e_{1,2}.
\ee
Thus, 
\be
\cA(z) &= \cA(z\e_1,z\e_2,\e_3,\e_4;s,zt) = z^{J_1+J_2} \cA(\e_1,\e_2,\e_3,\e_4;s,zt).
\ee
In terms of the boosted amplitude, the shock commutator becomes
\be
\cQ(0^+)-\cQ(0^-) &= \frac{2\pi t}{p^u}\int_{-\oo}^\oo dz\,\frac{\Disc_t\cA(z)}{z^{J_1+J_2}}.
\ee

Suppose that $\cA(z)$ grows like $z^{J_0}$ at large $z$, where $J_0$ is a theory-dependent Regge intercept. (The Regge intercept may be $s$-dependent, but we are suppressing that dependence for now.) We see that the shock commutator vanishes if and only if
\be
\label{eq:conditiongrowth}
J_1+J_2 > J_0 + 1.
\ee

In the particular case where particles $1$ and $2$ are gravitons, the
shock commutator vanishes if $J_0<3$. It was argued in \cite{Camanho:2014apa} that --- for physical $s$ --- $J_0>2$ leads to a violation of causality.\footnote{Note that kinematics \cite{Camanho:2014apa} leads to amplitudes that grow with $t$ and thus do not admit superconvergence sum rules that we consider here. We illustrate this in appendix (\ref{sec:scalargravitonscat}).} Assuming this argument, it follows that coincident gravitational shocks commute and the superconvergence sum rule (\ref{eq:superconvergencesumrule}) holds in any causal theory. Furthermore, a failure of commutativity signals a violation of causality. It is instructive to see how (\ref{eq:superconvergencesumrule}) is obeyed (or not) in various examples.

We can also consider higher-point scattering amplitudes, where momenta $p_3^{\mu}$ and $p_4^\mu$ in the argument above stand for a sum of momenta of many particles. Assuming that the same bound on the Regge behavior holds for higher-point amplitudes we get higher-point analogs of the superconvergence relation (\ref{eq:superconvergencesumrule}), where the integral is taken over the discontinuity with respect to $t$ of higher-point amplitudes. We expect that commutativity of shocks should be true as ``an operator equation,'' in other words, for any scattering amplitude. This is what we find in AdS, where commutativity of shocks is dual to an operator equation in CFT. It would be interesting to explore this possibility further.

\subsection{Shock commutativity in General Relativity}
\label{sec:shocksineinstein}

Let us re-derive commutativity of shocks in General Relativity using scattering amplitudes and equation (\ref{eq:commutatorintegral}). 
The on-shell condition for the intermediate particle implies that $p^v = 0$ and $p_X = (p_X^u, 0 , \vec q_X)$, where $\vec q$ depends on the order of shocks. We choose the polarization of the intermediate particle to be $\e_X = (0, - 2 {(\vec e_X . \vec q) \over p^u}, \vec e_X)$ so that the sum over intermediate states $\sum_X \e_X^{\mu} (\e^* _X)^\nu$ acts as an identity matrix in the transverse space, $\sum_X e_X^{i} (e^* _X)^j = \delta^{i j}$. We write all amplitudes below in all-incoming notation.

\subsubsection{Minimally-coupled scalar}
\label{sec:minimalcoupledscalars}

As the simplest example, consider a massless scalar field minimally coupled to gravity. The relevant scalar-scalar-graviton three-point amplitude takes the form
\be
{\cal A}_{\phi_3 \phi_X g_1} = \e_1\. p_3\,\e_1\.p_X =- (p^u)^2 , 
\ee
where $\e_1=(0,-2,0)$ is the polarization of the graviton and we evaluated the amplitude in the kinematics of the previous section, in particular $p_X =( - p^u, 0, \vec q_2 - \vec q_1)$. Note that the three-point amplitude does not depend on the transverse momentum of the shock $\vec q_{i}$ and therefore the shock commutator vanishes. We find
\be
\label{eq:minimalscalarQ}
{\cal Q}_\mathrm{scalar}(0^\pm) =-  {2 \pi \over p^u} {\cal A}_{\phi_3 \phi_X g_1} {\cal A}_{\phi_4 \phi_X g_2} =-  2 \pi (p^u)^3 .
\ee

Finally, we can reach the same conclusion by considering a scalar field propagating on a shockwave background
\be
d s^2 &= - d u\, d v +  \delta(u) h(u, \vec y) d u^2 + d \vec y^2 , \cr
h(u, \vec y) &=4 \pi( \e_1 e^{i \vec y\. \vec q_1} + \e_2 e^{i \vec y\. \vec q_2}  ),
\ee
where $\vec q_1^2 = \vec q_2^2 = 0$.
The wave equation $\nabla^2 \phi = 0$ takes the form
\be
\pa_u \pa_v \phi + \delta(u) h(u, \vec y) \pa_v^2 \phi - {1 \over 4} \pa_i^2 \phi = 0 .
\ee
We solve across the locus $u=0$ as
\be
\phi_{{\rm after}} = e^{- h(0,\vec y) \pa_v} \phi_{{\rm before}} .
\ee
Choosing the initial state $\phi_{{\rm before}} = e^{- i {1 \over 2} v p^u} e^{- i \vec y \vec q_2}$ as in the amplitude computation, and focusing on the term linear in $\e_1 \e_2$ we get 
\be
\phi_{{\rm after}} &=\delta_{PS} e^{- i {1 \over 2} v p^u} e^{i \vec y \vec q_1} , \\
\delta_\mathrm{PS} &= - 2 \pi^2 (p^u)^2 ,
\ee
where $\delta_\mathrm{PS}$ is the phase shift acquired by crossing a shock.

To compare with the amplitude computation we must compute 
\be
\la a_{p_4^u, \vec q_4} | \phi_{{\rm after}} \ra = 2 p^u ( 2 \pi)^{D-1} \delta(p_4^u + p^u) \delta^{(D-2)}(\vec q_4 + \vec q_1)  \delta_\mathrm{PS}\, ,
\ee
which leads to 
\be
{\cal Q} = { p^u \delta_\mathrm{PS} \over \pi},
\ee
in complete agreement with the amplitude computation. Note also that in this way we can compute the effect of propagation of a probe through an arbitrary number of shocks. Again, we see that the order of shocks does not matter in this case.

\subsubsection{Minimally-coupled photon}

Next, let us consider particles with spin. It is convenient to choose external polarizations as follows
\be
\label{eq:threefourpolarizations}
\e_3 &= \p{0, - 2 {\vec q_2 \. \vec e_3 \over p^u} , \vec e_3} , \\
\e_4 &=\p{0, 2 {\vec q_1 \. \vec e_4 \over p^u} , \vec e_4} .
\ee

In Einstein-Maxwell theory, the three-point amplitude is given by
\be
{\cal A}^{F^2}_{\gamma_3 \gamma_X g_1} &= \e_1\.p_3\, \e_1\.p_X\,\e_3 \. \e_X - \e_1\.\e_3\,\e_1\.p_X\,\e_X\.p_3 - \e_1\.\e_X\,\e_1\.p_3\,\e_3\.p_X = - (p^u)^2 \vec e_3 \. \vec e_X ,
\ee
where $\e_X = (0, - 2 {(\vec e_X \. \vec q_2 - \vec q_1) \over p^u}, \vec e_X)$ and we used that $\e_1 \. \e_3 = \e_1 \. \e_X = 0$.

Summing over intermediate states as in (\ref{eq:commutatorintegral}) trivially gives
\be
\label{eq:qphoton}
{\cal Q}_\mathrm{photon}(0^+) =- {2 \pi \over p^u} \sum_{X} {\cal A}^{F^2}_{\gamma_3 \gamma_X g_1}  {\cal A}^{F^2}_{\gamma_4 \gamma_X g_2} = - 2 \pi (p^u)^3 \vec e_3 \. \vec e_4, \ 
\ee
which obviously does not depend on the order of the shocks:
\be
\sum_{X} \left( {\cal A}^{F^2}_{\gamma_3 \gamma_X g_1}  {\cal A}^{F^2}_{\gamma_4 \gamma_X g_2} -  {\cal A}^{F^2}_{\gamma_3 \gamma_X g_2}  {\cal A}^{F^2}_{\gamma_4 \gamma_X g_1} \right) = 0.
\ee

Again we can reproduce this result using the Einstein-Maxwell equations of motion on a shockwave background \cite{Camanho:2014apa}
\be
\label{eq:einsteinmaxwelleom}
\pa_u F_{v i} + \de(u) h\, \pa_v F_{v i} = 0 .
\ee
Taking our initial state to be a plane wave, let us focus on the transverse polarizations $\vec A_{{\rm before}}  = \vec e_3 e^{- i {1 \over 2} v p^u} e^{- i \vec y \vec q_2}$. After crossing the shock, we have
\be
\label{eq:eomprop}
\vec A_{{\rm after}} =  e^{- h(0,\vec y) \pa_v} \vec A_{{\rm before}}.
\ee
Again, computing the overlap we recover (\ref{eq:qphoton}).

\subsubsection{Gravitons}

Finally, the three-point function of gravitons in General Relativity takes the form
\be
\label{eq:einsteinamplitude}
{\cal A}^{R}_{g_3 g_X g_1 } &=( \e_1 \. \e_X \e_3 \. p_1 + \e_1\. \e_3 \e_X\.p_3 + \e_X\. \e_3  \e_1\.p_X)^2 = (p^u)^2 (\vec e_3 \. \vec e_X)^2 .
\ee
Summing over intermediates states we find
\be
\sum_{X} {\cal A}^{R}_{g_1 g_3 g_X}  {\cal A}^{R}_{g_2 g_4 g_X} = (p^u)^4 (\vec e_3 \. \vec e_4)^2 . 
\ee
The result is independent of the shock ordering. This result can be reproduced via computing the propagation of a graviton through a shockwave background as above.

To summarize, minimally-coupled matter and gravitons lead to commuting gravitational shocks (at tree level). This can be verified by studying shock amplitudes, or by studying wave equations on a shockwave background.

\subsection{Non-minimal couplings}
\label{sec:nonminimalflat}

\subsubsection{Non-minimally coupled photons}

Let us now demonstrate that commutativity of shocks can be lost in theories with non-minimal couplings to gravity.
As a first example, consider a non-minimal coupling of photons to gravity of the schematic form $\alpha_2 R F F$. The relevant three-point amplitude takes the form
\be
{\cal A}^{R F F}_{\gamma_3 \gamma_X g_1}  = \e_1\. p_3\, \e_1\. p_X \,\e_3\. p_X\,\e_X \. p_3 = -(p^u)^2 \vec e_3 \. \vec q_1 \,\vec e_X \. \vec q_1,
\ee
where we used that $\e_X \. p_3 = - \vec e_X \. \vec q_1$.

Summing over intermediate states, we can compute the commutator (\ref{eq:commutatorintegral})
\be
\label{eq:nonminimalcomm}
&{1 \over (p^u)^4} \sum_{X} \left( {\cal A}^{F^2+ \alpha_2 R FF}_{\gamma_3 \gamma_X g_1}  {\cal A}^{F^2 + \alpha_2 R FF}_{\gamma_4 \gamma_X g_2} -  {\cal A}^{F^2 + \alpha_2 R FF}_{\gamma_3 \gamma_X g_2}  {\cal A}^{F^2 + \alpha_2 R FF}_{\gamma_4 \gamma_X g_1} \right) \nn\\
&=\alpha_2^2 \left( \vec e_3 \. \vec q_1 \vec e_4 \. \vec q_2 - \vec e_3 \. \vec q_2 \vec e_4 \. \vec q_1 \right) \vec q_1 \. \vec q_2,
\ee
which is clearly nonvanishing. Let us reproduce the same result using equations of motion. In the presence of the $RFF$ coupling, the equation of motion (\ref{eq:einsteinmaxwelleom}) gets modified to
\be
\pa_u F_{v i} + \de(u)(\delta_{i j} h + \alpha_2 \pa_i \pa_j h ) \pa_v F_{v j} = 0.
\ee
The solution takes the same form as (\ref{eq:eomprop}), except now the transfer matrix that describes how polarizations change when propagating though a shock is not diagonal. Instead it takes the form $\delta_{i j} h + \alpha_2 \pa_i \pa_j h$, which for a given shock is 
\be
M_{i j}(\vec q) = \delta_{i j} - \alpha_2 q_i q_j .
\ee
Different orderings of shocks now lead to different results. The commutator (\ref{eq:nonminimalcomm}) becomes $e_4^i [M(\vec q_1) , M(\vec q_2)]_{i j} e_3^j$. More generally, given multiple shocks ordered according to $u_1 > u_2 > \cdots > u_k$, the amplitude is given by a corresponding product of (noncommuting) shockwave transfer matrices 
\be
\label{eq:multiplephotonshocks}
{\cal Q} ={p^u \over \pi} (4 \pi)^k \left( {i p^u \over 2} \right)^k  \vec e_4 \.  M ( \vec q_1) \cdots M(\vec q_k) \. \vec e_3 .
\ee

\subsubsection{Higher derivative gravity}
\label{sec:higherdergravity}

In higher-derivative gravity, there are two additional graviton three-point amplitudes:
\be
\cA^{R^2}_{g_3 g_X g_1} &= (\e_1\.\e_X\, \e_3\.p_1 + \e_1 \. \e_3\, \e_X\.p_3 + \e_X\.\e_3\, \e_1\.p_X) \e_1\.p_X\, \e_X\.p_3\, \e_3\.p_1  \nn\\
&= -(p^u)^2 \vec e_3\.\vec e_X\,\vec e_3\.\vec q_1\,\vec e_X\.\vec q_1,
\nn\\
\cA^{R^3}_{g_3 g_X g_1} &= (\e_1\.p_X\, \e_X\.p_3\, \e_3\.p_1)^2 \nn\\
&= (p^u)^2 (\vec e_3\.\vec q_1)^2 (\vec e_X\.\vec q_1)^2.
\ee
Together with $\cA^R$ given in (\ref{eq:einsteinamplitude}), these form a complete list of three-point structures allowed by Lorentz invariance. To contract the three point amplitudes we substitute
\be
\sum_{X} (e_X^{i} e_X^j)  (e_X^k e_X^l)^* \to \Pi^{i j, k l} \equiv {1 \over 2} \left( \delta^{i k} \delta^{j l} + \delta^{i l} \delta^{j k} \right) - {1 \over D-2} \delta^{i j} \delta^{k l}.
\ee 
For a general linear combination $R + \a_2 R^2 + \a_4 R^3$, the shock commutator is
\be
\label{eq:gravitoncommutatorflat}
\frac{1}{(p^u)^4}&\sum_{X} \left( {\cal A}^{R+\a_2 R^2 + \a_4 R^3}_{g_3 g_X g_1}  {\cal A}^{R+\a_2 R^2 + \a_4 R^3}_{g_4 g_X g_2} -  {\cal A}^{R+\a_2 R^2 + \a_4 R^3}_{g_3 g_X g_2}  {\cal A}^{R+\a_2 R^2 + \a_4 R^3}_{g_4 g_X g_1} \right) \nn\\
&= 
{1 \over 2} \alpha_{2}^2 \vec{e}_{3}\cdot \vec{e}_{4} \vec{q}_{1}\cdot \vec{q}_{2} \left(\vec{e}_{3}\cdot \vec{q}_{1} \vec{e}_{4}\cdot \vec{q}_{2} - \vec{e}_{3}\cdot \vec{q}_{2} \vec{e}_{4}\cdot \vec{q}_{1} \right) 
\nn\\
&\quad
-\alpha_{2} \alpha_{4} \vec{q}_{1}\cdot \vec{q}_{2} \left(\vec{e}_{3}\cdot \vec{q}_{1} \vec{e}_{4}\cdot \vec{q}_{2} - \vec{e}_{3}\cdot \vec{q}_{2} \vec{e}_{4}\cdot \vec{q}_{1} \right)  \left(\vec{e}_{3}\cdot \vec{q}_{1} \vec{e}_{4}\cdot \vec{q}_{1}+\vec{e}_{3}\cdot \vec{q}_{2} \vec{e}_{4}\cdot \vec{q}_{2} \right)
\nn\\
&\quad
+ \left( \alpha_{4}^2 (\vec{q}_{1}\cdot \vec{q}_{2})^2 - {\alpha_2^2 \over D-2}\right) \left[ (\vec{e}_{3}\cdot \vec{q}_{1})^2 (\vec{e}_{4}\cdot \vec{q}_{2})^2 - (\vec{e}_{3}\cdot \vec{q}_{2})^2 (\vec{e}_{4}\cdot \vec{q}_{1})^2 \right].
\ee
It is easy to check that this vanishes if and only if $\a_2=\a_4=0$.

Again the same result can be obtained using the classical equations of motions as above, see \cite{Camanho:2014apa}. The difference compared to General Relativity is that the shockwave transfer matrix is polarization-dependent, which leads to non-commutativity or violations of superconvergence relations in theories with non-minimal coupling to gravity. Explicitly, the transfer matrix is
\be
M_{i j , k l}(\vec q) &=
\Pi_{i j, k l} +\frac{\alpha_2}{4}\left(\delta_{ik}P_{jl}+\delta_{il}P_{jk}+
\delta_{jl}P_{ik}+\delta_{jk}P_{il} -4 {\delta_{i j} P_{k l} + \delta_{k l} P_{i j} \over D-2} \right) + \alpha_4 P_{ijkl}, \nn \\
P_{ij} &=q_i q_j, \nn \\
P_{ijkl} &=	q_i q_j q_k q_l  .
\ee
To compute the transfer through $n$ shocks we simply multiply the corresponding transfer matrices, as in (\ref{eq:multiplephotonshocks}). 

Non-minimal couplings of gravitons to matter also contribute to noncommutativity. For example, consider an interaction of the schematic form $\f R^2$ where $\f$ is a (possibly massive) scalar. The only possible three-point structure is
\be
\cA^{\f R^2}_{g_3 \f_X g_1} &= ((\e_1\.\e_3)(p_1\.p_3) - (\e_1\.p_3)(\e_3\.p_1))^2 = (p^u)^2 (\vec q_1\.\vec e_3)^2.
\ee
The $\phi R^2$ coupling allows $\phi$ to appear as an intermediate state when a graviton propagates through two shocks. The corresponding contribution to the shock commutator is
\be
\sum_{X} \left( {\cal A}^{\f R^2}_{g_3 \f_X g_1} {\cal A}^{\f R^2}_{g_4 \f_X g_2} -  {\cal A}^{\f R^2}_{g_3 \f_X g_2}  {\cal A}^{\f R^2}_{g_4 \f_X g_1} \right) &= (p^u)^4((\vec q_1\.\vec e_3)^2(\vec q_2\.\vec e_4)^2 - (\vec q_2\.\vec e_3)^2(\vec q_1\.\vec e_4)^2).
\ee

\subsection{Graviton scattering in string theory}
\label{sec:gravitonamplitudestring}

Non-minimal couplings generically lead to non-commuting coincident shocks (at tree level).
In any theory with Regge intercept $J_0 < 3$, the contributions from non-minimal couplings to the shock commutator must be cancelled by high-energy states or loop effects.

As a concrete example, consider graviton scattering in tree-level string theory. The amplitude takes the form \cite{Kawai:1985xq,Cai:1986sa}
\be
\label{scatamp}
\cA^{(i,j)}_{gg\to gg}
&= (K^{(i)}_{\mu_1 \mu_2 \mu_3 \mu_4} \e_1^{\mu_1}\e_2^{\mu_2} \e_3^{\mu_3} \e_4^{\mu_4})
(K^{(j)}_{\mu_1 \mu_2 \mu_3 \mu_4} \e_1^{\mu_1}\e_2^{\mu_2} \e_3^{\mu_3} \e_4^{\mu_4}) 
 \cr
&\quad \x(- \pi^2 \kappa^2) {\Gamma(-s/2) \Gamma(-t/2) \Gamma(-u/2) \over \Gamma(s/2 + 1) \Gamma(t/2 + 1) \Gamma(u/2 + 1)} \ \qquad (i,j\in \{b,ss\}),
\ee
in units where $\a'=2$. (Recall that we have parametrized the graviton polarizations as $\e^{\mu\nu}=\e^\mu \e^\nu$.)

There are two basic tensors $K^{(i)}$ that can appear in the amplitude: $K^{(b)}$, which occurs in the open bosonic string, and $K^{(ss)}$, which appears in the open superstring. Accordingly, the four-point amplitude of closed strings has three possible tensor structures, corresponding to bosonic strings $K^{(b)} K^{(b)}$, heterotic strings $K^{(b)} K^{(ss)}$, and superstrings $K^{(ss)} K^{(ss)}$.

Plugging in the momenta (\ref{eq:momentumthreefour}) and polarization vectors (\ref{eq:polarizationvectors}) and (\ref{eq:threefourpolarizations}), we find
\be
\label{computerel}
K_{\mu_1 \mu_2 \mu_3 \mu_4}^{(ss)} \e_1^{\mu_1} \e_2^{\mu_2} \e_3^{\mu_3} \e_4^{\mu_4} &= (p^u)^2 s\, \vec e_3 \. \vec e_4 , \cr
K_{\mu_1 \mu_2 \mu_3 \mu_4}^{(b)} \e_1^{\mu_1} \e_2^{\mu_2} \e_3^{\mu_3} \e_4^{\mu_4}
&= (p^u)^2 s  
\Big(\vec e_3 \. \vec e_4 - \vec e_3 \. \vec q_1\, \vec e_4 \. \vec q_1  - \vec e_3 \. \vec q_2\, \vec e_4 \. \vec q_2\nn\\
& \qquad\qquad+ {t \over 2 + u} \vec e_3 \. \vec q_2\, \vec e_4 \. \vec q_1+ {u \over 2 + t} \vec e_3\.\vec q_1\, \vec e_4 \. \vec q_2
\Big) .
\ee
Including only the contribution from graviton exchange, the last two terms in (\ref{computerel}) give a nontrivial shock commutator. However, if we boost the energy $t\to z t$, the gamma functions in (\ref{scatamp}) ensure that the amplitude dies as $z^{-2+s}$. Thus, upon integrating the full discontinuity, we must find that the graviton contribution is cancelled by heavy modes. For example, in the heterotic string, we find the sum rule
\be
\label{eq:heteroticsumrule}
\int d t\, {\rm Disc}_t \cA &=  (p^u)^4 s \vec e_3 \. \vec e_4( \vec e_3\.\vec q_1 \vec e_4 \. \vec q_2 - \vec e_3 \. \vec q_2 \vec e_4 \. \vec q_1 ) \left( 1 + \sum_{n=1}^\oo r_n(s) \right) = 0 ,
\ee
where
\be
r_n(s) &= {(s-2)(s+4n) \over s (n+1) (s+2n-2)} \frac {(s/2)_n^2}{n!^2},
\ee
with $(a)_n=\frac{\G(a+n)}{\G(a)}$ the Pochhammer symbol.
In equation (\ref{eq:heteroticsumrule}), $1$ represents the graviton multiplet contribution and the sum over $n\geq 1$ is the contribution of heavy modes.
It is possible to decouple almost all the heavy modes by setting $s=0$, in which case one finds
\be
\label{simple}
1 + r_1(0) = 0 .
\ee
In general, the sum over heavy modes converges like $\sum_n n^{s-3}$.

\subsection{Shock $S$-matrix in string theory}
\label{sec:shocksmatrix}

In addition to studying four-point amplitudes, we can equivalently analyze the equation of motion for a string on a shockwave background, analogous to our discussions of equations of motion for scalars, photons, and gravitons. Propagation of strings on shockwave backgrounds was discussed in \cite{AmatiWW,Horowitz:1989bv,Horowitz:1990sr,deVega:1990nr,Horowitz:1990ns,deVega:1990kk,deVega:1990gq,deVega:1991nm}. This leads to an understanding of shock commutativity in terms of the string $S$-matrix for propagation through a shock. 

Let us follow the conventions of \cite{AmatiWW}. Recall that the string mode operators obey
\be
[\alpha_n^i, \alpha_m^j ] = n \delta_{n+m,0} \delta^{ij},
\ee
where negative modes $n<0$ create string excitations, while positive modes $n >0$ annihilate the vacuum
\be
\alpha_{n>0} | 0 \ra = 0 .
\ee
We choose the conformal gauge for the worldsheet metric $h^{\alpha \beta} = \eta^{\alpha \beta}$, and fix the light-cone gauge
\be
u(\sigma, \tau) = P^{u} \tau .
\ee
The closed string mode expansion takes the form
\be
X^{i} (\sigma, \tau) = x^i + p^i \tau + {i \over 2} \sqrt{\alpha'} \sum_{n \neq 0} \left[ \tilde \alpha^i_n e^{- 2 i n \tau} - \alpha_n^{i \dagger} e^{2 i n \tau} \right] e^{- 2 i n \sigma},\quad (i=2,\dots,D-1),
\ee
where $ \alpha_n^{i \dagger} =  \alpha_{-n}^{i}$.

Before and after a shock, the string propagates freely. If the shock geometry has the metric
\be
ds^2 =- d u d v + \delta(u) f(\vec x) d u^2 + d \vec x^2 \, ,
\ee
the transition through the shock is described by the $S$-matrix \cite{AmatiWW},
\be
S_\mathrm{shock} = e^{{i \over 2 \pi} P^u \int_0^{\pi} f(\vec X(\sigma,0)) d \sigma }.
\ee
As an example, consider a shock created by a fast-moving particle at position $\vec x_a$,
\be
f_a (\vec X(\sigma,0)) =   \frac{\G(\frac{D-4}{2})}{4\pi^{\frac{D-2}{2}}}{1 \over ((\vec X(\sigma,0)- \vec x_a)^2 )^{\frac{D-4} 2} }  = \int { d^{D-2} \vec q  \over \vec q^2}  e^{i \vec q\. (\vec X(\sigma, 0)- \vec x_a)}.
\ee
In writing $f_a$ above, there is an ambiguity in the ordering of operators $X^i (\sigma, 0)$. However, this ambiguity is proportional to $\vec{q}_1^2$ and is localized at zero impact parameter upon doing the Fourier transform. It is therefore irrelevant for our purposes.

Note that the the operator $S_\mathrm{shock}$ is diagonal in the position basis $X^i(\s,0)$ for the transverse oscillators. Thus, it instantaneously changes the momenta of the oscillators without affecting their positions. Overall, the effect of the shock on the string is the same as in the geodesic calculation (\ref{eq:geodesiccalculation}): the center of mass of the string moves in the $v$ direction (Shapiro time delay), and the transverse modes receive an instantaneous kick that depends on the profile $f_a(\vec X)$. Essentially, each part of the string individually follows a geodesic through the shock. Thus, coincident shocks commute because they commute for geodesics.
 
To see this in more detail, consider the matrix element for propagation through two shocks
\be
\la \Psi | \left( {i \over 2 \pi} P^u \right)^2 \int_0^{\pi}  d \sigma f_1(\vec X(\sigma,0))\int_0^{\pi}  d \sigma'  f_2(\vec X(\sigma',0)) | \Psi \ra .
\ee
We are interested in states $|\Psi \ra$ that are eigenstates of $\vec x$ (fixed center of mass position in the transverse plane), with a finite number of oscillator excitations above the vacuum. As in the previous section, let us choose $\vec x | \psi_\mathrm{cm} \ra= \vec 0$. The relevant correlator  takes the form
\be
 &\int { d^{D-2} \vec{q}_1  \over \vec{q}_1^2} e^{- i \vec{q}_1\. \vec x_1}  \int { d^{D-2} \vec{q}_2
   \over \vec{q}_2^2} e^{- i \vec{q}_2 \.\vec x_2}
     \int_0^{\pi}  d \sigma \int_0^{\pi}  d \sigma'  \la \psi_\mathrm{osc} |   e^{i \vec{q}_1 \.\vec X_\mathrm{osc}(\sigma, 0)}   e^{i \vec{q}_2 \.\vec X_\mathrm{osc}(\sigma', 0)}   | \psi_\mathrm{osc} \ra .
\label{XtwopointB}
\ee
The same formula is valid for superstrings as well \cite{AmatiWW}. 

Let us first compute the correlator in the oscillator vacuum. We have
\be
\la 0_\mathrm{osc} |  e^{i \vec{q}_1 \.\vec X_\mathrm{osc}(\sigma, 0)}  e^{i \vec{q}_2 \. \vec X_\mathrm{osc}(\sigma', 0)}   | 0_\mathrm{osc} \ra = | 2 \sin (\sigma - \sigma')  |^{{1 \over 2} \alpha' \vec{q}_1 \. \vec {q}_2} .
\ee
The integral over the worldsheet coordinate gives \cite{Hofman:2008ar}
\be
\int_0^{2\pi}  {d \sigma \over 2 \pi}  \left| 2 \sin {\sigma  \over 2} \right|^{\alpha' \vec{q}_1 \. \vec {q}_2} = {2^{\alpha' \vec{q}_1 \. \vec {q}_2} \over \sqrt \pi} {\Gamma( {1 \over 2} + {\alpha' \vec{q}_1 \. \vec {q}_2 \over 2}) \over \Gamma(1+ {\alpha' \vec{q}_1 \. \vec {q}_2 \over 2})} .
\ee
Expanding at small $\alpha'$ and plugging back into (\ref{XtwopointB}) reproduces the result from \cite{Hofman:2008ar}. 
In general states, we can use the formula
\be
 : e^{i \vec{q}_1 \. \vec X_\mathrm{osc}(\sigma, 0)} :  : e^{i \vec{q}_2 \. \vec X_\mathrm{osc}(\sigma', 0)} : = | 2 \sin (\sigma - \sigma')  |^{{1 \over 2} \alpha' \vec{q}_1 \. \vec {q}_2}   : e^{i \vec{q}_1 \. \vec X_\mathrm{osc}(\sigma, 0) + i \vec{q}_2 \. \vec X_\mathrm{osc}(\sigma', 0)} : 
\ee
and then Taylor expand inside the normal ordering. In this way, commutativity is manifest.

Let us now understand how commutativity is achieved in more detail. For concreteness, consider bosonic string theory and take a graviton as the external state
\be
| \psi_\mathrm{osc} \ra = \tilde \alpha_{-1}^{a}  \alpha_{-1}^{b} | 0 \ra .
\ee
It is sufficient to keep only the leading $\alpha'$ correction to see the effect:
\be
\int_0^{\pi}  {d \sigma \over \pi}  e^{i \vec q\. \vec X_\mathrm{osc}(\sigma, 0)} \ & \sim\   1+ \frac {\alpha'} 4  q_i q_j T^{i j},
\label{eq:shockalphaprimexpansion}
\ee
where
\be
T^{i j} &= \sum_{n>0} (\tilde \alpha_{-n}^{(i} - \alpha_n^{(i}) (\tilde \alpha_{n}^{j)} - \alpha_{-n}^{j)}),
\ee
where $(i j)$ stands for symmetrization. It is easy to check that $[T^{ij},T^{kl}]=0$.

Let us act with the operator (\ref{eq:shockalphaprimexpansion}) on a graviton state
\be
\left( 1 + \frac{\alpha'}{4} q_i q_j T^{i j} \right) \tilde \alpha_{-1}^{a}  \alpha_{-1}^{b} | 0 \ra  .
\ee
Inside the operator $T^{i j}$, there are two types of terms. Firstly, the terms $\tilde \alpha_{-1}^{(i} \tilde \alpha_{1}^{j)} +  \alpha_{-1}^{(i} \alpha_{1}^{j)} $ shuffle massless modes among each other. These by themselves lead to non-commutative shocks.  However, crucially there are also terms $- \tilde \alpha_{n}^{(i} \alpha_{n}^{j)} -  \alpha_{-n}^{(i} \tilde \alpha_{-n}^{j)}$ which move the state across the string levels. In particular, the first term leads to mixing with the tachyon, whereas the second term produces higher level states. For example, the $n=1$ term leads to mixing between the graviton and a spin-4 particle.  As expected, the extra states restore commutativity.

\subsection{A stringy equivalence principle}

We have seen that in string theory, extra states restore commutativity of coincident shocks and satisfy the corresponding superconvergence sum rule.  This phenomenon should occur in any gravitational theory with $J_0 < 3$, where $J_0$ is the Regge intercept. We give more details on the corresponding superconvergence sum rule in a generic tree-level theory of gravity in appendix~\ref{sec:gravitationalsumrulegeneric}. In a non-tree-level theory with $J_0 < 3$, the superconvergence sum rule can receive contributions from loops or nonperturbative effects. 

Reference~\cite{Camanho:2014apa} argued using causality that a tree-level theory of gravity should have $J_0 \leq 2$.  Their argument applies for scattering at nonzero impact parameter. As far as we are aware, there is currently no flat-space argument that the same should be true away from tree level, or at zero impact parameter.\footnote{It was proven in the 60's that scattering amplitudes in gapped QFTs satisfy dispersion relations with at most two subtractions for $|t|<R$ \cite{Martin:1965jj}, where $R$ depends on the mass of the lightest particle in the spectrum. A similar statement holds for scattering of particles with spin \cite{Mahoux:1969um,deRham:2017zjm}. The corresponding superconvergence sum rules for spinning particles were studied in \cite{Mahoux:1969um}. It would be interesting to understand the relation between their work and the commutativity of gravitational shocks discussed here.}

However, as we will see in the next section, shock commutativity in AdS can be proved rigorously, nonperturbatively, and for all values of the impact parameter. This leads us to conjecture that the same is true in flat space and dS as well. More precisely, we propose
\begin{conjecture}[Stringy equivalence principle]
Coincident gravitational shocks commute in any nonperturbative theory of gravity in AdS, dS, or Minkowski spacetime.
\end{conjecture}
We use the term ``equivalence principle" because this is a modified version of the statement that all particles follow geodesics. The word ``stringy" comes from the fact that mixing with stringy states can restore shock commutativity that would otherwise be lost.

\subsection{Gravitational DIS and ANEC commutativity}
\label{sec:gravitationalDIS}

We can easily repeat the same discussion in the context of a gapped QFT (or, more generally, a QFT that is free in the IR), where it becomes the statement about commutativity of the ANEC operators when evaluated in one-particle states. The virtual graviton $g^*$ of the previous sections couples to the QFT stress-energy tensor as $h_{\mu \nu} T^{\mu \nu}$ and therefore the scattering process $g^* X \to g^* X$ is described by the following matrix element
\be
\label{eq:matrixDIS}
{\cal A}(s,t) = \la p_4 | {\rm T} \left( T_{vv}(p_1) T_{vv}(p_2) \right) | p_3 \ra ,
\ee
where as usual (\ref{eq:matrixDIS}) describes the nontrivial part that multiplies $\delta^{(D)}(\sum_i p_i)$.\footnote{For a related discussion see \cite{Hofman:2008ar, Komargodski:2012ek, Komargodski:2016gci}.}
Unlike gravitational theories, there is no problem in defining off-shell observables in QFT and decoupling the probe that creates $g^*$ from the rest of the system (by considering the $G_{N} \to 0$ limit). One consequence of that is that we can formulate the problem using real momenta and keep $g^*$ off-shell. As before we will be interested in the following kinematics
\be
\label{eq:momentumDIS}
p_1 &= (0,-p^v+{\vec q_1^2 \over p^u},\vec q_1), ~~~
p_2 = (0,p^v -{\vec q_2^2 \over p^u},\vec q_2), \nn\\
p_3 &= ( p^u, {\vec q_2^2 \over p^u} , -\vec q_2), ~~~
p_4 = (- p^u, -{\vec q_1^2 \over p^u}, -\vec q_1) ,
\ee
where $\vec q_1^2$ and $\vec q_2^2$ are non-zero. We chose the probe particle to be massless but it does not have to be the case. Mandelstam invariants take the form $s=-(p_1+p_2)^2 = (\vec q_1 + \vec q_2)^2$ and $t = -(p_2+p_3)^2 = p^u p^v$. The matrix element that describes $g^* X \to g^* X$ is time-ordered as in the usual description of deep inelastic scattering. 
The discontinuities of the time-ordered matrix element ${\cal A}(s,t)$ are computed by the corresponding Wightman functions
\be
{\rm Disc}_{t>0} {\cal A}(s,t) = \la p_4 | T_{vv}(p_1) T_{vv}(p_2) | p_3 \ra , \nn\\
-{\rm Disc}_{t<0} {\cal A}(s,t) = \la p_4 | T_{vv}(p_2) T_{vv}(p_1) | p_3 \ra .
\ee
Integrating the discontinuity over $t$ we get
\be
\label{eq:commutatorANECQFT}
{1 \over p^u} \int_{- \infty}^{\infty} dt {\rm Disc}_t {\cal A}(s,t) = \int d^{d-2} \vec y_1 e^{i \vec q_1 \cdot \vec y_1} \la p_4 | \left[ \int dv T_{vv}(u=0,v, \vec y_1) , T_{vv}(0) \right] | p_3 \ra ,
\ee
where we used momentum conservation to rewrite the result in position space. We see that the integral over the discontinuity of ${\cal A}(s,t)$ is related to the commutator of ANEC operators inserted at the same time $u=0$.

As in the scattering amplitude considerations of the previous section, causality considerations apply to the matrix element (\ref{eq:matrixDIS}).
In particular, for physical $s$ we expect ${\cal A}(s,t)$ to obey
\be
\label{eq:ReggeDIS}
|{\cal A}(s,t) | < {1 \over | t |} , ~~~ |t| \to \infty .
\ee

The Regge boundedness condition (\ref{eq:ReggeDIS}) together with the usual assumptions about the analyticity of ${\cal A}(s,t)$ implies commutativity of ANECs in a gapped QFT via the superconvergence relations
\be
0 = \oint {d t \over 2 \pi i} {\cal A}(s,t) =  \int_{- \infty}^{\infty} dt {\rm Disc}_t {\cal A}(s,t) + \oint_{{\cal C}_{\infty}} {d t \over 2 \pi i} {\cal A}(s,t) = \int_{- \infty}^{\infty} dt {\rm Disc}_t {\cal A}(s,t) .
\ee
We conclude that for every $\vec q_1$, the commutator 
\be
\int d^{d-2} \vec y_1 e^{i \vec q_1 \cdot \vec y_1} \la p_4 | [ \int dv T_{vv}(u=0,v, \vec y_1) , T_{vv}(0)] | p_3 \ra
\ee vanishes. This implies that coincident ANEC operators commute inside one-particle states.

\section{Event shapes in CFT and shocks in AdS}
\label{sec:shocksinads}

In this section, we study shock commutativity and superconvergence sum rules in AdS, interpreting them in CFT language. We focus on shocks created by integrating a local CFT operator along a null line on the boundary of AdS.\footnote{These backgrounds will already be rich enough to develop an analogy to the flat-space story. It would be interesting to study other types of shock backgrounds \cite{Afkhami-Jeddi:2017rmx} in the future.} The simplest example is a gravitational shock created by the average null energy (ANEC) operator $\cE=\int dv\, T_{vv}$ \cite{Hofman:2008ar}. We will argue that ANEC operators on the same null plane commute, and this leads to nontrivial superconvergence sum rules that must be satisfied by CFT data. One of the nice properties of such superconvergence sum rules is that in large-$N$ theories, they get contributions only from single-trace operators and non-minimal bulk couplings.

We start by introducing null integrals (``light-transforms") of local operators in section~\ref{sec:lighttransform}. In section~\ref{sec:eventshapreview}, we review ``event shapes," which are certain matrix elements of light-transformed operators. In section~\ref{sec:computinginbulk}, we compute some simple event shapes in AdS, emphasizing the similarities to our shock amplitude calculations in section~\ref{sec:shocksinflatspace}.

\subsection{Review: the light transform}
\label{sec:lighttransform}

We will be interested in integrals of a local CFT operator along a null line on the boundary of AdS. For example, let $\cO^{\mu_1\cdots\mu_J}$ be a traceless symmetric tensor, and consider the integral
\be
\label{eq:nullintegratedoperator}
\int_{-\oo}^\oo dv\,\cO_{v\cdots v}(u=0,v,\vec y).
\ee
Here, we use lightcone coordinates (\ref{eq:lightconeycoordinates}), except we are now in $d=D-1$ dimensions (so that $\vec y\in \R^{d-2}$). In holographic theories, such operators create shocks in the bulk.
An example is the average null energy (ANEC) operator
\be
\label{eq:anecdefinition}
\int_{-\oo}^\oo dv\,T_{vv}(u=0,v,\vec y).
\ee
The ANEC operator on the boundary creates a gravitational shock in the bulk of the form described by the AdS-Aichelburg-Sexl metric.

In the examples (\ref{eq:nullintegratedoperator}) and (\ref{eq:anecdefinition}), the integration contour starts at the point $(u,v,\vec y)=(0,-\oo,\vec y)\in\mathscr{I}^-$  at past null infinity and ends at the point $(u,v,\vec y)=(0,\oo,\vec y)\in\mathscr{I}^+$ at future null infinity. More generally, we can perform a conformal transformation to bring the initial point of the null integral to some generic point $x$. The result is an integral transform called the light-transform \cite{Kravchuk:2018htv},
\be
\label{eq:lighttransformdefinition}
\wL[\cO](x,z) &= \int_{-\oo}^\oo d\a (-\a)^{-\De-J} \cO\p{x-\frac{z}{\a},z}.
\ee
Here, $z\in \R^{1,d-1}$ is a null vector, $\De$ and $J$ are the dimension and spin of $\cO$, and we use index-free notation
\be
\cO(x,z) &= \cO^{\mu_1\cdots\mu_J}(x)z_{\mu_1}\cdots z_{\mu_J}.
\ee
The light-transformed operator depends on an initial point $x$ and a null direction $z$. The integration contour runs from $x$, along the $z$-direction, to the point in the next Poincare patch on the Lorentzian cylinder with the same Minkowski coordinates as $x$. 

An advantage of this language is that it makes the conformal transformation properties of null-integrated operators manifest. $\wL$ is a conformally-invariant integral transform that changes the quantum numbers as follows:
\be
\wL:(\De,J) &\to (1-J,1-\De).
\ee
In other words, $\wL[\cO](x,z)$ transforms like a primary operator at $x$ with dimension $1-J$ and (non-integer) spin $1-\De$. As we will see, this simplifies several computations involving these operators. We see from (\ref{eq:lighttransformdefinition}) that the light-transform is well-defined whenever $\De+J>1$. An important property is that $\wL[\cO]$ annihilates the vacuum whenever it's defined. This can be established formally using the conformal transformation properties of $\wL[\cO]$, or by deforming the $\a$ contour inside a correlation function involving $\wL[\cO]$ \cite{Kravchuk:2018htv}.

\subsection{Review: event shapes}
\label{sec:eventshapreview}

We will be interested in matrix-elements of light-transformed operators called ``event shapes." First, consider a three-point function
\be
\<\Omega|\f_1(x_1) \wL[\cO](x,z) \f_2(x_2)|\Omega\>,
\ee
where we take $\f_1,\f_2$ to be primary scalars for simplicity. This transforms like a three-point function of primary operators, which is fixed by conformal invariance up to an overall constant. Without loss of generality, we can place $x$ at spatial infinity $x\to \oo$ so that the light-transform contour runs along $\mathscr{I}^+$,
\be
\label{eq:lighttransformatinfinity}
\<\Omega|\f_1(x_1) \wL[\cO](\oo,z) \f_2(x_2)|\Omega\>
&\equiv
\lim_{x\to \oo} x^{2(1-J)} \<\Omega|\f_1(x_1) \wL[\cO](x,I(x)z) \f_2(x_2)|\Omega\>\nn\\
&= \cM(x_1-x_2).
\ee
The factor $x^{2(1-J)}$ ensures a finite result as $x\to \oo$. (Recall that $\wL[\cO]$ transforms like a primary with dimension $1-J$.) We must act on the polarization vector $z$ with the inversion matrix $I^\mu{}_\nu(x)=\de^{\mu}_\nu - \frac{2x^\mu x_\nu}{x^2}$, so that $x_1,x_2$ and $z$ transform in the same way under Lorentz transformations.

The operator $\wL[\cO](\oo,z)$ transforms like a primary inserted at spatial-infinity, which means it is annihilated by momentum generators
\be
[P^\mu,\wL[\cO](\oo,z)] = 0.
\ee
Hence, the matrix element (\ref{eq:lighttransformatinfinity}) is translationally-invariant, which is why we have written $\cM(x_1-x_2)$ in (\ref{eq:lighttransformatinfinity}). Recall that $\f_1,\f_2$ are operator-valued distributions, so $\cM(x_1-x_2)$ is really a translationally-invariant integral kernel that can be paired with test functions $f_1(x_1), f_2(x_2)$. This kernel can be diagonalized by going to momentum space. Let us define the Fourier-transformed states
\be
|\f(p)\> &= \int d^d x\, e^{ip\.x} \f(x)|\Omega\>.
\ee
Positivity of energy implies that $|\f(p)\>$ is nonvanishing only if $p$ is inside the forward lightcone. The event shape is
\be
\<\f_1(q)|\wL[\cO](\oo,z)|\f_2(p)\> &= (2\pi)^d \de^d(p-q) \tl \cM(p).
\ee
We often abuse notation and write
\be
\tl \cM(p) &= \<\f_1(p)|\wL[\cO](\oo,z)|\f_2(p)\>,
\label{eq:onepointeventshape}
\ee
where it is understood that we have stripped off the momentum-conserving $\de$-function. 

The physical interpretation of $\tl \cM(p)$ is that the state $|\f_2(p)\>$ acts like a source in Minkowski space. Excitations from the source fly out to $\mathscr{I}^+$, where they hit the ``detector" $\wL[\cO](\oo,z)$. Here, $z$ specifies a particular direction on the celestial sphere $S^{d-2}$ where the detector sits. Finally, we take the overlap of the resulting state with the sink $\<\f_1(p)|$. The correlator (\ref{eq:onepointeventshape}) is called a one-point ``event shape" because it involves a single detector.

More generally, we can consider an expectation value of multiple detectors at $\mathscr{I}^+$, i.e.\ a multi-point event shape
\be
\<\f_1(p)|\wL[\cO_1](\oo,z_1)\cdots \wL[\cO_n](\oo,z_n)|\f_2(p)\>.
\ee
Multi-point event shapes involve a limit where the initial and final points of the light-transform contours become coincident. Specifically, all detector integration contours start at spatial infinity and end at future infinity. We discuss the conditions under which this limit is well-defined in section~\ref{sec:commutativityone}.

Future null infinity $\mathscr{I}^+$ is conformally equivalent to a null plane. For example, by performing a null inversion, we can bring $\mathscr{I}^+$ to the plane $u=0$ in the $(u,v,\vec y)$ coordinates
\be
\label{eq:nullinversion}
(u,v,\vec y) &= \p{-\frac{1}{x^-},x^+-\frac{\vec x^2}{x^-},-\frac{\vec x}{x^-}}.
\ee
In these coordinates, the event shape takes the form
\be
\label{eq:eventshapey}
\<\f_1(p)|\frac{1}{(1+n_1^{d-1})^{\De_1-1}}\int dv_1\, \cO_{1\,v\cdots v}(0,v_1,\vec y_1)\cdots \frac{1}{(1+n_n^{d-1})^{\De_n-1}}\int dv_n\, \cO_{n\,v\cdots v}(0,v_n,\vec y_n)|\f_2(p)\>.
\ee
Note that the sink and source states are still defined via a Fourier transform with respect to $x$. The transverse position of the detectors $\vec y_i$ are related to the null vectors $z_i$ by stereographic projection,
\be
\label{eq:dictionarybetweenframes}
z &= \p{1,\vec n},\quad \vec n = (\vec n_\perp, n^{d-1}) = \p{\frac{2\vec y}{1+\vec y^2}, \frac{1-\vec y^2}{1+\vec y^2}} \in S^{d-2}\, .
\ee

A consequence of writing the event shape in the form (\ref{eq:eventshapey}) is that it makes clear that the operators $\cO_i$ remain spacelike-separated along their integration contours. This ostensibly implies that detectors should commute. However, this argument ignores the fact that the operators become coincident at the ends of their integration contours. The question of commutativity of detectors is more subtle, and is related to singularities of four-point functions as we discuss in section~\ref{sec:commutativityone}.

\subsection{Computing event shapes in the bulk}
\label{sec:computinginbulk}

To understand the connection between event shapes and flat-space shock amplitudes, let us review how event shapes are computed in theories with a gravity dual \cite{Hofman:2008ar,Buchel:2009sk, Camanho:2009vw, Myers:2010jv}.\footnote{We thank Xián Camanho, Jose Edelstein, Diego Hofman and Juan Maldacena for discussions on this topic.} For simplicity, we focus on energy detectors. A multi-point event shape for ANEC operators $\cE$ is called an ``energy correlator." We will see that commutativity of energy detectors is essentially equivalent to the coincident shock commutativity discussed in section~\ref{sec:shocksinflatspace}.

To begin, let us separate the detectors in the $u$ direction, so that $\wL[\cO_i]$ is at position $u_i$ with the ordering $u_1>u_2>\dots>u_k$. The insertion of an integrated stress tensor $\sum_i\lambda_i \int_{- \infty}^{\infty} d v\, T_{vv}(u_i,v, \vec y_i)$ in a CFT is dual to the shockwave geometry
\be
\label{eq:adsshock}
d s^2 &={- d u \,d v + \sum_{i=1}^{d-2} d \vec y^2 +  d z^2   + \sum_i \delta(u - u_i) h_i(\vec y,z ) (d u)^2\over z^2} \, , \\
h_i(\vec y,z) &=\lambda_i {2^{d-1} \over \vol S^{d-2}}  {z^{d} \over (z^2 + |\vec y - \vec y_i|^2 )^{d-1} } \, .
\label{eq:shockprofile}
\ee
To derive the proportionality constant in (\ref{eq:shockprofile}), recall that according to the standard AdS/CFT dictionary, the source for the stress tensor is encoded in the ${1 \over z^2}$ deformation of the metric as $z \to 0$. One can check that (\ref{eq:shockprofile}) corresponds to a source $\l_i \delta(u - u_i) \delta^{(d-2)}(\vec y - \vec y_i)$ for $T_{vv}(y)$.

A remarkable property of the metric (\ref{eq:adsshock}) (or any superposition of such shock waves) is that it is an exact solution of Einstein's equations, even when arbitrary higher derivative corrections are included \cite{Horowitz:1999gf}. Here, we used the $y=(u,v,\vec y)$ coordinate system, which is related to the coordinate $x$ by null inversion~(\ref{eq:nullinversion}). The locus $u=0$ is the Poincare horizon of the original Poincare patch in $x$.  
We define energy detectors as
\be
\cE(\vec n) &\equiv 2 L[T](\oo,(1,\vec n)), \quad \vec n \in S^{d-2}.
\ee
In the $\vec y$ conformal frame, this is 
\be
{\cal E}(\vec n) = {2 \over (1+n^{d-1})^{d-1} } \int_{- \infty}^{\infty} d v\, T_{vv}(u_i,v, \vec y_i) ,
\ee
where $\vec y = {\vec n_\perp \over 1+ n^{d-1}}$ and $\vec n = (\vec n_\perp , n^{d-1})$.

To compute energy correlators, we need to compute an overlap between states before and after propagation through a series of shocks
\be
\cA(\lambda_i) = \la \Psi | \Psi' \ra,
\ee
where $|\Psi'\>$ represents the state after propagation through shocks.
We compute the amplitude with shocks at locations $u_1>u_2>\dots>u_k$, and then subsequently take $u_i\to 0$. The energy correlator is given by
\be
\label{eq:bulkcomp}
\la \Psi | {\cal E}(\vec n_1) \cdots {\cal E}(\vec n_k)|\Psi \ra = \prod_{i=1}^k \left.{2 \over (1+n_{i}^{d-1})^{d-1}  } (-i \pa_{\lambda_1} ) \cdots (-i \pa_{\lambda_k} ) \cA(\lambda_i) \right|_{\lambda_i = 0} .
\ee

We are interested in states dual to a single-trace operator insertion with momentum $q$:
\be
| \Psi \ra = \int d x\, e^{i q x} {\cal O}(x) | \Omega \ra.
\ee
This corresponds to a single-particle state in the bulk. When the particle crosses a shock, it could lead to particle production. However, particle-production is suppressed to leading order in $1/c_T$. Instead, the leading-order effect is mixing with other single-particle states. 

Mixing is captured by one-point energy correlators $\la \Psi_2 | {\cal E}(\vec n_1) |\Psi_1 \ra$. To understand these, let us study in more detail the propagation of fields on a shockwave background. This problem was considered in \cite{Hofman:2008ar}. 
Consider a shock located at $u=0$. Without loss of generality, we can set $q=(q^0,0,\dots,0)$. Using the bulk-to-boundary propagator, we can compute the wave-function of the particle in the bulk. In the vicinity of $u=0$, it takes the form
\be
\label{eq:particlewavefunction}
\phi(u,v,\vec y,z) \sim e^{-i \frac{q^0}{2} v} \delta(z - 1) \delta^{(d-2)}(\vec y).
\ee
In other words, the particle crosses $u=0$ at a fixed transverse location. In (\ref{eq:particlewavefunction}) we used the AdS $y$-coordinates, see formula (3.3) in \cite{Hofman:2008ar}. The location of the probe particle in the radial direction is related to its momentum. For $\vec q = 0$ we get $z=1$.

Before and after the shockwave, the scalar field propagates freely. At the location of the shock, however, it changes discontinuously. The 
change is dictated by the equations of motions, which for the minimally coupled scalar field $\phi$ become
\be
\label{eq:scalareom}
\pa_u \pa_v \phi + \delta(u) h(\vec y,z) \pa_v^2 \phi + ... = 0 .
\ee
where we only kept the terms that contribute to the $\delta(u)$ discontinuity. The matching condition for the field across the shock is obtained by integrating (\ref{eq:scalareom}) over a small interval $u \in [-\e,\e]$, with the result
\be
\label{eq:matchingcondition}
\phi_\mathrm{after}(0,v,\vec y,z) = e^{- h \pa_v} \phi_\mathrm{before}(0,v,\vec y,z).
\ee
Plugging in the wavefunction (\ref{eq:particlewavefunction}), we see that (\ref{eq:matchingcondition}) becomes multiplication by a phase
\be
\phi_\mathrm{after}(0,v,\vec y,z) \sim e^{i \frac{q^0}{2} h(\vec y=\vec 0,z=1)} \f_\mathrm{before}(0,v,\vec y,z).
\ee
Finally, using (\ref{eq:bulkcomp}) and the expression (\ref{eq:shockprofile}) for $h$, we obtain
\be
\label{eq:bulkonepointscalar}
\<\cE(\vec n)\>_{\cO(q)} &\equiv \frac{\<\Psi|\cE(\vec n)|\Psi\>}{\<\Psi|\Psi\>} = \frac{q^0}{\vol S^{d-2}}.
\ee
This calculation is almost identical to our calculation of the amplitude for a minimally-coupled scalar to cross a shock in flat-space in section~\ref{sec:minimalcoupledscalars}.

To compute a $k$-point energy correlator, we can imagine propagating the particle through a series of shocks at $u_1 > u_2 > \dots > u_k$, and then taking the limit $u_i\to 0$. Alternatively, we can simply multiply a series of one-point correlators (\ref{eq:bulkonepointscalar}). Either way, the result is
\be
\label{eq:ansGR}
\la{\cal E}(\vec n_1) \cdots {\cal E}(\vec n_k) \ra_{\cO(q)} \equiv {\la \Psi | {\cal E}(\vec n_1) \cdots {\cal E}(\vec n_k)|\Psi \ra \over \la \Psi | \Psi \ra } =  \left({q^0 \over \vol S^{d-2}  } \right)^k .
\ee

The situation becomes more interesting if we include higher derivative corrections or mixing between fields. For example, consider a massless vector field $A_\mu$. As discussed in \cite{Camanho:2014apa}, the most general equation of motion for $A_\mu$ takes the form
\be
\nabla^{\mu} F_{\mu \nu} - {a_2 \over d (d-1) } \hat R_{\nu}^{\ \mu \alpha \beta} \nabla_{\mu} F_{\alpha \beta} = 0  ,
\ee
where $a_2$ is a non-minimal coupling constant, and following \cite{Horowitz:1999gf} we introduced the effective shockwave Riemann tensor $\hat R_{\nu}^{\ \mu \alpha \beta}$ via
\be
R_{\mu \nu \rho \sigma} = - g_{\mu \rho} g_{\nu \sigma} + g_{\nu \rho} g_{\mu \sigma} + \hat R_{\mu \nu \rho \sigma}.
\ee
The shockwave Riemann tensor is given by
\be
\hat R_{\mu \nu \rho \sigma} = l_{[\mu} K_{\nu][\rho}l_{\sigma]},
\ee
where $l_{\mu} = \pa_{\mu} u$ and $K_{\mu \nu}$ is symmetric, satisfies $K_{\mu \nu} l^{\mu} = 0$, and takes the form
\be
K_{z z} &={1 \over 2} \left(  z^{-2} \pa_{z}^2 h - z^{-3} \pa_{z} h  \right) ,\nn\\
K_{z i} &={1 \over 2}{\pa_i \pa_{z} h \over z^2} ,\nn \\
K_{i j} &={1 \over 2} \left(z^{-2} \pa_i \pa_j h - \delta_{i j} z^{-3} \pa_z h \right) .
\ee
For us, the relevant component of $\hat R_{\mu \nu \rho \sigma}$ is
\be
\hat R_{+ i j +} |_{\vec y = 0, z = 1} = {d (d-1) \over 2} \left( {1+n^{d-1} \over 2} \right)^{d-1} \left( n_i n_j - {\delta_{i j} \over d-1}\right).
\ee
Note that the indices $i$ and $j$ above go from $i,j = 1, ... , d-2$, whereas the problem naturally has $\SO(d-1)$ symmetry. It is easy to make this symmetry manifest by going from planar coordinates to the coordinates of \cite{Horowitz:1999gf}.

Consider now a perturbation to the transverse components of $A_\mu$,
\be
A_i  (u,v,\vec y,z) &\sim \e_i e^{-i {q^0 v \over 2}} \delta^{d-2} (\vec y) \delta(z - 1),
\ee
where $\e_i$ are components of a polarization vector $\e_\mu$.
The equations of motion become
\be
\label{eq:transfervector}
& \pa_u \pa_v A_i + \delta(u) \hat H_{i j} \pa^2_v A_j + ... =0, \\
\hat H_{i j} &=\sum_{a=1}^{k} {\lambda_a  \over \Omega_{S^{d-2}} }  \left( 1+n_a^{d-1} \right)^{d-1} (\hat H_{a})_{i j} , \\
( H_{a})^{i j} &\equiv \left[ \delta^{i j} + a_2 \left( n_a^i n_a^j - {\delta^{i j} \over d-1} \right) \right].
\label{eq:transfermatrixA}
\ee
The solution is simply
\be
A_i^\mathrm{after} = \left[ e^{- \hat H \pa_v} A^\mathrm{before} \right]_{i} . 
\ee
The formulas above are sufficient to compute a one-point energy correlator for the transverse components of $A_\mu$. However, note that the full problem additionally involves the component $A_z$, which also gets excited by crossing the shock. 
As we remarked above, its behavior is fixed by $\SO(d-1)$ symmetry. The effect of including $A_z$ should be to enlarge the transfer matrix $H_{ij}$ to include the $(d-1)$'th component of $n^{\mu}$. The same formula (\ref{eq:transfermatrixA}) applies, except now the indices $ij$ run from $1$ to $d-1$. 

We can now compute the energy correlator (\ref{eq:bulkcomp}). As before, let us order the shocks such that $u_1 > u_2 > \dots > u_k$ and then take the limit $u_i\to 0$. This introduces a corresponding ordering of the transfer matrices (\ref{eq:transfermatrixA}) that govern propagation through each shock. The result becomes
\be\label{eq:Jeventshapebulk}
\la {\cal E}(\vec n_1) \cdots {\cal E}(\vec n_k)  \ra_{\e\.J (q)} &= \left( {q^0 \over \vol S^{d-2}}  \right)^k {\e^\dag  H_1 \cdots  H_k \e \over \e^\dag \e} , 
\ee
where the subscript on the left-hand side indicates that we compute the event shape in the state created by $\e\.J^\mu$, where $J^\mu$ is a current. A commutator $[\cE(n_1),\cE(n_2)]$ is simply related to a commutator of transfer matrices $[ H_1,  H_2]$.

Let us next consider the case of gravity. As discussed in \cite{Camanho:2014apa}, the most general equation of motion for a gravitational perturbation takes the form
\be
\label{eq:gravitonEoMAdS}
\delta R_{\mu \nu} + \alpha_2 \hat R_{(\mu}^{\ \ \rho \alpha \beta} \delta R_{\nu) \rho \alpha \beta} +\alpha_4 \left[ \nabla_{( \mu} \nabla_{\nu )} \hat R^{\alpha \beta \rho \sigma} \right] \delta R_{\alpha \beta \rho \sigma} = 0.
\ee
As before, it is enough to consider transverse perturbations $h_{ij}$ and keep only the dependence on $u$ and $v$ (the rest is fixed by symmetries). Our equation becomes
\be
\label{eq:shockequationgravity}
 &\pa_{u} \pa_{v} h_{i j}+ \delta(u) \hat H_{i j, m n} \pa^2_{v} h_{m n} + ... = 0. 
\ee
The form of the shockwave matrix $\hat H_{i j, m n}$ is fixed by the symmetries of the problem to be\footnote{The precise relation between $\alpha_2$ and $\alpha_4$ that appear in (\ref{eq:gravitonEoMAdS}) and more familiar three-point structures $t_2$ and $t_4$ can be found in section 5 of \cite{Camanho:2014apa}.}
\be
\hat H_{i j , m n}&=\sum_{a=1}^k {\lambda_a  \over \Omega_{S^{d-2}} }  \left( 1+n_a^{d-1} \right)^{d-1} ( H_a)_{i j , m n} , \nn\\
H_{i j , k l}(\vec n) &\equiv \frac12\left(\delta_{ik}\delta_{jl}+
\delta_{il}\delta_{jk} \right) - {1 \over d-1} \delta_{i j} \delta_{k l}+\frac{t_2}{4}\left(\delta_{ik}P_{jl}+\delta_{il}P_{jk}+
\delta_{jl}P_{ik}+\delta_{jk}P_{il}\right) \nn\\
&\quad + t_4 P_{ijkl}-\frac{t_2 + t_4}{d-1}\left(\delta_{ij}P_{kl}+\delta_{kl}P_{ij}\right),
\ee
where $H_a = H(\vec n_a)$ and we have introduced
\be
P_{ij} &=n_in_j-\frac{\delta_{ij}}{d-1}, \cr
P_{ijkl} &=n_in_jn_kn_l-\frac{\delta_{ij}\delta_{kl}
+\delta_{ik}\delta_{jl}+\delta_{il}\delta_{jk}}{(d+1)(d-1)} ,
\ee
and used the known result for the one-point energy correlator. 

The solution of (\ref{eq:shockequationgravity}) is
\be
h_{i j}^\mathrm{after} = \left[ e^{- \hat H \pa_{v}} h^\mathrm{before} \right]_{i j } . 
\ee
The energy correlator (\ref{eq:bulkcomp}) again depends on the ordering of the shocks. We get
\be
\label{eq:energycorrelatorsbulk}
\la {\cal E}(\vec n_1) \cdots {\cal E}(\vec n_k)  \ra_{\e\.T(q^0)} = \left( {q^0 \over \Omega_{S^{d-2}}}  \right)^k {\e^\dag  H_1 \cdots  H_k \e \over \e^\dag \e}.
\ee

In these calculations, it was crucial to compute the propagation amplitude through separated shocks $u_1 > u_2 > \cdots > u_k$, and afterwards take the limit $u_i\to 0$. This is guaranteed to produce the ordering of operators $\cE(\vec n_1)\cdots \cE(\vec n_k)$. By comparing different orderings, we can obtain interesting consistency conditions on the theory. A different procedure is to first set $u_i=0$, compute the propagation through a shock background $h(\vec y, z) = \sum_i h_i(\vec y,z)$, and then take derivatives with respect to $\l_i$. Instead of producing an ordered product of transfer matrices, this latter procedure produces a symmetrized product
\be
\label{eq:energycorrelatorsbulkB}
\e^\dag  H_{(1}  H_{2\phantom)}\! \cdots  H_{k)} \e.
\ee
If we only have access to this object, we cannot study commutativity. However, one can study positivity of energy correlators \cite{CamanhoUN}.

\section{Products of light-ray operators and commutativity}
\label{sec:commutativityone}

In this section, we discuss in detail the question of when light-transformed operators on the same null plane commute. We find that commutativity requires nontrivial conditions on the OPE limit, lightcone limit, and Regge limit of CFT four-point functions, all of which we verify in the case of ANEC operators in a nonperturbative CFT. In the Regge limit, we find that $J_0 < 3$ is a sufficient condition for commutativity of ANEC operators. One can argue using the light-ray OPE that it is also necessary \cite{AnecOPE}. Thus, superconvergence sum rules hold also in planar theories which do not satisfy nonperturbative bounds on the Regge limit, but do satisfy the bound on chaos $J_0 \leq 2$ \cite{Maldacena:2015waa}.

\subsection{Existence vs.\ commutativity}

Our goal is to answer two related questions:
\begin{itemize}
\item Is a product of light-transforms at coincident points $\wL[\cO_1](x,z_1) \wL[\cO_2](x,z_2)$ well-defined?
\item Do light-transforms at coincident points commute, $[\wL[\cO_1](x,z_1), \wL[\cO_2](x,z_2)] = 0$?
\end{itemize}
In event shapes we set $x=\oo$, but the answer is the same for any $x$, by conformal invariance. We take $z_1\neq z_2$, so that the integration contours for $\wL[\cO_1]$ and $\wL[\cO_2]$ are not identical.

The coincident limit of light-transformed operators is defined by
\be
\label{eq:coincidentpointlimit}
\lim_{y\to x} \wL[\cO_1](x,z_1) \wL[\cO_2](y,z_2).
\ee
That is, we first perform the light-transforms starting from distinct $x$ and $y$ and then take the limit where the initial points $x$ and $y$ coincide. To determine when the operator (\ref{eq:coincidentpointlimit}) exists, let us study a matrix element between states created by operators $\cO_3$ and $\cO_4$\footnote{The positions of $\cO_3,\cO_4$ can be smeared to create normalizable states, though that will not be important in our analysis.}
\be
\label{eq:amatrixelement}
&\lim_{y\to x} \<\Omega|\cO_4 \wL[\cO_1](x,z_1) \wL[\cO_2](y,z_2) \cO_3|\Omega\> \nn\\
&= \lim_{y\to x} \int_{-\oo}^\oo d\a_1 d\a_2 (-\a_1)^{-\De_1-J_1} (-\a_2)^{-\De_2-J_2} \<\Omega|\cO_4 \cO_1(x-z_1/\a_1,z_1) \cO_2(y-z_2/\a_2,z_2) \cO_3|\Omega\>.
\ee
If the above integral converges absolutely in the limit $y\to x$, then we can commute the limit and the integral to obtain
\be
\label{eq:integraloveralpha}
&= \int_{-\oo}^\oo d\a_1 d\a_2 (-\a_1)^{-\De_1-J_1} (-\a_2)^{-\De_2-J_2} \<\Omega|\cO_4 \cO_1(x-z_1/\a_1,z_1) \cO_2(x-z_2/\a_2,z_2) \cO_3|\Omega\>.
\ee
When (\ref{eq:integraloveralpha}) converges, commutativity is manifest because $\cO_1$ and $\cO_2$ remain spacelike-separated everywhere in the region of integration. In the next subsection, we analyze in detail when (\ref{eq:integraloveralpha}) converges.

However, it can happen that the integral (\ref{eq:integraloveralpha}) is not absolutely convergent. This does not necessarily mean that the product of operators (\ref{eq:coincidentpointlimit}) is ill-defined, but it does mean that commutativity can be lost. To understand how this works, let us start with the first line of (\ref{eq:amatrixelement}) and use the fact that $\wL[\cO_1]$ and $\wL[\cO_2]$ annihilate the vacuum to rewrite the correlator in terms of a double-commutator:
\be
&\lim_{y\to x} \<\Omega|[\cO_4, \wL[\cO_1](x,z_1)] [\wL[\cO_2](y,z_2), \cO_3]|\Omega\> \nn\\
&= \lim_{y\to x} \int_{-\oo}^\oo d\a_1 d\a_2 (-\a_1)^{-\De_1-J_1} (-\a_2)^{-\De_2-J_2} \<\Omega|[\cO_4, \cO_1(x-z_1/\a_1,z_1)] [\cO_2(y-z_2/\a_2,z_2), \cO_3]|\Omega\>.
\label{eq:thisintegral}
\ee
If this integral is absolutely convergent in the limit $y\to x$, then we can commute the limit and integration to obtain
\be
\label{eq:doublecommutatorintegral}
&=\int_{-\oo}^\oo d\a_1 d\a_2 (-\a_1)^{-\De_1-J_1} (-\a_2)^{-\De_2-J_2} \<\Omega|[\cO_4, \cO_1(x-z_1/\a_1,z_1)] [\cO_2(x-z_2/\a_2,z_2), \cO_3]|\Omega\>,
\ee
which differs from (\ref{eq:integraloveralpha}) only in that the Wightman function has been replaced by a double-commutator. The key point is that (\ref{eq:doublecommutatorintegral}) might converge in a wider variety of situations than (\ref{eq:integraloveralpha}). This is because the double-commutator might be better behaved than the Wightman function in singular limits. Thus, the integral (\ref{eq:doublecommutatorintegral}) gives a more general way to define matrix elements of the product (\ref{eq:coincidentpointlimit}), but it does not manifest commutativity. The fact that coincident light-transformed operators sometimes fail to commute has been noticed previously and is sometimes called ``detector cross-talk" \cite{Belitsky:2013ofa,Belitsky:2013xxa}. We give an example of this phenomenon in appendix~\ref{app:lighttransformscalarsubtleties}.

To summarize,
\begin{itemize}
\item Absolute convergence of the double commutator integral (\ref{eq:doublecommutatorintegral}) is a sufficient condition for the existence of $\wL[\cO_1](x,z_1) \wL[\cO_2](x,z_2)$.
\item Absolute convergence of the Wightman function integral (\ref{eq:integraloveralpha}) is a sufficient condition for the existence of $\wL[\cO_1](x,z_1) \wL[\cO_2](x,z_2)$ and commutativity $[\wL[\cO_1](x,z_1),\wL[\cO_2](x,z_2)]=0$.
\end{itemize}
Note that when the integrals do not converge absolutely, it may be still possible to prescribe values to them, but these values may suffer from ambiguities in how the integral is computed. In the following sections, we discuss in detail when the above conditions hold.

\subsection{Convergence of the Wightman function integral}

\begin{figure}[t]
	\centering
		\begin{tikzpicture}[scale=1.3]
		
		\draw (-3,0) -- (0,3) -- (3,0) -- (0,-3) -- cycle;
		
		\draw[fill=black] (-3,0) circle (0.05); 
		\draw[fill=black] (3,0) circle (0.05);  
		
		\draw[fill=black] (-1,0) circle (0.05); 
		\draw[fill=black] (1,0) circle (0.05);  
		
		\draw[fill=black] (-1.5,-1.5) circle (0.05);  
		\draw[fill=black] (1.5,1.5) circle (0.05);  
				
		\begin{scope}[decoration={
        markings,
        mark=at position 0.25 with {\arrow{>}}}
        ] 
		\draw[postaction={decorate},blue] (-1.5,-1.5) -- (1.5,1.5);
		\end{scope}
		\begin{scope}[decoration={
        markings,
        mark=at position 0.45 with {\arrow{>}}}
        ]
		\draw[postaction={decorate},blue] (-1.5,-1.5) -- (1.5,1.5);
		\end{scope}
		
		\node[right] at (3,0) {$\oo$};
		\node[left] at (-3,0) {$\oo$};
		
		\node[left] at (-1.05,0) {$4$};
		\node[right] at (1.05,0) {$3$};	
		
		\node[left] at (-1.55,-1.55) {$x$};
		\node[right] at (1.55,1.55) {$x^+$};
		
		\node at (-1.05,-0.74) {$1$};
		\node at (0.05,-0.25) {$2$};
		
		\end{tikzpicture}
		\caption{We consider the causal configuration where $4>x$ and $3<x^+$ with $3$ and $4$ spacelike from each other. The points $1$ and $2$ are integrated over parallel null lines in the same null plane, with nonzero transverse separation. Here, we have suppressed the transverse direction, so the null plane appears as a single diagonal line (blue) from $x$ to $x^+$. (The conformal completion of the null plane also includes the left-moving diagonal line from $x$ to $\oo$ on the left, and then from $\oo$ to $x^+$ on the right.)}
		\label{fig:regionS}
\end{figure}
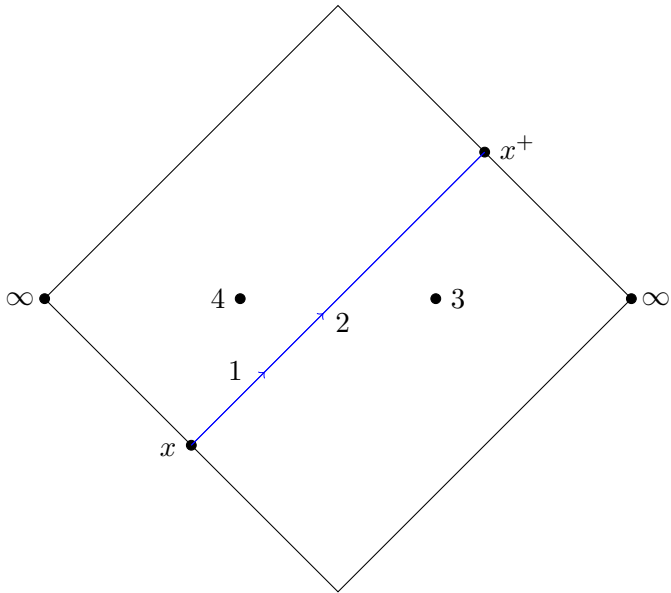

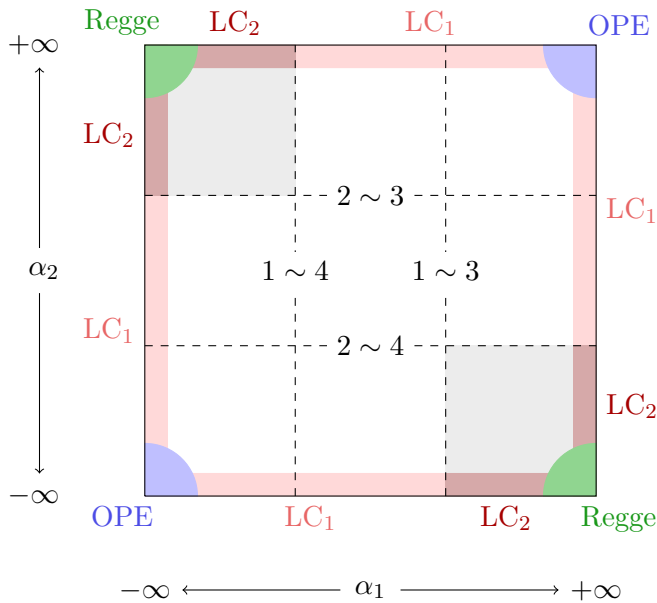
\begin{figure}[ht!]
	\centering		
	\begin{tikzpicture}
		\draw[draw=gray!15,fill=gray!15] (0,6) -- (2,6) -- (2,4) -- (0,4) -- cycle;
		\draw[draw=gray!15,fill=gray!15] (6,0) -- (6,2) -- (4,2) -- (4,0) -- cycle;
		\draw[draw=red!15,fill=red!15] (0,0) -- (0.3,0) -- (0.3,4) -- (0,4) -- cycle;
		\draw[draw=red!15,fill=red!15] (0,0) -- (0,0.3) -- (4,0.3) -- (4,0) -- cycle;
		\draw[draw=red!15,fill=red!15] (6,6) -- (5.7,6) -- (5.7,2) -- (6,2) -- cycle;
		\draw[draw=red!15,fill=red!15] (6,6) -- (6,5.7) -- (2,5.7) -- (2,6) -- cycle;
		\draw[draw=red!35!gray!50,fill=red!35!gray!50] (6,0) -- (5.7,0) -- (5.7,2) -- (6,2) -- cycle;
		\draw[draw=red!35!gray!50,fill=red!35!gray!50] (6,0) -- (6,0.3) -- (4,0.3) -- (4,0) -- cycle;
		\draw[draw=red!35!gray!50,fill=red!35!gray!50] (0,6) -- (0,5.7) -- (2,5.7) -- (2,6) -- cycle;
		\draw[draw=red!35!gray!50,fill=red!35!gray!50] (0,6) -- (0.3,6) -- (0.3,4) -- (0,4) -- cycle;
		\draw[draw=blue!25,fill=blue!25] (0,0) -- (0,0.7) to[out=0,in=90] (0.7,0) -- cycle;
		\draw[draw=blue!25,fill=blue!25] (6,6) -- (6,5.3) to[out=180,in=-90] (5.3,6) -- cycle;		
		\draw[draw=green!45!gray!60,fill=green!45!gray!60] (0,6) -- (0.7,6) to[out=-90,in=0] (0,5.3) -- cycle;
		\draw[draw=green!45!gray!60,fill=green!45!gray!60] (6,0) -- (6,0.7) to[out=180,in=90] (5.3,0) -- cycle;
		\draw (0,0) -- (6,0) -- (6,6) -- (0,6) -- (0,0);
		\draw[->] (3.4,-1.25) -- (5.5,-1.25);
		\draw[->] (2.6,-1.25) -- (0.5,-1.25);
		\node[below] at (0,-1) {$-\oo$};
		\node[below] at (6,-1) {$+\oo$};
		\node[below] at (3,-1) {$\a_1$};
		\draw[->] (-1.4,3.3) -- (-1.4,5.7);
		\draw[->] (-1.4,2.7) -- (-1.4,0.3);
		\node[left] at (-1,0) {$-\oo$};
		\node[left] at (-1,6) {$+\oo$};
		\node[left] at (-1,3) {$\a_2$};
		\node[below,text=blue!75!gray!80] at (-0.3,0) {OPE};
		\node[above,text=green!55!black!90] at (-0.3,6) {Regge};
		\node[above,text=blue!75!gray!80] at (6.3,6) {OPE};
		\node[below,text=green!55!black!90] at (6.3,0) {Regge};
		\node[below,text=red!85!black!60] at (2.2,0) {LC$_1$};
		\node[below,text=red!65!black!100] at (4.8,0) {LC$_2$};
		\node[left,text=red!85!black!60] at (0,2.2) {LC$_1$};
		\node[left,text=red!65!black!100] at (0,4.8) {LC$_2$};
		\node[above,text=red!85!black!60] at (3.8,6) {LC$_1$};
		\node[above,text=red!65!black!100] at (1.2,6) {LC$_2$};
		\node[right,text=red!85!black!60] at (6,3.8) {LC$_1$};
		\node[right,text=red!65!black!100] at (6,1.2) {LC$_2$};
		\draw[dashed] (2,0) -- (2,6);
		\draw[dashed] (0,4) -- (6,4);
		\draw[dashed] (0,2) -- (6,2);
		\draw[dashed] (4,0) -- (4,6);
		\node[fill=white] at (3,2) {$2\sim 4$};
		\node[fill=white] at (3,4) {$2\sim 3$};
		\node[fill=white] at (4,3) {$1\sim 3$};
		\node[fill=white] at (2,3) {$1\sim 4$};
	\end{tikzpicture}
	\caption{Integration region for the double light-transform of a Wightman function (\ref{eq:integraloveralpha}). The $1\x 2$ OPE converges in the white region (the ``first sheet"), but it does not necessarily converge in the gray-shaded region (the ``second sheet"). The OPE limit is indicated in blue, the Regge limit in green, the lightcone limit on the first sheet (LC$_1$) in light red, and the lightcone limit on the second sheet (LC$_2$) in dark red.}
	\label{fig:integrationregiondoublelight}
\end{figure}

Let us analyze in detail the conditions under which the Wightman function integral (\ref{eq:integraloveralpha}) is absolutely convergent. For now, we consider the causal configuration $4>x$, $x^+>3$, with $4$ spacelike from $3$, see figure~\ref{fig:regionS}. We comment on other configurations in section~\ref{sec:summaryanddiscussionconvergence}. The region of integration is shown in figure~\ref{fig:integrationregiondoublelight}. Let us describe some of its important features.

The dashed lines indicate when $1$ and $2$ become lightlike-separated from $3$ and $4$. When $1,2$ are spacelike from $4$, or $1,2$ are spacelike from $3$, we can rearrange the operators in the Wightman function so that $1$ and $2$ both act on the vacuum. The $1\x2$ OPE is guaranteed to converge in this case \cite{Mack:1976pa,Pappadopulo:2012jk}. We refer to this region as the ``first sheet" because the conformal cross-ratios need not move around branch points to get there.

Meanwhile, when $4>1$ and $2>3$, or $4>2$ and $1>3$, the $1\x 2$ OPE is not guaranteed to converge. (Other OPEs do converge in these regions.) We call these regions the ``second sheet" (shaded gray in figure~\ref{fig:integrationregiondoublelight}) because the cross ratios must move around branch points to get there.

We only need to analyze the convergence of the integral near the boundary of the integration region. Indeed, in the bulk of the integration region the only singularities are due to $1$ or $2$ becoming lightlike from $3$ or $4$. These singularities are removed by the $i\e$ prescriptions for operators $\cO_3$ and $\cO_4$. After this we can split the integral into the near-boundary region and the bulk region. The bulk region is compact and free of singularities, so the convergence of the integral there is straightforward. 

On the boundary of the region of integration, there are several types of singularities.
\begin{itemize}
\item Firstly, when $\a_1,\a_2 \to \pm \oo$ simultaneously, operators $1$ and $2$ become close on the first sheet. This singularity is described by the OPE.
\item When $\a_1\to +\oo,\a_2\to -\oo$ simultaneously or $\a_1\to -\oo, \a_2\to +\oo$ simultaneously, this is the Regge limit, which lies on the second sheet. This singularity is described by conformal Regge theory \cite{Cornalba:2007fs,Costa:2012cb,Kravchuk:2018htv}.
\item Another type of singularity occurs when either $\a_1$ or $\a_2$ approach $\pm \oo$, with the other variable held fixed. This is the lightcone limit, where $1$ and $2$ become lightlike separated. The lightcone limit on the first sheet is described by the $1\x 2$ OPE, while the lightcone limit on the second sheet is not necessarily described by the $1\x 2$ OPE.\footnote{The lightcone limit on the second sheet has been conjectured to have an asymptotic expansion in terms of $1\x 2$ OPE data \cite{Hartman:2015lfa}. We comment on the implications of this conjecture in section~\ref{sec:asymptoticLC}.}
\end{itemize}

Our strategy will be to first analyze the singularities on the first sheet. Then we will use Rindler positivity and the Cauchy-Schwarz inequality to bound singularities on the second sheet in terms of singularities on the first.

\subsubsection{OPE limit on the first sheet}
\label{sec:OPEfirstsheet}

Let us begin by studying the OPE singularity on the first sheet. Without loss of generality, we take $\a_1,\a_2\to -\oo$. We can choose a conformal frame where $\cO_1$ and $\cO_2$ are both approaching the origin. For simplicity, consider a traceless symmetric tensor $\cO_0\in \cO_1\x \cO_2$ with dimension $\De_0$ and spin $J_0$. The contribution of $\cO_0$ in the OPE takes the form
\be
\label{eq:opefirstsheet}
&\cO_1\p{x_1=-\frac{z_1}{\a_1},z_1} \cO_2\p{x_2=-\frac{z_2}{\a_2},z_2}
\nn\\
 &\supset \sum_{m,n,k} c_{mnk} (x_{12}^2)^{\frac{\De_0-\De_1-\De_2}{2}} (z_1\.\hat x_{12})^{J_1-n} (z_2\.\hat x_{12})^{J_2-k}\nn\\
 &\qquad\qquad \x (\hat x_{12})_{\mu_1}\cdots (\hat x_{12})_{\mu_m} z_{1\nu_1}\cdots z_{1\nu_n} z_{2\rho_1}\cdots z_{2\rho_k} \cO_0^{\mu_1\cdots\mu_m \nu_1\cdots\nu_n \rho_1\cdots \rho_k}(0),
\ee
where $m+n+k=J_0$ and $\hat x_{12} = x_{12}/(x_{12}^2)^{1/2}$. The factors of $x_{12}^2$ come from dimensional analysis. The factors of $z_i\.\hat x_{12}$ come from demanding the correct homogeneity in $z_i$.

Let us make the change of variables
\be\label{eq:alphasigma}
-\frac{1}{\a_1} = r \s,\quad -\frac{1}{\a_2} = r(1-\s).
\ee
Supplying the appropriate factors $(-\a_i)^{-\De_i-J_i}$, the double light-transform of a single term above becomes
\be
\label{eq:opeinnewvariables}
&\int \frac{dr}{r^3} \int_0^1 \frac{d\s}{\s^2(1-\s)^2} r^{\De_1+J_1+\De_2+J_2} \s^{\De_1+J_1} (1-\s)^{\De_2+J_2} \nn\\
&\quad \x r^{\De_0-\De_1-\De_2} (\s(1-\s))^{\frac{\De_0-\De_1-\De_2}{2}} \p{\frac{1-\s}{\s}}^{\frac{J_1-n}{2}} \p{\frac{\s}{1-\s}}^{\frac{J_2-k}{2}} \p{\frac{\s z_1 - (1-\s)z_2}{\sqrt{-2 z_1\.z_2 \s(1-\s)}}}^m \nn\\
&\quad \x \dots,
\ee
where ``$\dots$" indicates quantities independent of $r$ and $\s$, and we have written $(\hat x_{12})_{\mu_1}\cdots (\hat x_{12})_{\mu_m}$ schematically as $(\hat x_{12})^m$.

Requiring that the $r$-integral converge near $r=0$ gives the condition
\be
\label{eq:OPEcondition}
J_1+J_2 > 2 - \De_0.
\ee
This is a consequence of dimensional analysis. To derive it more succinctly, recall that $\wL[\cO_i]$ has dimension $\De_i^L \equiv  1-J_i$. If $\cO_0$ appears in the OPE, then the coincident limit of $\wL[\cO_1]$ and $\wL[\cO_2]$ can only be finite if $\De_0 > \De_1^L + \De_2^L$, which is equivalent to (\ref{eq:OPEcondition}). In particular, this shows that (\ref{eq:OPEcondition}) must hold even if $\cO_0$ is not a traceless symmetric tensor.

Requiring that the $\s$-integral converge near $\s=0,1$ gives the conditions
\be
J_1 + J_2 +\De_1-\De_2 &> 2-\tau_0 - 2n, \nn\\
J_1 + J_2 +\De_2-\De_1 &> 2-\tau_0 - 2k,
\ee
where $\tau_0\equiv \De_0-J_0$ is the twist of $\cO_0$. These conditions are strongest when $n=k=0$, in which case they together become
\be
\label{eq:lightconecondition}
J_1+J_2 - |\De_1-\De_2| &> 2-\tau_0.
\ee

The conditions (\ref{eq:OPEcondition}) and (\ref{eq:lightconecondition}) can be weakened slightly using the special kinematics of the light transform. Because the polarization vectors $z_i$ are aligned with the positions $x_i=-z_i/\a_i$, the two-point function $\<\cO_1\cO_2\>$ vanishes except when $\cO_1$ and $\cO_2$ are scalars. Consequently, the unit operator does not appear in the $\cO_1\x\cO_2$ OPE, and we can write
\be\label{eq:OPELCconditions}
J_1+J_2 &> 2-\De_0' && (J_1>0\textrm{ or }J_2>0),\nn\\
J_1+J_2 - |\De_1-\De_2| &> 2-\tau_0' && (J_1>0\textrm{ or }J_2>0),
\ee
where $\De_0'$ and $\tau_0'$ are the smallest {\it nonzero} dimension and twist, respectively, appearing in the $1\x 2$ OPE.

If the conditions~\eqref{eq:OPELCconditions} are satisfied, then the integral of each term in the OPE expansion over the region $r<r_0$ (for some sufficiently small $r_0$, indicated by the quarter-disc in the lower-left or upper-right corner of figure~\ref{fig:integrationregiondoublelight}) converges, including the boundaries where it probes the light-cone regime. Furthermore, the OPE expansion converges absolutely and exponentially in this region, and integrating each term doesn't change this --- the convergence rate only improves as we approach the OPE or light-cone boundary. In other words, the sum of integrals of absolute values converges. The Fubini-Tonelli theorem then establishes absolute convergence of the Wightman function integral over the OPE corners in figure~\ref{fig:integrationregiondoublelight} under conditions~\eqref{eq:OPELCconditions}.

\subsubsection{Lightcone limit on the first sheet}
\label{sec:LCfirstsheet}

We do not need to do additional work to analyze the lightcone limit on the first sheet. Studying the lightcone limit is equivalent to studying convergence of (\ref{eq:opeinnewvariables}) when $\s\to 0,1$, with $r$ held fixed. Because the $r$-dependence of the integrand  (\ref{eq:opeinnewvariables}) factors out from the $\s$-dependence, this again gives (\ref{eq:lightconecondition}). We don't have to worry about the null-cone singularities when $1$ or $2$ become light-like from $3$ or $4$, because these are avoided by $i\e$-prescriptions for the operators $3$ and $4$.

In this section we will re-derive (\ref{eq:lightconecondition}) in a way that works when $\cO_0$ is not a traceless symmetric tensor. This approach will also be helpful for the discussion of the lightcone limit on the second sheet.

Consider the product
\be
\label{eq:theproduct}
	\cO_1(-z_1/\a_1, z_1)\cO_2(-z_2/\a_2, z_2)= \sum_{k} f_{k}(\a_1)\cO_k(0),
\ee
in the limit $\a_1\to -\oo$ with fixed $\a_2$. There exists a boost generator $M$ such that $e^{\l M}z_1=e^{-\l}z_1$ and $e^{\l M}z_2=e^{\l}z_2$. Let us define $V(\l)=e^{-\l D}e^{\l M}$, where $D$ is the dilatation generator. Acting on the left-hand side of (\ref{eq:theproduct}), we have
\be
	&V(\l)\cO_1(-z_1/\a_1, z_1)\cO_2(-z_2/\a_2, z_2)V(\l)^{-1}
	\nn\\
	&=e^{-\l(\De_1+\De_2-J_2+J_1)}\cO_1(-e^{-2\l}z_1/\a_1,z_1)\cO_2(-z_2/\a_2, z_2).
\ee
Acting on a single term on the right-hand side, we have
\be
	V(\l) f_0(\a_1)\cO_0(0)V(\l)^{-1}= e^{-\l\tau_0}f_0(\a_1)\cO_0(0),
\ee
where $\cO_0$ has eigenvalue $\tau_0$ under $D-M$. Comparing both sides gives
\be\label{eq:f0result}
	f_0(\a_1)\propto (-\a_1)^{\frac{-\tau_0+\De_1+\De_2-J_2+J_1}{2}}.
\ee
Requiring that $\int d\a_1 (-\a_1)^{-\De_1-J_1} f_0(\a_1)$ converges gives the condition
\be
	J_1+J_2+\De_1-\De_2 >2-\tau_0.
\ee
From a similar analysis with $1\leftrightarrow 2$, we recover (\ref{eq:lightconecondition}).

We will also need a simple generalization of this result. Consider the same OPE~\eqref{eq:theproduct} but with more general polarization vectors, i.e.
\be\label{eq:thegenericproduct}
	\cO_1(-z_1/\a_1, z'_1)\cO_2(-z_2/\a_2, z'_2)= \sum_{k} f_{k}(\a_1)\cO_k(0).
\ee
For generic values of $z'_i$, both operators will not be eigenstates of $M$, but instead contain a mixture of different eigenstates. Suppose we isolate eigenstates with eigenvalues $m_1$ and $m_2$. The corresponding piece of $f_0(\a_1)$ will be then, by a straightforward generalization of the above argument
\be\label{eq:f0general}
	f_0(\a_1)\propto (-\a_1)^{\frac{-\tau_0+\De_1+\De_2-m_1-m_2}{2}}.
\ee
When $z'_i=z_i$ as above, we have $m_1=-J_1$ and $m_2=J_2$, recovering~\eqref{eq:f0result}. For generic $z'_i$ the dominant contribution to~\eqref{eq:thegenericproduct} will be determined by $m_1=-J_1$ and $m_2=-J_2$, i.e.
\be
	f_0(\a_1)\propto (-\a_1)^{\frac{-\tau_0+\De_1+\De_2+J_1+J_2}{2}}.
\ee

On the other hand, in order for the stronger result~\eqref{eq:f0result} to be true it suffices to have $z_1'=z_1$, $(z_2\.z_2')=O(\a_1^{-1})$ and $z'_2$ has finite limit as $\a_1\to -\oo$. In other words, we can allow $z'_2$ to vary with $\a_1$. The condition on $z'_1$ implies $m_1=-J_1$, and is as good as generic $z'_1$. To understand the condition on $z'_2$, assume without any loss of generality that $z_2=(0,1,\vec 0)$ in $(u,v,\vec y)$ coordinates. This implies that 
\be
	z_2'=((-\a_1)^{-1} u_z(\a_1), u_z^{-1}(\a_1)y^2_z(\a_1), (-\a_1)^{\frac{1}{2}}\vec y_z(\a_1)),
\ee
where $u_z(\a_1)$ and $\vec y_z(\a_1)$ are $O(1)$ as $\a_1\to-\oo$. We have then 
\be
	\cO_2(-z_2/\a_2, z'_2)=\sum_{n+k+l=J_2}(-\a_1)^{-n-l/2} u_z(\a_1)^{n-k}y^{2k}_z(\a_1)y_z^{i_1}(\a_1)\cdots y_z^{i_l}(\a_1)\cO_{2,u\ldots uv\ldots vi_1\ldots i_{l}}(-z_2/\a_2)
\ee
where $\cO_{2,u\ldots uv\ldots vi_1\ldots i_{l}}$ has $n$ $u$-indices and $k$ $v$-indices. If we were contracting with $z_2$, we would only get $v$-indices, and $m_2=J_2$. Thus, $v$-indices carry positive charge under $M$, and we have $m_2=k-n=J_2-l-2n$. We thus see that for non-zero $n$ or $l$ we depart from the optimal eigenvalue $m_2=J_2$. However, such terms are additionally suppressed by $(-\a_1)^{-n-l/2}$. Combining these two effects with the help of~\eqref{eq:f0general} we see that all terms contribute as~\eqref{eq:f0result}.

\subsubsection{Rindler positivity}

Rindler positivity enables us to bound second-sheet correlators in terms of first-sheet correlators. Its implications are simple for four-point functions of scalar primaries, and this case has been analyzed previously in \cite{Hartman:2015lfa,Caron-Huot:2017vep}. However, its implications are more subtle for spinning correlators, so let us discuss them in more detail.

Any CFT has an anti-unitary symmetry $J=\mathsf{CRT}$ satisfying $J^2=1$. This symmetry acts on local operators as
\be
	J \cO(x,s) J^{-1} = i^F[\cO (\bar x, e^{-\pi M_E^{01}}s)]^\dagger,
\ee
where $\bar x = e^{-\pi M_E^{01}}x=(-x^0,-x^1,x^2,\ldots)$, $s$ is a polarization variable appropriate for the Lorentz irrep of $\cO$, and $M_E^{01}=iM^{01}$ is the Euclidean rotation in plane $01$ which rotates positive $x^1$ into positive $ix^0$. Fermion number is $F=0$ for bosonic $\cO$ and $F=1$ for fermionic $\cO$.
In this paper we will only study its action on traceless-symmetric operators, which reduces to
\be
J \cO(x,z) J^{-1} = \cO^\dagger (\bar x, \bar z),
\ee
where $\bar z=(-z^0,-z^1,z^2,\ldots)$.

The statement of Rindler positivity is \cite{Casini:2010bf}
\be
	i^{-F}\<\O|J \cO_1\cdots \cO_n J\cO_1\cdots \cO_n |\O\> \geq 0,
\ee
where all operators $\cO_i$ lie in the right Rindler wedge $x^1>|x^0|$, and $F$ is $1$ if the number of fermions among $\cO_1\cdots\cO_n$ is odd and $0$ otherwise.\footnote{For $F=1$ this form of Rindler positivity was proven in~\cite{Casini:2010bf} only under an additional assumption of ``wedge ordering'' of the coordinates of $\cO_i$. The version without this assumption was proven using Tomita-Takesaki modular theory and assuming $F=0$. While we have not proven this, we expect that the general version also holds for $F=1$. This can probably be checked explicitly in CFT using conformal block expansions. We thank Nima Lashkari for discussions on this point.} This is a bit of an oversimplification, the general statement is that the operators $\cO_i$ should be smeared with test functions, and one can take arbitrary linear combinations of such smeared products. When the correlation function is well-defined as a number (rather than just as a distribution), the smearing is not necessary.

Now let $a$ and $b$ be sums of products of (possibly smeared) operators contained in the right Rindler wedge, and define 
\be
\label{eq:rindlerinnerproduct}
(a,b)=i^{-F(a)}\<\O|J a J b|\O\>,
\ee
where $F(a)\in\{0,1\}$ is defined as above. Then we have
\be
	(a,b)^*=(-i)^{-F(a)}\<\O| a J b J|\O\>=i^{-F(a)}\<\O|J b J a |\O\>=(b,a),
\ee
where we used anti-unitarity of $J$, $J|\O\>=|\O\>$, and the fact that $JbJ$ and $a$ are space-like separated. Rindler positivity implies $(a,a)>0$. Thus, $(\.,\.)$ is a Hermitian inner product and we have the Cauchy-Schwarz inequality 
\be
	|(a,b)|^2\leq (a,a)(b,b).
\ee

Let us develop more conformally-invariant versions of these statements. The geometry that defines $J$ is given by the codimension-1 planes $x^0\pm x^1=0$. These planes can be described more invariantly as the past null cones of points $A$ and $B$ at future null infinity. Given these points, the right Rindler wedge is given by $B>x>A^-$ and the left is given by $A>x>B^-$.\footnote{Here, $A^-$ ($B^-$) represents the point obtained by sending lightrays in all past directions from $A$ ($B$) and finding the point where they converge. See, e.g.\ \cite{Kravchuk:2018htv} for details.} In general, for any spacelike-separated pair of points $A$ and $B$, there exists an anti-unitary Rindler conjugation $J_{AB}$ that depends on these two points and exchanges the two wedges. A positive-definite Hermitian inner product analogous to (\ref{eq:rindlerinnerproduct}) can be defined for each $J_{AB}$.

It is convenient to describe the action of $J_{AB}$ using the embedding formalism. Let $X_A$ and $X_B$ be the embedding space coordinates of $A$ and $B$. Then $J_{AB}$ acts on spacetime as a Euclidean rotation by $\pi$ in the plane spanned by $X_A$ and $X_B$. It can be written as\footnote{We abuse notation and write $J_{AB}$ for both the anti-unitary operator on Hilbert space and a linear transformation in the embedding space.}
\be\label{eq:JABXaction}
	J_{AB}(X)=X-2\frac{(X\.X_A)}{(X_A\.X_B)}X_B-2\frac{(X\.X_B)}{(X_A\.X_B)}X_A.
\ee
The action of $J_{AB}$ on local operators is then
\be
	J_{AB} \cO(X,Z) J_{AB}^{-1}=\cO^\dagger(J_{AB}(X), J_{AB}(Z)).
\ee

Consider now a configuration of points $1,2,3,4$ with the causal relationships $4>1$ and $2>3$, where all other pairs of points are spacelike-separated. We can find a conformal transformation that brings these points into a  configuration where the pair $1,2$ and the pair $3,4$ are each symmetric with respect to the standard Rindler reflection $J$. Thus, there must exist $A,B$  such that $J_{AB}$ maps $1 \leftrightarrow 2$ and $3\leftrightarrow 4$. In embedding-space language, 
\be
\label{eq:ABequations}
	J_{AB}(X_1)=\l_{12}X_2,\quad
	J_{AB}(X_3)=\l_{34}X_4,
\ee
where we must introduce scaling factors $\l_{ij}$ because the $X$'s are projective coordinates. Here, $X_A,X_B$ and the coefficients $\l_{12},\l_{34}$ are all functions of $X_1,\dots,X_4$. These functions are somewhat complicated in general, but we will only need them in certain limits.

In our configuration, the Wightman correlator
\be
	\<\O|\cO_4\cO_1\cO_2\cO_3|\O\>
\ee
is on the second sheet. However, we can write
\be
	\<\O|\cO_4\cO_1\cO_2\cO_3|\O\>=(J_{AB}^{-1}\cO_4\cO_1 J_{AB}, \cO_2\cO_3),
\ee
and use the Rindler Cauchy-Schwarz inequality to write
\be\label{eq:schwarzbound}
	|\<\O|\cO_4\cO_1\cO_2\cO_3|\O\>|^2\leq (\cO_2\cO_3,\cO_2\cO_3)(J_{AB}^{-1}\cO_4\cO_1 J_{AB},J_{AB}^{-1}\cO_4\cO_1 J_{AB}).
\ee

Let us focus on the first factor (the second can be treated equivalently)
\be
	(\cO_2\cO_3,\cO_2\cO_3)=\<\O|\bar\cO_2\bar\cO_3\cO_2\cO_3|\cO\>
\ee
Here we use the notation $\bar\cO=J_{AB}\cO J_{AB}^{-1}$. Note that $\bar\cO_2$ is inserted at $X_1$ and $\bar\cO_3$ is inserted at $X_4$. Both of these points are in the opposite Rindler wedge from $X_2$ and $X_3$. This implies that $\bar \cO_3$ and $\cO_2$ commute so we can reorder the operators to obtain
\be
\label{eq:afterrearranging}
	=\<\O|\bar\cO_2\cO_2\bar\cO_3\cO_3|\cO\>.
\ee
The operators $\bar{\cO}_2$ at $X_1$ and $\cO_2$ at $X_2$ now act on the vacuum. By the results of \cite{Mack:1976pa,Pappadopulo:2012jk}, the correlator (\ref{eq:afterrearranging}) is on the first sheet and we can use the OPE to control its behavior, and hence bound the left-hand side of~\eqref{eq:schwarzbound}. It is convenient that the correlators in the right hand side of~\eqref{eq:schwarzbound} have the same insertion points $X_i$ as the correlator in the left hand side, and hence the same cross-ratios.

Let us briefly mention how one can determine the points $X_{A}$ and $X_{B}$ as functions of $X_i$. To do this, we must solve the equations~\eqref{eq:ABequations}, together with the conditions
\be\label{eq:ABnull}
	X_A^2=X_B^2=0.
\ee
Using~\eqref{eq:JABXaction} we can see that $X_A$ and $X_B$ must have the form
\be
	X_A&=c_{A1}(X_1-\l_{12}X_2)+c_{A3}(X_3-\l_{34}X_4),\nn\\
	X_B&=c_{B1}(X_1-\l_{12}X_2)+c_{B3}(X_3-\l_{34}X_4),
\ee
for some coefficients $c_{ai}$. We have 2 scalar equations coming from each equation in~\eqref{eq:ABequations}, by projecting on $X_1-\l_{12}X_2$ or $X_3-\l_{34}X_4$. We also have 2 scalar equations~\eqref{eq:ABnull}, which adds up to 6 equations for 6 unknowns $c_{ai}, \l_{ij}$. 

It is easy to solve these equations by making use of the conformal symmetry. For this, recall that all coordinates $X$ are projective, and hence our unknowns $c_{ai}$ and $\l_{ij}$ have non-trivial projective weights as well. We can construct combinations such as
\be
	c_{A1}\frac{(X_1\.X_2)}{(X_A\.X_2)},
\ee
which are projective invariants. Projective invariants are the same as conformal invariants, and thus must be expressible in terms of cross-ratios, i.e.
\be
	c_{A1}\frac{(X_1\.X_2)}{(X_A\.X_2)}=f_{A1}(z,\bar z).
\ee
The function $f_{A1}(z,\bar z)$ can be computed by using the expressions for $X_i$, $X_A,\,X_B$ for the standard Rindler reflection $J$. As soon as we know the function $f_{A1}(z,\bar z)$, we can find
\be
	c_{A1}=f_{A1}(z,\bar z)\frac{(X_A\.X_2)}{(X_1\.X_2)}.
\ee
Note that the expression in the right-hand side depends on $X_A$ but only through an overall coefficient $(X_A\.X_2)$. The same coefficient can be factored out from $c_{A3}$, and thus $X_A$ is determined up to an overall rescaling, which is irrelevant. For example, we can simply set $(X_A\.X_2)=1$ to get a concrete solution. All the other coefficients $c_{ai}$ and $\l_{ij}$ can be determined in the same way. We will not need the complete solution, but it is helpful for explicitly checking our arguments.

\subsubsection{Regge limit}
\label{sec:reggelimit}

In the previous section we saw that Rindler positivity implies a bound on the correlator of the form
\be\label{eq:finalCSbound}
	|\<\O|\cO_4\cO_1\cO_2\cO_3|\O\>|^2\leq& \<\O|\bar\cO_2\bar\cO_3\cO_2\cO_3|\O\>\<\O|\cO_4\cO_1 \bar\cO_4\bar\cO_1|\O\>\nn\\
	=&\<\O|\bar\cO_2\cO_2\bar\cO_3\cO_3|\O\>\<\O|\cO_4 \bar\cO_4\cO_1\bar\cO_1|\O\>.
\ee
We can now use the first-sheet bounds from sections~\ref{sec:OPEfirstsheet} and~\ref{sec:LCfirstsheet} to bound the correlators in the right hand side. Before doing so, let us write out these correlators a bit more carefully. For example,
\be
	\<\O|\bar\cO_2\cO_2\bar\cO_3\cO_3|\O\>=\<\O|\cO_2^\dagger(\l_{12}^{-1}X_1,\tl Z_2)\cO_2(X_2,Z_2)\cO_3^\dagger(\l_{34}X_4,\tl Z_3)\cO_3(X_3,Z_3)|\O\>,
\ee
where
\be
	\tl Z_i\equiv J_{AB}(Z_i).
\ee
Writing this in terms of real space operators, we find
\be\label{eq:realspace2233}
	\<\O|\bar\cO_2\cO_2\bar\cO_3\cO_3|\O\>=\l_{12}^{\De_2}\l_{34}^{-\De_3}\<\O|\cO_2^\dagger(x_1,\tl z_2)\cO_2(x_2,z_2)\cO_3^\dagger(x_4,\tl z_3)\cO_3(x_3,z_3)|\O\>.
\ee
Let's focus on the Regge limit when $\a_1\to +\oo$ and $\a_2\to -\oo$. Similarly to section~\ref{sec:OPEfirstsheet} it is convenient to work in the frame in which $x_2$ is approaching the origin, 
\be
	x_2=-\frac{z_2}{\a_2}.
\ee
In this frame point $x_1$ is in the next Poincare patch, so it is convenient to work with $x_1^-$, which is the image of $x_1$ in the Poincare patch of $x_2$,
\be
	x_1^-=-\frac{z_1}{\a_1}.
\ee
In fact, since $\cO(x_1)$ acts on future vacuum, the correlator~\eqref{eq:realspace2233} changes only by a constant phase upon replacement $x_1\to x_1^-$, so the analysis is very similar to section~\ref{sec:OPEfirstsheet}. If the coefficients $\l_{ij}$ and polarization vectors $\tl z_2$ and $\tl z_3$ were constant, we could simply reuse the results of that section.

Instead, since $\l_{ij}$ and $\tl z_2$, $\tl z_3$ depend on the $x_i$ (and thus $\a_1,\a_2$), we need to analyze their behavior in the limit $\a_1\to +\oo$ and $\a_2\to -\oo$. Let us parameterize, similarly to~\eqref{eq:alphasigma}
\be
\frac{1}{\a_1} = r \s,\quad -\frac{1}{\a_2} = r(1-\s).
\ee
We have checked that in the limit $r\to 0$ the coefficients $\l_{12},\l_{34}$ stay finite and
\be
	\tl z_2\to z_2,\quad \tl z_3\to z_3',
\ee
where $z_3'$ is finite and depends on $z_3$ and relative positions of $0,x_3,x_4$. This means that in Regge limit at fixed $\s$~\eqref{eq:realspace2233} is bounded in absolute value by
\be
	C\left\vert\<\O|\cO_2^\dagger\p{x_1=-\frac{z_1}{\a_1},z_2}\cO_2\p{x_2=-\frac{z_2}{\a_2},z_2}\cO_3^\dagger(x_4,z_3')\cO_3(x_3,z_3)|\O\>\right\vert,
\ee
for some constant $C>0$. The same analysis applies to the second correlator in~\eqref{eq:finalCSbound}.

We can now reuse the arguments of section~\ref{sec:OPEfirstsheet} to conclude that the Wightman function integral converges near Regge limit, at fixed $\sigma$, provided that
\be
	J_1+J_2>2-\De_\text{vac},
\ee
where $\De_\text{vac}$ is the smallest scaling dimension appearing in the $\cO_2\times\cO_2^\dagger$ (or $\cO_1\times\cO_1^\dagger$) OPE. Since $\cO_2\times\cO_2^\dagger$ always contains the identity operator, $\De_\text{vac}=0$. Note that the polarizations of both $\cO_2$ and $\cO_2^\dagger$ are $z_2$, so we cannot exclude the identity contribution using kinematics as we did in section~\ref{sec:OPEfirstsheet}. So we finally obtain the sufficient condition
\be
	J_1+J_2>2.
\ee
Note that we have only shown that this is sufficient for convergence of the integral at fixed $\s$. We will discuss the case of $\s$ approaching the light-cone boundaries, and thus of the two-variable integral, in the next section.

This latter condition is sufficient, but may turn out to be not necessary, since we cannot prove that the Cauchy-Schwartz bound~\eqref{eq:finalCSbound} is tight. To allow for this possibility, let us introduce a parameter $J_0$ which is defined as the smallest real number such that
\be\label{eq:J0definition}
	\frac{\<\O|\cO_4\cO_1\cO_2\cO_3|\O\>}{\sqrt{\<\cO_3(x_3)\cO_3^\dagger(x_4)\>\<\cO_4(x_4)\cO_4^\dagger(x_3)\>\<\cO_1(x_1)\cO_1^\dagger(x_2)\>\<\cO_2(x_2)\cO_2^\dagger(x_1)\>}}\in O(r^{1-J_0}),
\ee
where all polarization vectors in denominator are generic. Then the Wightman function integral converges in the Regge limit if 
\be
	J_1+J_2>J_0+1.
\ee
The result of this section can in turn be summarized by saying that
\be
	J_0\leq 1.
\ee

\subsubsection{Lightcone limit on the second sheet}

Consider now the lightcone limit on the second sheet. The situation is very similar to the Regge regime, and we can use the same frame as above to analyze it. The only difference is that now we consider either $\a_1\to+\oo$ or $\a_2\to-\oo$, corresponding to the right or the lower boundary in figure~\ref{fig:integrationregiondoublelight} respectively. The other two boundaries can be treated in the same way. For concreteness, let us focus on the limit $\a_2\to -\oo$.

For simplicity, we will assume that $x_1$ is in the same Poincare patch as $x_2$, and write 
\be
	x_1=-\frac{z_1}{\a_1},\quad x_2=-\frac{z_2}{\a_2}.
\ee
This time, we will have to discuss both correlators in the bound~\eqref{eq:finalCSbound}. Modulo factors of $\l_{ij}$ in~\eqref{eq:realspace2233} and their analogues for the second correlator in~\eqref{eq:finalCSbound}, we are essentially interested in the behavior of the OPEs
\be
	\bar\cO_2\cO_2&\sim \cO_2^\dagger(x_1,\tl z_2)\cO_2(x_2,z_2)\nn\\
	\cO_1\bar\cO_1&\sim \cO_1(x_1,z_1)\cO_1^\dagger(x_2,\tl z_1).
\ee
The $\l$-factors and polarizations entering $\bar\cO_3$ and $\bar\cO_4$, similarly to Regge limit, can be ignored because they all tend to some generic finite limits.\footnote{This is true as long as we stay in the interior of the second sheet. These coefficients may diverge as we approach $1\sim 3$ lightcone limit. We comment on this subtlety below.} 

It is easy to determine the direction of $\tl z_1$ in the strict lightcone limit $\a_2=-\oo$. In this limit we find $x_2=0$, and $z_1$ lies along the unique null ray which connects $x_1$ and $x_2$. This is a conformally-invariant statement, and so it should hold after we apply $J_{AB}$. Applying $J_{AB}$ sends $x_1$ to $x_2$, $x_2$ to $x_1$, and $z_1$ to $\tl z_1$. Therefore, we find that $\tl z_1$ should also lie along the unique null ray connecting $x_1$ and $x_2$, and thus it is proportional to $z_1$. In other words,
\be
	\tl z_1\to c z_1,
\ee
for some finite $c$. This implies that 
\be
	(z_1\.\tl z_1) = O(\a_2^{-1}).
\ee
Using the results of section~\ref{sec:LCfirstsheet} (after swapping $\a_1$ and $\a_2$) we find that 
\be
	\cO_1(x_1,z_1)\cO_1^\dagger(x_2,\tl z_1)\sim (-\a_2)^{\frac{2\De_1-2J_1-\tau_0}{2}}.
\ee
Here $\tau_0$ is the smallest twist that appears in $\cO_1^\dagger \cO_1$ OPE.

There is nothing special we can say about $\tl z_2$, and it simply tends to some generic finite value in the limit. This and results of section~\ref{sec:LCfirstsheet} imply that 
\be
	\cO_2^\dagger(x_1,\tl z_2)\cO_2(x_2,z_2)\sim (-\a_2)^{\frac{2\De_2+2J_2-\tau_0}{2}}.
\ee

Combining these results we find that
\be
|\<\O|\cO_4\cO_1\cO_2\cO_3|\O\>|\leq C' (-\a_2)^{\frac{-\tau_0+\De_1+\De_2-J_1+J_2}{2}}
\ee
near the second sheet null-cone limit $\a_2\to -\oo$. This bound is uniform away from the boundary between first and second sheets $1\sim 3$. Including the light-transform weight $(-\a_2)^{-\De_2-J_2}$, we conclude that the Wightman function integral converges absolutely in this region provided
\be
	J_1+J_2-\De_1+\De_2+\tau_0>2.
\ee
Combining this with the condition from $\a_1\to \pm\oo$ we find the sufficient condition
\be\label{eq:secondsheetLCconditino}
	J_1+J_2>2+|\De_1-\De_2|-\tau_0.
\ee
Similarly to the Regge limit, we cannot exclude the identity operator from $\cO_1^\dagger\cO_1$ (or $\cO_2^\dagger\cO_2$) OPE, and so we must set $\tau_0=0$, resulting in the final condition
\be
	J_1+J_2>2+|\De_1-\De_2|.
\ee

There are two subtleties which we still need to address. One is the convergence of the two-variable integral near Regge limit -- we have established the convergence of the radial integral in the previous section and of the angular integral in this section, but we have not yet proved that the double integral converges. To see that it does, note that we have succeeded in bounding both the Regge limit and the second-sheet lightcone limit by using Rindler positivity. A closer look at our arguments shows that for $r<r_0$ and all $\sigma$ the second-sheet correlator is bounded in absolute value by the product of first-sheet correlators times a uniform constant $C''$, where $r_0$ is sufficiently small. Convergence of the double-integral of the product of first-sheet correlators can be established by the same methods as in section~\ref{sec:reggelimit}, and then the convergence of the integral on the second sheet follows immediately.

The second subtlety is that near the lightcone limit, either on the first or on the second sheet, we have only established the convergence of the integral provided we exclude a region near the boundary between the first and the second sheets. We now turn to a discussion of this subtlety.

\subsubsection{Asymptotic ligthcone expansion}
\label{sec:asymptoticLC}
The discussion of previous sections provides us with rigorous bounds on the growth of the Wightman function on the first and the second sheets. The situation is, however, qualitatively different for the two sheets. 

On the first sheet we have a tight bound on the growth in OPE and lightcone limits -- we can use the convergent OPE expansion to see that the Wightman function actually does saturate the bound. This means that unless the conditions of sections~\ref{sec:OPEfirstsheet} and~\ref{sec:LCfirstsheet} are satisfied, the Wightman function integral diverges (in absolute value sense).

On the second sheet we have a potentially non-optimal bound on the growth in the Regge and lightcone limits, which we derived from the Cauchy-Schwarz inequality for Rindler reflection positivity. We have already pointed out that the growth in the Regge limit may be weaker than the bound, and we parametrized the true growth by an exponent $J_0$. Similarly, there is no a-priori reason to believe that the lightcone bound is tight. 

In fact, there is a reason to believe that the growth of the correlator on the second sheet is the same as on the first sheet. Indeed, it is natural to expect that the lightcone OPE expansion, even though not convergent on the second sheet, is still valid asymptotically \cite{Hartman:2015lfa}. Schematically,
\be\label{eq:lightconeasymptotic}
	\cO_1(x_1)\cO_2(x_2)=\sum_{\tau_\cO\leq\tau} (x_{12}^2)^{\frac{\tau_\cO-\De_1+m_1-\De_2+m_2}{2}}\cO(x_2)+o\p{(x_{12}^2)^{\frac{\tau-\De_1+m_1-\De_2+m_2}{2}}},
\ee
where $m_i$ is the boost eigenvalue of $\cO_i$ (similarly to section~\ref{sec:LCfirstsheet}), and the sum on the right-hand side is over spin components of primaries and descendants. Such an asymptotic expansion is sufficient to establish the growth rate of the Wightman function near the second sheet lightcone limit, and gives the same results as on the first sheet. In particular, the contribution of the identity operator to~\eqref{eq:lightconeasymptotic} is excluded in the same way as on the first sheet, and we don't have to set $\tau_0=0$ in~\eqref{eq:secondsheetLCconditino} anymore.

By using the asymptotic expansion~\eqref{eq:lightconeasymptotic}, we can also prove that the Wightman function integral converges in the ligthcone limit near the boundary between the first and the second sheets (with $i\e$ prescription employed for $\cO_3,\cO_4$), closing the loophole in our arguments.

While~\eqref{eq:lightconeasymptotic} is a natural expectation, we don't have a general proof that it holds. An argument was given in~\cite{Hartman:2015lfa} for the case of scalar $\cO_1=\cO_2$ and $\cO_3=\cO_4$, with real coordinates. In their argument, one splits the $t$-channel ($\cO_1\times\cO_4$ OPE) expansion of the Wightman function into two parts. The first part $\mathcal{I}_1$ consists roughly of double-traces $[\cO_1\cO_2]_{n,\ell}$, and is responsible for reproducing a finite number of terms in~\eqref{eq:lightconeasymptotic} in the $s$-channel. The second part $\mathcal{I}_2$ contains all other contributions. By going sufficiently far into the $s$-channel lightcone regime, one can make sure that $\cI_1$ completely dominates over $\cI_2$. Continuation of $\cI_1$ to the second sheet is straightforward, since it is simply equal to a finite number of terms in~\eqref{eq:lightconeasymptotic}. The expansion~\eqref{eq:lightconeasymptotic} then follows if we can show that the second part $\cI_2$ remains subleading on the second sheet. In the setup of~\cite{Hartman:2015lfa} this is easy to show, since all the terms in the $t$-channel channel expansion (and hence in $\cI_2$) are positive, and continuation to the second sheet merely adds some phases, which cannot increase the total sum. 

This last step is problematic in more general setups. The positivity of $t$-channel contributions is due to the fact that the Wightman function considered in~\cite{Hartman:2015lfa} is Rindler-reflection positive on the first sheet. As soon as we consider non-identical operators or operators with spin, the Rindler-reflection positivity ceases to be generic, and the argument cannot be applied. Still, the fact that~\eqref{eq:lightconeasymptotic} is valid at least for scalar $\cO_1=\cO_2$, in some states, strongly suggests that it can be valid more generally. For the argument of~\cite{Hartman:2015lfa} to fail in the general setup, it must be that the phases in $\cI_2$ on the first sheet conspire to give an abnormally small result for all values of $\bar z$, which seems rather unlikely.

In view of this discussion, we will assume that the asymptotic expansion~\eqref{eq:lightconeasymptotic} holds.

\subsection{Convergence of the double commutator integral}
\label{sec:doublecommutatorconvergence}

\begin{figure}[t!]
	\centering
	\begin{tikzpicture}
	
	\draw (-3,-2) -- (-3,2) -- (3,2) -- (3,-2) -- cycle;
	\draw[dashed] (0,0) -- (3,0);
	\draw (-3,0) -- (0,0);
	\node[below] at (0,0) {$1$};
	\draw[fill=black] (0,0) circle (0.05);
	
	\draw[thick,->] (-3,0.1) -- (-0.5,0.1) to[out=0, in=180] (0,0.5);
	\draw[thick] (0,0.5) to[out=0,in=90] (0.5,0) ;
	\draw[thick, draw=gray!70,->] (0.5,0) to[out=-90,in=0] (0,-0.5);
	\draw[thick,draw=gray!70] (0,-0.5) to[out=180, in=0] (-0.5,-0.1) -- (-3,-0.1);
	
	\draw (-2.5,1.9) -- (-2.5,1.5) -- (-2.9,1.5);		
	\node[right] at (-2.95,1.7) {$\bar z$};
	
	\end{tikzpicture}
	\caption{Trajectory of $\bar z$ near the boundary between the first and the second sheets.}
	\label{fig:zbtrajectory}
\end{figure}
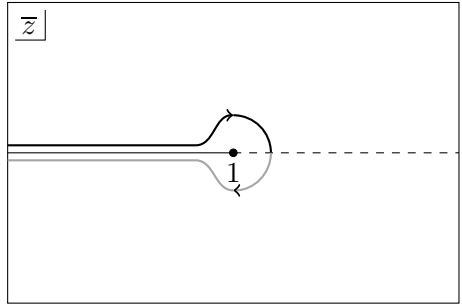

We now consider the convergence of the double commutator integral~\eqref{eq:doublecommutatorintegral}. The first observation is that the double-commutator vanishes when $1\approx 4$ or $3\approx 2$. Our integration region is therefore restricted to $1<4$ and $2>3$. This corresponds to the upper left shaded square in figure~\ref{fig:integrationregiondoublelight}. Note, however, that it is now meaningless to say that in this square the correlator is on the second sheet. The double-commutator consists of four Wightman correlators, and it is not guaranteed that they all have cross-ratios on the same sheet. In fact, by using the causal relations between the points and microcausality, we can write in this region
\be\label{eq:doublecommutatorexpanded}
	\<\O|[\cO_4,\cO_1][\cO_2,\cO_3]|\O\>=&\<\O|\cO_4\cO_1\cO_2\cO_3|\O\>+\<\O|\cO_3\cO_1\cO_2\cO_4|\O\>\nn\\
	&-\<\O|\cO_1\cO_2\cO_3\cO_4|\O\>-\<\O|\cO_3\cO_4\cO_1\cO_2|\O\>.
\ee
The first two Wightman functions are of the same type as studied in the previous subsection, and are on the second sheet. In the last two Wightman functions, the operators $\cO_1$ and $\cO_2$ act on the vacuum, and so they are on the first sheet.

When analyzing the convergence of the double commutator integral~\eqref{eq:doublecommutatorintegral}, once again we only need to consider the convergence near the boundaries of the integration region. Furthermore, we don't have to worry about the boundaries where $2\sim 3$ or $1\sim 4$. Near these boundaries the integral is defined by $i\e$ prescriptions. More precisely, the double-commutator is obtained by folding integration contours in the Wightman function integral~\eqref{eq:integraloveralpha}. Due to this folding, near these boundaries the four terms above pair up and form integration contours similar to figure~\ref{fig:zbtrajectory}. Overall, the double-commutator integral is an integral of a single Wightman function over a folded complex integration cycle in cross-ratio space. Everywhere away from the $2\sim 3$ or $1\sim 4$ boundaries this integration cycle can be split into four layers, and each layer can be interpreted as an integral of a Wightman function from~\eqref{eq:doublecommutatorexpanded}. Near these boundaries the layers merge, wrapping around branch cuts and providing a canonical regularization of the integral.

We will continue to refer to the remaining two boundaries as the lightcone limits, and to the corner where they meet as the Regge limit. We can use the methods of the previous subsections to bound the growth of each Wightman function in~\eqref{eq:doublecommutatorexpanded} in these limits and find that the same conditions as we derived for~\eqref{eq:integraloveralpha} are also sufficient for convergence of the double-commutator integral.\footnote{With the same caveat about the boundary between the first and second sheets as before.}

It is easy to see that weaker conditions are in fact sufficient for convergence of the double-commutator integral. Let us consider the lightcone limits first. The asymptotic expansion~\eqref{eq:lightconeasymptotic} essentially implies that we can approximate the double-commutator near these limits by replacing the Wightman functions in~\eqref{eq:doublecommutatorexpanded} by a finite number of $s$-channel conformal blocks for the leading twist operators, analytically continued to the desired Wightman orderings. However, it is known~\cite{Caron-Huot:2017vep} that $s$-channel conformal blocks cancel in the combination~\eqref{eq:doublecommutatorexpanded}. This means that the asymptotic expansion~\eqref{eq:lightconeasymptotic} does not contribute to the double-commutator.
If this expansion were valid for any $\tau$, then we would conclude that the double-commutator decays faster than any power of $\a_1$ or $\a_2$ near the lightcone limits.

However, it is well-known that the spectrum of primaries in any OPE has accumulation points in twist~\cite{Fitzpatrick:2012yx,Komargodski:2012ek}. Let us denote the first twist accumulation point in $\cO_1\times\cO_2$ OPE by $\tau_*$. We can only trust~\eqref{eq:lightconeasymptotic} for $\tau<\tau_*$, since for $\tau\geq \tau_*$ we would have to include infinitely many terms in the sum. Therefore, we can only conclude that the double-commutator grows in the ligthcone regime no faster than at the rate determined by $\tau_*$. This leads to the following sufficient condition,
\be
	J_1+J_2-|\De_1-\De_2|>2-\tau_*.
\ee

For the Regge limit, we don't have an analogue of~\eqref{eq:lightconeasymptotic}, and we will simply introduce a growth exponent $J_0^\text{dDisc}$ for the double-commutator in complete analogy with~\eqref{eq:J0definition}. The condition for absolute convergence of the double-commutator integral near the Regge limit is then
\be
	J_1+J_2>1+J_0^\text{dDisc}.
\ee
Similarly to $J_0$, Rindler positivity bounds imply
\be
	J_0^\text{dDisc}\leq 1.
\ee

\subsection{Summary of non-perturbative convergence conditions}
\label{sec:summaryanddiscussionconvergence}

Let us summarize the various conditions we have obtained thus far. Assuming the asymptotic light-cone expansion~\eqref{eq:lightconeasymptotic} we have shown that the Wightman function integral~\eqref{eq:integraloveralpha} converges absolutely if and only if the conditions 
\be
	J_1+J_2&>\max\p{2-\De_0',1+J_0}, \label{eq:deltaprimecondition}\\
	J_1+J_2&>2-\tau_0'+|\De_1-\De_2| \label{eq:tauconditionone}
\ee
are satisfied. Here $\tau_0'\geq \frac{d-2}{2}$ and $\De'_0\geq \frac{d-2}{2}$ are the smallest non-zero twist and dimension that appear in $\cO_1\times\cO_2$ OPE,\footnote{When $J_1=J_2=0$, then we cannot exclude unit operator contributions in the OPE and lightcone limits and $\De_0'$ and $\tau_0'$ should be the lowest dimension and twist in their respective OPEs. In other words, they do not have to be nonzero in that case.} $J_0$ is the growth exponent in the Regge limit, defined by~\eqref{eq:J0definition}. Using Rindler positivity, we have shown that 
\be
	J_0\leq 1.
\ee
We have also shown that the above conditions with $\tau_0'=0$ are sufficient even if we don't assume the asymptotic expansion~\eqref{eq:lightconeasymptotic}, but simply use bounds from Rindler positivity.\footnote{This argument has a small loophole discussed, e.g., in section~\ref{sec:asymptoticLC}.}

For the double-commutator integral~\eqref{eq:doublecommutatorintegral} we have shown that the sufficient conditions for its convergence are
\be
	J_1+J_2&>1+J_0^\text{dDisc},\\
	J_1+J_2&>2-\tau_*+|\De_1-\De_2|,
	\label{eq:tauconditiontwo}
\ee
where $\tau_*\geq d-2$ is the first twist accumulation point in the $\cO_1\times\cO_2$ OPE, and $J_0^\text{dDisc}$ parameterizes the Regge growth of the double-commutator in the same way that $J_0$ parametrizes growth of the Wightman function through~\eqref{eq:J0definition}. Note that we have
\be
\De_0' &\geq \tau_0' \geq \frac{d-2}{2}, \nn\\
\tau_* &\geq \tau_0' \geq \frac{d-2}{2}. 
\ee

Let us briefly discuss the values of $J_0$ and $J_0^\text{dDisc}$. First of all, from the expansion~\eqref{eq:doublecommutatorexpanded} if follows that
\be
	J_0^\text{dDisc}\leq \max\p{1-\De_0',J_0}.
\ee
Both $J_0$ and $J_0^\text{dDisc}$ can be studied using conformal Regge theory~\cite{Cornalba:2007fs,Costa:2012cb,Caron-Huot:2017vep,Kravchuk:2018htv}.  Conformal Regge theory implies that the Regge limit of the four-point function behaves as $1+r^{1-j(0)}$, where $j(0)$ is the spin of the leading Regge trajectory at dimension $\De=d/2$. Here, the $1$ comes from the identity operator in the $\cO_1\x\cO_2$ OPE. However, we are considering special kinematics where the unit operator does not contribute unless $\cO_1$ and $\cO_2$ are identical scalars. In these special kinematics, the four-point function behaves as $r^{1-j(0)}$. The double-discontinuity also does not get a contribution from the unit operator, so it also behaves as $r^{1-j(0)}$. Thus, we expect
\be
J_0 = 1\textrm{ and } J_0^\mathrm{dDisc}=j(0) \leq 1 &&& \textrm{if $\cO_1=\cO_2$ are identical scalars},\nn\\
J_0 = J_0^\mathrm{dDisc} = j(0) \leq 1&&& \textrm{if $\cO_1,\cO_2$ are not identical scalars}.
\ee

Let us consider the case where $\cO_1,\cO_2$ are not identical scalars. Note that if $J_0$ is the intercept of the stress-tensor trajectory, then $1\geq J_0\geq 2-\frac{d}{2}$ by Nachtmann's theorem \cite{Nachtmann:1973mr,Komargodski:2012ek,Costa:2017twz}. Furthermore, by unitarity $1-\De_0'\leq 1-\tau_0' \leq 2-\frac{d}{2}$, so the condition $J_1+J_2 > 2-\De_0'$ in (\ref{eq:deltaprimecondition}) is redundant with $J_1+J_2>1+J_0$. If $\De_1=\De_2$, then the conditions (\ref{eq:tauconditionone}) and (\ref{eq:tauconditiontwo}) are redundant as well. 

As an example, consider the case where $\cO_1=\cO_2=T$ (the stress tensor). By the above reasoning, a product of ANEC operators on the same null plane is well-defined and commutative if
\be
3 > J_0 = J_0^\mathrm{dDisc}.
\ee

In our analysis so far, we have considered the causal configuration $4>x$ and $x^+>3$ with $3$ and $4$ spacelike. One can use $\cO_3,\cO_4$ to generate a dense subspace of the Hilbert space while staying in this causal configuration. Thus, our analysis establishes well-definedness and commutativity (when applicable) acting on this dense subspace. However, this subspace does not include important states like momentum eigenstates, so we might hope to establish commutativity on a larger dense subspace.

Some different causal configurations can be reached by acting on $\cO_3$ and $\cO_4$ with the operator $\cT$ that translates operators to image points in other Poincare patches (see \cite{Kravchuk:2018htv} for details). This operation simply introduces a phase, leaving our analysis unchanged. However, there exist other causal configurations that cannot be reached in this way: for example, if $4>x$ and $x^+>3$ but $3$ and $4$ are timelike.
In this case, one cannot use Rindler positivity and we have not found a way to rigorously bound the behavior of the correlator. We can argue non-rigorously as follows. To establish convergence of the Wightman or double-discontinuity integrals in the lightcone limit, we can invoke the asymptotic lightcone assumption of \cite{Hartman:2015lfa} described in section~\ref{sec:asymptoticLC}. Under this assumption, the analysis of the lightcone limit between $1$ and $2$ is independent of the positions of points $3$ and $4$, and thus identical to what we have already done. To establish convergence in the Regge limit, we can assume that the leading Regge behavior $r^{1-J_0}$ predicted by conformal Regge theory continues to hold in this different causal configuration. It is possible that conformal Regge theory can be used to rigorously establish both of these assumptions. We leave this problem for future work.

In upcoming work \cite{AnecOPE}, we establish a connection between commutativity and the Regge limit in a different way. We show that the commutator $[\wL[\cO_1],\wL[\cO_2]]$ is nonvanishing if and only if the four-point function $\<\cO_4\cO_1\cO_2\cO_3\>$ has a Regge pole at $J=J_1+J_2-1$. This again shows that the commutator vanishes if $J_1+J_2>J_0+1$, where $J_0$ is the position of the rightmost Regge pole.

\subsection{Convergence in perturbation theory}
\label{sec:convergenceinperturbationtheory}

Although we have shown that $J_0\leq 1$ in a nonperturbative theory, this bound can be violated at a fixed order in perturbation theory (e.g.\ either weak coupling or large-$N$). In large-$N$ perturbation theory, the bound on chaos \cite{Maldacena:2015waa} implies that
\be
J_0^\mathrm{planar}\leq 2,
\ee
where $J_0^\mathrm{planar}$ characterizes the Regge growth of the leading nontrivial term in $1/N$. This bound is saturated in holographic theories with a large gap, where $J_0^\mathrm{planar}=2$ comes from tree-level graviton exchange in the bulk. At one-loop in the bulk, we get contributions from two graviton exchange, and hence we expect $J_0^\textrm{1-loop}=3$ in such theories.

From our discussion above, it follows that in these theories, products of ANEC operators on the same null plane should be well-defined and the ANEC operators should commute at planar level, since $J_0^\mathrm{planar} < 3$. However, a product of ANEC operators on the same null plane is not well-defined at 1-loop order in the bulk, since $J_0^\textrm{1-loop} \geq 3$. To recover a well-defined observable, one would have to appropriately re-sum $1/N$ effects.

Note that there is no contradiction with the product of ANEC operators being defined nonperturbatively. The four-point function has an expansion in powers of $1/N$ at fixed values of cross-ratios, but this expansion does not commute with taking the Regge limit. Since the ANEC integrals probe the Regge limit, we find that these integrals do not commute with $1/N$ expansion. Let us consider a (not necessarily physical) toy model of such a situation, 
\be
	I_N=\int_1^\oo ds s^{-3} f_N(s),\quad f_N(s)=\frac{1}{1+s/N^2}.
\ee
Here the $I_N$ is the analogue of the energy-energy correlator, and $f_N(s)$ is the analogue of the correlation function, where $s\to+\oo$ is the Regge limit. The integral for $I_N$ converges if $f_N(s)$ grows as $s^{j-1}$ with $j<3$. This is true for $N^{0}$ and $N^{-2}$ terms in large-$N$ expansion of $f_N(s)$, which predict
\be
	I_N=\frac{1}{2}-N^{-2}+\cdots.
\ee
The exact answer is
\be
	I_N=\frac{1}{2}-N^{-2}+\frac{\log(N^2+1)}{N^4}.
\ee
We see that term-wise integration is reliable for the orders at which it converges, but at higher orders $I_N$ may cease to have a simple $1/N$ expansion. We expect that something similar happens in energy-energy correlators, i.e.\ 
\be\label{eq:EEexpansion}
	\<\cE(n_1)\cE(n_2)\>_{\e\.T}=f^\text{planar}(n_1,n_2)+f^\text{rest}_N(n_1,n_2),
\ee
where $f^\text{planar}(n_1,n_2)$ is computed from the planar part of the four-point function, and
\be
	\lim_{N\to \oo}f^\text{rest}_N(n_1,n_2)=0,
\ee
but $f^\text{rest}_N(n_1,n_2)$ does not admit an expansion in integer powers of $1/N^2$. In $\cN=4$ SYM, at finite t'Hooft coupling $J_0$ will be less than $2$ and~\eqref{eq:EEexpansion} may have more $1/N$ terms in it.

\subsection{Other types of null-integrated operators}

The works \cite{Casini:2017roe,Cordova:2018ygx} have considered other examples of operators integrated over null rays. For example, the authors of \cite{Casini:2017roe} introduced
\be
\label{eq:morenullintegrals}
L^{n}(\vec y) &\equiv \int_{-\oo}^\oo dv\,v^{n+1}\, T_{vv}(u=0,v,\vec y).
\ee
They studied the algebra of such operators (under the assumption that products can be suitably renormalized) and found that it resembles a separate Virasoro algebra for each transverse point $\vec y$. The work \cite{Cordova:2018ygx} did a similar analysis of other null integrated operators and found that they generate a BMS algebra.

A first comment about the expression (\ref{eq:morenullintegrals}) is that it is generically divergent in any correlation function when $n\geq d$. A definition of $L^n(\vec y)$ which is not divergent is in terms of descendants of $\wL[T]$, as we explain below.

Let us make two additional comments about such operators. Firstly, the additional insertions of $v^{n+1}$ in the integrand make it more difficult to argue that products are well-defined and commutative, even at nonzero transverse separation $\vec y_{12}$. The required analysis is similar to the previous subsections. For example, suppose we would like to establish that $\<\cO_4|[L^n(\vec y_1) , L^m(\vec y_2)]|\cO_3\>=0$, for nonzero $\vec y_{12}$. The integral of the Wightman function $\<\Omega|\cO_4 T T \cO_3|\Omega\>$ is absolutely convergent in the Regge limit if
\be
J_1+J_2 > J_0 + 3 + n + m,
\ee
where $J_1=J_2=2$. If $0<J_0<1$ (as expected for the 3d Ising model), we can only prove commutativity at nonzero $\vec y_{12}$ for the cases $n+m\leq 0$. If $J_0=1$ (as expected in a gauge theory), we can only prove commutativity at nonzero $\vec y_{12}$ for $n+m<0$. One should also consider constraints coming from the lightcone limit. A full analysis of well-definedness and commutativity of more general light-ray operators is outside the scope of this work.

If Wightman function integrals are not absolutely convergent, then it may still be possible to renormalize products  $L^n(\vec y_1) L^m(\vec y_2)$, but the renormalized product may not be commutative at nonzero $\vec y_{12}$. 

This discussion assumes that the Regge limit is always dominated by a fixed Regge intercept $J_0$. More generally, the Regge limit of a four-point function is related to an integral over the leading Regge trajectory $J_0(\nu)$ where $\De=\frac{d}{2}+ i\nu$ and $\nu$ ranges from $-\oo$ to $\oo$ \cite{Cornalba:2007fs,Costa:2012cb}.  The $J_0$ we have discussed so far is shorthand for the maximum value of $J_0(\nu)$ along this trajectory. However, it is possible to isolate different values of $\nu$ by performing an integral transform in the transverse positions $\vec y$. Thus, perhaps by passing to $\nu$ space and choosing a $\nu$ such that $J_0(\nu)$ is sufficiently small, one could alleviate the problems with defining products $L^n(\vec y_1) L^m(\vec y_2)$. We briefly discuss this possibility again in section~\ref{sec:bootstrappingstuff}.

Finally, let us explain how operators like $L^n(\vec y)$ and those in \cite{Cordova:2018ygx} can be described using light transforms. The significance of $\wL[T](x,z)$ is that it transforms like a conformal primary. By contrast, the operators  $L^n(\vec y)$ can be understood as descendants of $\wL[T](x,z)$ --- i.e.\ derivatives with respect to $x$ and/or $z$. As usual in conformal field theory, correlators of descendants are determined by correlators of primaries.

Let us understand how this works for the case of 
\be
L^0(\vec y) &= \int_{-\oo}^\oo dv\, v\, T_{vv}(u=0,v,\vec y).
\ee
This expression looks like a light-transform, except that the object $v T_{vv}$ in the integrand is not a conformal primary. To understand its conformal transformation properties, it is helpful to think of $v T_{vv}$ as a component of a larger object
\be
X^m T(X,Z),
\ee
where 
\be
X^m=(X^+,X^-,X^\mu) = (1,x^2,x^\rho)\in \R^{d,2}
\ee
is an embedding-space vector. Here, $T(X,Z)$ is the embedding-space lift of $T_{\mu\nu}(x)$, which we describe in more detail in section~\ref{sec:lighttransformgeneralthreept}. We recover $v T_{vv}$ by setting $m=v$ and $Z=(0,0,z)$ with $z\.x=v$.

 The product $X^m T(X,Z)$ is an example of acting on $T(X,Z)$ with a weight-shifting operator \cite{Karateev:2017jgd}.  Weight-shifting operators are conformally-covariant differential operators that shift the conformal weights of the objects they act on, in addition to introducing a free index for a finite-dimensional representation $W$ of the conformal group. In this case, $X^m$ is a 0-th order differential operator (since it does not involve any derivatives $\pdr{}{X}$ or $\pdr{}{Z}$). It transforms in the vector representation $W=\Box$ of $\SO(d,2)$, and shifts weights by $(\De,J) \to (\De-1,J)$.
 
Some weight-shifting operators for $W=\Box$ are
\be\label{eq:vectoroperators}
\cD_m^{-0} &= X_m, \nn\\
\cD_m^{0+} &= (J+\De)Z_m + X_m Z\.\pdr{}{X}.
\ee
Here $m=+,-,0,\dots,d-1$ is a vector index for $\SO(d,2)$. The superscripts indicate how the operators shift dimension and spin, respectively:
\be
\cD^{\a\b}_m : (\De,J) \to (\De+\a,J+\b).
\ee
The representation $W=\Box$ possesses other weight-shifting operators that will not be important for our discussion.

Weight-shifting operators and conformally-invariant integral transforms satisfy a natural algebra. We can guess the form of this algebra simply by inspecting quantum numbers. For example, because $\wL:(\De,J)\to (1-J,1-\De)$, we must have\footnote{In general, a conformally-invariant integral transform $\mathbf{I}_r$ is associated to a Weyl reflection $r$ of $\SO(d,2)$ \cite{KnappStein1,KnappStein2,Kravchuk:2018htv}. When we commute a weight-shifting operator $D^w$ with weight $w$ past the integral transform, the weight gets reflected, $\mathbf{I}_r D^w  = D^{r(w)} \mathbf{I}_r$.}
\be
\wL \cD_m^{\a,\b} \propto \cD_m^{-\b,-\a} \wL.
\ee
In particular,
\be
\wL[X_m T(X,Z)] &= \wL [\cD_m^{-0} T(X,Z)]\nn\\
&\propto \cD_m^{0+} \wL[T](X,Z) \nn\\
&=\p{-d\,Z_m + X_m Z\.\pdr{}{X}} \wL[T](X,Z).
\ee
The operator $L^0(\vec y)$ can be obtained by appropriately specializing $X,Z$ above:
\be
L^0(\vec y) &\propto \left.\p{-d\,Z_m + X_m Z\.\pdr{}{X}} \wL[T](X,Z)\right|_{\substack{
X_0 = -(0,0,\tfrac 1 2,\tfrac 1 2,\vec 0)\\
Z_0 = (1,\vec y^2,0,0,\vec y)
}}.
\ee

\section{Computing event shapes using the OPE}
\label{sec:computingeventshapes}

In this section, we discuss how event shapes can be computed using the OPE of the boundary CFT. Simple examples of event shapes have been computed before, for example in~\cite{Hofman:2008ar}. Our goal here is to provide tools for the calculation of $n$-point event shapes with general intermediate and external operators. For the purposes of this paper, this will allow us to match the results of section~\ref{sec:shocksinads} and to express the commutativity of shocks as an exact constraint on the CFT data. We will also compute in closed form all conformal blocks which appear in scalar two-point event shapes, which may be useful in other contexts.

Let us give a brief overview of this section. We start by introducing the appropriate conformal blocks in section~\ref{sec:tchannelblocks}. After describing their general structure, we explain in detail how they can be computed in subsections~\ref{sec:wightmantwopt}-\ref{sec:three-pt-fourier}. We then use these results in section~\ref{sec:boundarymultipoint} to match the bulk results of section~\ref{sec:shocksinads}, and in section~\ref{sec:tsumrule} to describe the constraints that shock commutativity implies for the CFT data. Finally, in section~\ref{sec:scalarblocks} we give a closed-form expression for a general conformal block appearing in a scalar event shape, and demonstrate how the expansion works in a simple generalized free theory example.

\subsection{$t$-channel blocks}
\label{sec:tchannelblocks}

Let us consider a general two-point event shape\footnote{We are suppressing possible Lorentz indices for $\cO_3$ and $\cO_4$.}
\be\label{eq:tchanneleventshape}
	\<\cO_4(p)|\wL[\cO_1](\oo,z_1)\wL[\cO_2](\oo,z_2)|\cO_3(p)\>.
\ee
We can rewrite it by inserting a complete set of states between the two light transforms
\be\label{eq:tchannelpreexpansion}
	\sum_\Psi\<\cO_4(p)|\wL[\cO_1](\oo,z_1)|\Psi\>\<\Psi|\wL[\cO_2](\oo,z_2)|\cO_3(p)\>.
\ee
By the operator-state correspondence, the Hilbert space is spanned by states created by a single insertion of a local operator, and thus the above sum can be interpreted as an OPE expansion of $\cO_4\wL[\cO_1]$ or $\wL[\cO_2]\cO_3$ in terms of local operators. However, since the states $\<\O|\cO_4\wL[\cO_1]$ and $\wL[\cO_2]\cO_3|\O\>$ are not, strictly speaking, of finite norm, it is not obvious that this expansion converges. In this paper we will assume that it does, modulo some remarks that we defer to section~\ref{sec:tchannelconvergence}. We also discuss there some informal arguments in favor of convergence.

To compute the expansion~\eqref{eq:tchannelpreexpansion}, we first reorganize it into conformal families
\be\label{eq:tchannelpreexpansion2}
\sum_\cO\sum_{\Psi_\cO}\<\cO_4(p)|\wL[\cO_1](\oo,z_1)|\Psi_\cO\>\<\Psi_\cO|\wL[\cO_2](\oo,z_2)|\cO_3(p)\>,
\ee
where  we sum over primary operators $\cO$, and $\Psi_\cO$ run over an orthonormal basis of descendants of $\cO$. Let us focus on the contribution of a single primary operator $\cO$. To simplify the discussion, assume for the moment that $\cO$ is a scalar. 

Normally, the sum over the descendants $\Psi_\cO$ is rather complicated, since in order to perform it one needs to find an orthonormal basis in the conformal multiplet of $\cO$. A simple, although a bit unorthodox, way to perform such orthogonalization is to consider the momentum eigenstates
\be
|\cO(p)\>\equiv \int d^dx e^{ipx} \cO(x)|\O\>.
\ee
Since we are in Lorentzian signature, these states are perfectly well-defined (as distributions in $p$). They also form an orthogonal set thanks to momentum conservation,
\be\label{eq:scalartwopt}
\<\cO(q)|\cO(p)\>=\cA(\De)(2\pi)^d\de^d(p-q) (-p^2)^{\De-\frac{d}{2}}\theta(p),
\ee
where the right hand side is completely fixed by momentum conservation, Lorentz and scale invariance, and energy positivity, up to an overall factor $\cA(\De)$. Here, $\De$ is the scaling dimension of $\cO$.
Moreover, these states are complete in the conformal multiplet of $\cO$, which follows from completeness of the states 
\be
\int d^dx f(x)\cO(x)|\O\>,
\ee
where $f$ ranges over Schwartz test functions~\cite{Mack:1975je}. Using this basis, we can write\footnote{A version of~\eqref{eq:scalaridentityresolution} with spin was recently used in~\cite{Gillioz:2018mto}. We will use a slightly different generalization to spin.}
\be\label{eq:scalaridentityresolution}
\sum_{\Psi_\cO}|\Psi_\cO\>\<\Psi_\cO|=\cA(\De)^{-1}\int_{p>0} \frac{d^dp}{(2\pi)^d} (-p^2)^{\frac{d}{2}-\De}|\cO(p)\>\<\cO(p)|,
\ee
This expression is morally equivalent to the shadow integral approach to conformal blocks (see, e.g.\ \cite{SimmonsDuffin:2012uy}). An important difference is, however, that~\eqref{eq:scalaridentityresolution} is a rigorous identity in the Hilbert space, and does not require subtraction of any analogs of shadow contributions.

Expression~\eqref{eq:scalaridentityresolution} turns out to be perfectly suited for our needs. Indeed, in~\eqref{eq:tchannelpreexpansion2} we are taking an inner product of $\<\Psi_\cO|$ with momentum eigenstates,\footnote{Recall that $\wL[\cO_i]$ is inserted at infinity and is thus translationally-invariant} and this localizes the $p$ integral in~\eqref{eq:scalaridentityresolution}. Using this observation, we find the following expression for the contribution of $\cO$ to the event shape~\eqref{eq:tchanneleventshape}
\be\label{eq:tchannelscalarblock}
&\cA(\De)^{-1}\int_{q>0} \frac{d^dq}{(2\pi)^d} (-q^2)^{\frac{d}{2}-\De}\<\cO_4(p)|\wL[\cO_1](\oo,z_1)|\cO(q)\>\<\cO(q)|\wL[\cO_2](\oo,z_2)|\cO_3(p)\>\nn\\
&=\cA(\De)^{-1} (-p^2)^{\frac{d}{2}-\De}\<\cO_4(p)|\wL[\cO_1](\oo,z_1)|\cO(p)\>\<\cO(p)|\wL[\cO_2](\oo,z_2)|\cO_3(p)\>
\ee
Above, we used momentum conservation to go to the last line, and as usual abused the notation by implicitly removing the momentum-conserving $\de$-functions in the final three-point functions. 

Let us immediately note an important feature of this expansion. Since the light transform $\wL[\cO_2]$ annihilates the vacuum, we can write
\be\label{eq:tchanneldoubletracebyebye}
\<\cO(p)|\wL[\cO_2](\oo,z_2)|\cO_3(p)\>=\<\cO(p)|[\wL[\cO_2](\oo,z_2),\cO_3(p)]|\O\>,
\ee
which, analogously to the situation with $t$-channel conformal blocks in the Lorentzian inversion formula \cite{Caron-Huot:2017vep}, implies that the contribution of $\cO$ vanishes if it is a double-trace of $\cO_2\cO_3$.\footnote{The reason is that in this case the commutator vanishes, as can be checked from explicit expressions for the three-point tensor structures.} 

In~\eqref{eq:tchannelscalarblock} we have essentially computed (in the case of scalar $\cO$) what we will call the $t$-channel conformal blocks for the event shape~\eqref{eq:tchanneleventshape}. We use the name $t$-channel block, because we would like to reserve $s$-channel to mean the OPE of $\wL[\cO_1]\wL[\cO_2]$, which we discuss in~\cite{AnecOPE}. More precisely, the conformal block corresponding to~\eqref{eq:tchannelscalarblock} is given by stripping off the OPE coefficients,
\be\label{eq:scalartchannelblockfinal}
G^{t,ab}_{\De,0}(p,z_1,z_2)=\cA(\De)^{-1} (-p^2)^{\frac{d}{2}-\De}\<\cO_4(p)|\wL[\cO_1](\oo,z_1)|\cO(p)\>^{(a)}\<\cO(p)|\wL[\cO_2](\oo,z_2)|\cO_3(p)\>^{(b)},
\ee
where we used a superscript ${}^{(a)}$ to indicate that we are working not with a physical three-point function, but with a standard conformally-invariant three-point tensor structure with label $a$. We will denote the conformal block for exchange of $\cO$ in a general Lorentz representation $\rho_\cO$ by
\be
	G^{t,ab}_{\De_\cO,\rho_\cO}(p,z_1,z_2).
\ee
Again, the indices of operators $\cO_3$ and $\cO_4$ are implicit in this notation. The analog of~\eqref{eq:scalartchannelblockfinal} for these more general blocks is a bit more involved, owing to the spin indices of $\cO$, and we defer its discussion to section~\ref{sec:wightmantwopt}.

With this notation, the event shape can be written as
\be\label{eq:tchannelblockexpansion}
\<\cO_4(p)|\wL[\cO_1](\oo,z_1)\wL[\cO_2](\oo,z_2)|\cO_3(p)\>=
\sum_{\cO,a,b} \l_{14\cO,a}\l_{23\cO,b}G^{t,ab}_{\De_\cO,\rho_\cO}(p,z_1,z_2),
\ee
where $\rho_\cO$ is the Lorentz representation of $\cO$, and $\l$ are the OPE coefficients dual to the chosen basis of three-point tensor structures. 

In principle, the $t$-channel event shape conformal blocks can be computed from the usual conformal blocks by applying the light- and Fourier transforms. However, the kinematics of event shapes are very special, and it is easier to directly use~\eqref{eq:scalartchannelblockfinal}. In particular, as we will soon see, any $t$-channel event shape block can be written in terms of simple functions, which is not true for general conformal blocks. 

To conclude this section, let us note that the above discussion can be straightforwardly generalized to multi-point event shapes such as
\be
	\<\cO_{1}'(p)|\wL[\cO_1](\oo,z_1)\cdots \wL[\cO_n](\oo,z_n)|\cO'_{n+1}(p)\>.
\ee
We can insert a complete set of states in between each consecutive pair of light transforms.\footnote{Multi-point conformal blocks of this topology are sometimes called ``comb-channel'' blocks.} The conformal block is obtained by restricting the sums over states to conformal multiplets of some primary operators $\cO_i'$,
\be
	\sum_{\Psi_{\cO'_i}}\<\cO_{1}'(p)|\wL[\cO_1](\oo,z_1)|\Psi_{\cO'_2}\>\<\Psi_{\cO'_2}|\cdots |\Psi_{\cO'_n}\>\<\Psi_{\cO'_n}|\wL[\cO_n](\oo,z_n)|\cO'_{n+1}(p)\>.
\ee
Assuming again that all operators $\cO_2',\ldots,\cO_n'$ are scalars, and repeating the arguments leading to~\eqref{eq:scalartchannelblockfinal}, we obtain the expression for the conformal block
\be
&G^{t,a_2\cdots a_n}_{\De,0}(p,z_1,\ldots,z_n)=\nn\\
&\cA(\De_{n+1})(-p^2)^{-\frac{d}{2}+\De_{n+1}}\prod_{i=1}^n\cA(\De_{i+1})^{-1} (-p^2)^{\frac{d}{2}-\De_{i+1}}\<\cO_i'(p)|\wL[\cO_i](\oo,z_i)|\cO_{i+1}'(p)\>^{(a_i)}.
\ee
Again, the generalization to $\cO_i'$ of non-trivial spin is straightforward.

\subsubsection{Convergence of $t$-channel expansion}
\label{sec:tchannelconvergence}

In this section, we discuss in more detail the convergence of the expansion~\eqref{eq:tchannelpreexpansion}. On general grounds,~\eqref{eq:tchannelpreexpansion} converges absolutely if the states $\wL[\cO_2](\oo,z_2)\cO_3|\O\>$ and $\wL[\cO_1^\dagger](\oo,z_1)\cO_4^\dagger|\O\>$ have finite norm. This is certainly not the case since we have, for example,
\be
||\wL[\cO_2](\oo,z_2)|\cO_3(p)\>||^2=\<\O|[\cO_3(p)]^\dagger\wL[\cO^\dagger_2](\oo,z_2)\wL[\cO_2](\oo,z_2)\cO_3(p)|\O\>,
\ee
which is in general divergent because of the momentum-conserving delta-function and the fact that polarizations of the two detectors are the same. We can try to avoid both problems by considering the smeared ket state
\be
\int D^{d-2}z_2 d^dp f(z_2,p)\wL[\cO_2](\oo,z_2)|\cO_3(p)\>,
\ee
and similarly for the bra. If we only had $\cO_3$ and not the light-transform $\wL[\cO_2]$, this state would be finite-norm by the usual Wightman axioms. In order to have a finite-norm state with the insertion of $\cO_2$ we would like to also have smearing over the coordinates of $\cO_2$ with an appropriate test function. Smearing over $z_2$ is in principle equivalent to smearing of $\cO_2$, but the effective smearing function is not a test function -- it only has support on future null infinity, which is codimension 1. Whether this smearing yields a finite-norm state is not obvious. One instance in which it does is given by uniform smearing of $z_2$ over the celestial sphere and $\cO_2=T$. In this case we get
\be
&\int D^{d-2}z_2 d^dp f(z_2,p)\wL[T](\oo,z_2)|\cO_3(p)\>\nn\\
&\propto \int d^dp f(p)H|\cO_3(p)\>=\int d^dp\, p^0f(p)|\cO_3(p)\>,
\ee
which is finite-norm. There are several other smearing functions which give different components of momentum generator $P^\mu$.\footnote{For $\cO_2=T$ a general smearing over $z_2$ can be interpreted as a difference of modular Hamiltonians for two particular spatial regions~\cite{Casini:2017roe}. It is unclear whether $\int d^d p f(p)|\cO_3(p)\>$ is in the domain of this difference. We thank Nima Lashkari for discussion on this point.} More generally, we can also smear the coordinate $x_2$ of $\wL[\cO_2](x_2,z_2)$, which yields a finite-norm state and thus a convergent expansion~\eqref{eq:tchannelpreexpansion}, and take the limit of localized $x_2=\oo$. If the event shape is well-defined in the first place (c.f.\ discussion in section~\ref{sec:commutativityone}), then one can expect this limit to commute with the expansion~\eqref{eq:tchannelpreexpansion}, thus showing that smearing in polarization vectors is sufficient.\footnote{Additional smearing in $p$ should not be important since the dependence of event shape on $p$ is essentially fixed by Lorentz invariance.}

In \cite{AnecOPE}, we will relate the event shape (\ref{eq:tchanneleventshape}) to the Lorentzian OPE inversion formula at spin $J_1+J_2-1$, with $\De=\frac d 2 + i\nu$ on the principal series. In particular, the coefficient function $C(\De=\frac d 2 + i\nu,J=J_1+J_2-1)$ appearing in the inversion formula is equal to a smearing of the two-point event shape with a particular test function that depends on $\nu$. In $\nu$-space, the question of convergence of the OPE expansion (\ref{eq:tchannelpreexpansion}) is thus equivalent to the question of convergence of the conventional $t$-channel conformal block expansion, when inserted into the Lorentzian inversion formula. Thus, the $t$-channel expansion for the event shape converges in $\nu$-space if $J_1+J_2-1>J_0$ \cite{Caron-Huot:2017vep,Simmons-Duffin:2017nub}. This is equivalent to the condition for the event shape to make sense in the first place.

In what follows we will mostly be interested in event shapes in the space of spherical harmonics, as opposed to $\nu$-space. We will study in section~\ref{sec:tchanexpample} a simple example in which~\eqref{eq:tchannelpreexpansion} converges after smearing with a test function, provided this test function vanishes sufficiently quickly near the collinear limit $z_1\propto z_2$. Smearing with spherical harmonics does not have this property, but it can be achieved by taking appropriate finite linear combinations. The number of such ``subtractions'' in the example of section~\ref{sec:tchanexpample} depends on the scaling dimensions of $\cO_1$ and $\cO_2$. We will take it as an assumption that this is the general picture. Furthermore, we will assume that no subtractions are necessary if $\cO_1=\cO_2=T$, and smearing polarizations $z_1$ and $z_2$ against spherical harmonics already leads to a convergent expansion~\eqref{eq:tchannelpreexpansion}.\footnote{If this turns out not to be the case, our results can be straightforwardly modified to account for the subtractions. Note that the discussion of convergence is irrelevant for applications to planar theories with a finite number of single-trace exchanges, since the sum over $\cO$ in~\eqref{eq:tchannelpreexpansion2} is finite in such theories.} It would be interesting to examine this question more rigorously.

Let us finally comment on a related subtlety. In the preceding discussion we showed that the ANEC operators commute in the sense
\be
	[\wL[T](\oo,z_1),\wL[T](\oo,z_2)]=0
\ee 
for non-collinear $z_1$ and $z_2$. In principle we have not excluded the possibility of contact terms at $z_1\propto z_2$ in the right-hand side. Since we only study the $t$-channel expansion after smearing with test functions, we might worry that the smeared commutators do not vanish because of these potential contact terms. It was argued in~\cite{Cordova:2018ygx} that under natural assumptions there are no contact terms in this commutator, and we will work under this assumption. Even if there are contact terms, one can still perform the same subtractions as above to avoid them.


\subsubsection{Fourier transform of Wightman two-point function}

\label{sec:wightmantwopt}

In this section we discuss the generalization of~\eqref{eq:scalaridentityresolution} to $\cO$ with non-trivial spin, and compute the coefficients $\cA(\De)$ and their generalizations in the case of traceless-symmetric $\cO$.

The identity~\eqref{eq:scalaridentityresolution} is essentially dual to the two-point function in momentum space~\eqref{eq:scalartwopt}. Thus, in order to find its generalization to $\cO$ with spin, we should study the general Wightman two-point function in momentum space. 

The two-point function is constrained by scale, translation, Lorentz, and special conformal invariance. Let us set special conformal invariance aside for the moment and consider the implications of the other symmetries. First of all, scale and translation invariance imply
\be
	\<\cO^\a(p)|\cO^\b(q)\> = (2\pi)^d \de^d(p-q)(-p^2)^{\De-\frac{d}{2}}\theta(p)F^{\a\b}(p/|p|)
\ee
for some function $F$. Lorentz invariance further constrains the form of $F$. Suppose we have defined 
\be\label{eq:Fvalue}
	F^{\a\b}(\hat e_0),
\ee
where $\hat e_0$ is the unit vector along the time direction. Then Lorentz invariance allows us to determine $F^{\a\b}(v)$ for any unit-normalized timelike $v$, and thus also the complete two-point function. The value of~\eqref{eq:Fvalue} is only constrained by invariance under $\SO(d-1)$ rotations. In other words, the allowed values of~\eqref{eq:Fvalue} are in one-to-one correspondence with 
\be
	(\rho_\cO^\dagger\otimes \rho_\cO)^{\SO(d-1)}.
\ee
Under reduction to $\SO(d-1)$, any Lorentz irrep $\rho_\cO$ decomposes into $\SO(d-1)$ irreducible components without multiplicities. The complex conjugate irrep $\rho_\cO^\dagger$ decomposes into dual $\SO(d-1)$ irreps. This implies that a natural basis of invariants is given by
\be
	\Pi_\l^{\a\b}(v)
\ee
where $\l$ is an $\SO(d-1)$ irrep which appears in the decomposition of $\rho_\cO$. These invariants are defined, up to a constant multiple, by the following property: $\Pi_\l^{\a\b}(\hat e_0)$ is the $\SO(d-1)$ invariant which has non-zero components only along the irrep $\l$ in index $\b$ and $\l^*$ in index $\a$. We will see explicit examples of such invariants below. Using this basis we can write
\be
	F^{\a\b}(v)=\sum_{\l\in\rho_\cO}\cA_\l(\De,\rho_\cO)\Pi_\l^{\a\b}(v),
\ee
for some coefficients $\cA_\l$ and thus~\cite{Mack:1975je}
\be
\<\cO^\a(p)|\cO^\b(q)\>=(2\pi)^d\de^d(p-q)(-p^2)^{\De-\frac{d}{2}}\theta(p)\sum_{\l\in \rho_\cO} \cA_\l(\De,\rho) \Pi^{\a\b}_\l(p/|p|).
\ee
Invariance under special conformal transformations now fixes the relation between coefficients $\cA_\l$ with different $\l$~\cite{Mack:1975je}, yielding a unique solution for the momentum-space two-point function. We will determine these coefficients for traceless-symmetric $\cO$ below.

To proceed, we will need the dual invariants $\Pi_{\a\b,\l}(v)$, defined by the completeness relation
\be\label{eq:dualinvdefn}
	\sum_{\l\in\r_\cO}\Pi_{\a\b,\l}(v)\Pi^{\b\s}_\l(v)=\de^\s_\a.
\ee
It is an easy exercise to establish the existence of $\Pi_{\a\b,\l}(v)$ from basic representation-theoretic arguments. Using these invariants, we can write the general form of~\eqref{eq:scalaridentityresolution} as\footnote{Here and below we abuse the notation by writing $\Pi(p)$ instead of $\Pi(p/|p|)$. In other words, we assume that $\Pi(p)=\Pi(p/|p|)$, i.e.\ $\Pi$ is a scale-invariant function. This is consistent because we have only defined $\Pi(v)$ for $v^2=-1$.}
\be
\sum_{\Psi_\cO}|\Psi_\cO\>\<\Psi_\cO|=\sum_{\l\in \rho_\cO}\cA_\l(\De,\rho)^{-1}\int_{p>0} \frac{d^dp}{(2\pi)^d} (-p^2)^{\frac{d}{2}-\De}\Pi_{\a\b,\l}(p)|\cO^\a(p)\>\<\cO^\b(p)|.
\ee
The general $t$-channel conformal block is then given by
\be\label{eq:tchannelgeneralblock}
&G^{t,ab}_{\De,\rho}(p,z_1,z_2)\nn\\
&=\sum_{\l\in \rho_\cO}\cA_\l(\De,\rho)^{-1} (-p^2)^{\frac{d}{2}-\De}\<\cO_4(p)|\wL[\cO_1](\oo,z_1)|\cO^\a(p)\>^{(a)}\Pi_{\a\b,\l}(p)\<\cO^\b(p)|\wL[\cO_2](\oo,z_2)|\cO_3(p)\>^{(b)}.
\ee

The simplest ingredients which enter into~\eqref{eq:tchannelgeneralblock} are the coefficients $\cA_\l(\De,\rho)$ and the invariants $\Pi_{\a\b,\l}(p)$. In the case when $\rho_\cO$ is traceless-symmetric tensor, $\l$ is a traceless-symmetric tensor of spin $s=0,1,\ldots J$. The invariant $\Pi_s(p)$ has two sets of traceless-symmetric indices,
\be
\Pi^{\mu_1\ldots \mu_J;\nu_1\ldots \nu_J}_s(p).
\ee
We can view $\Pi_s(p)$ as a linear operator on traceless-symmetric spin-$J$ tensors, and define $\Pi_s(p)$ as the orthogonal projectors onto the spin-$s$ $\SO(d-1)$ irrep inside the spin-$J$ traceless-symmetric irrep of $\SO(d-1,1)$. Note that $\Pi_s(p)$ have to be proportional to these projectors; requiring them to be equal to the projectors gives a convenient normalization with which $\Pi_{\a\b,\l}$ and $\Pi^{\a\b}_\l$ are equal. In particular, equation~\eqref{eq:dualinvdefn} follows from, in operator notation,
\be
	\sum_{s=0}^J \Pi_{s}(p)\Pi_{s}(p)=\sum_{s=0}^J \Pi_s(p)=1,
\ee
where we have used the standard properties of projectors.

It is convenient to contract the indices with null polarization vectors $z_1$ and $z_2$ to define
\be
\Pi_{J,s}(p;z_1,z_2)\equiv z_{1,\mu_1}\cdots z_{1,\mu_J}\Pi^{\mu_1\ldots \mu_J;\nu_1\ldots \nu_J}_s(p)z_{2,\nu_1}\cdots z_{2,\nu_J}.
\ee
We have included an explicit $J$ label to keep track of the Lorentz irrep when working in index-free formalism. By Lorentz invariance and the homogeneity properties of $\Pi_s(p)$, we must have
\be	\label{eq:tstprojectorcovariant}
\Pi_{J,s}(p;z_1,z_2)\equiv (-p^2)^{-J}(-z_1\.p)^{J}(-z_2\.p)^{J}\Pi_{J,s}(\eta),
\ee
where $\Pi_{J,s}(\eta)$ is a polynomial of degree at most $J$ and
\be
	1-\eta=\frac{p^2(z_1\.z_2)}{(z_1\.p)(z_2\.p)}.
\ee
In particular, if we set $p=(1,0,\ldots,0)$ and $z_1=(1,n_i)$, where $n_i$ are unit vectors in $\R^{D-1}$, then $\eta=(n_1\.n_2)$. Since $\Pi_{J,s}$ is projecting the spin-$J$ $\SO(d)$ irrep onto the spin-$s$ $\SO(d-1)$ irrep, we should have
\be
	\Pi_{J,s}(\eta)\propto C_s^{(\tfrac{d-3}{2})}(\eta),
\ee
where $C_s^{(\tfrac{d-3}{2})}$ is a Gegenbauer polynomial.\footnote{This follows, for example, from the quadratic Casimir equation for $\SO(d-1)$.} We can fix the coefficients by requiring, as a linear operator,
\be
	\sum_{s=0}^J\Pi_{J,s}(p)=1,
\ee
or in other words
\be
	\sum_{s=0}^J\Pi_{J,s}(\eta)=(z_1\.z_2)^J=(\eta-1)^J.
\ee
This leads to
\be\label{eq:sod-1projector}
	\Pi_{J,s}(\eta)=\frac{2^{-J}J!(d+J-2)_J(d-2)_J}{(\tfrac{d-1}{2})_J}\frac{(-1)^{s+J}(d+2s-3)}{(J-s)!(d-3)_{J+s+1}}C_s^{(\tfrac{d-3}{2})}(\eta).
\ee

We will normalize the time-ordered two-point function for spacelike separation by 
\be\label{eq:twoptnorm}
	\<\cO(x_1,z_1)\cO(x_2,z_2)\>=\frac{(z_1\.I(x_{12})\.z_2)^J}{x_{12}^{2\De}},
\ee
where 
\be
	I_{\mu\nu}(x)=h_{\mu\nu}-2\frac{x_\mu x_\nu}{x^2}.
\ee
With this normalization, one can compute~\cite{Dobrev:1977qv}
\be\label{eq:twoptfourier}
	&\<\cO(p,z_1)|\cO(p,z_2)\>\nn\\
	&=\frac{2^{d-2\De}2\pi^{\frac{d+2}{2}}\Gamma(\De+2-d)}{\Gamma(\De-\tfrac{d-2}{2})\Gamma(\De+J)\Gamma(\De+2-d-J)}
	\sum_{s=0}^J \frac{(-1)^s (\De-1)_s}{(d-\De-1)_s}(-1)^{J+s}\Pi_{J,s}(p;z_1,z_2)(-p^2)^{\De-\frac{d}{2}}\theta(p).
\ee
We reproduce the calculation in appendix~\ref{app:twoptforier}. We then read off
\be\label{eq:twoptcoeffs}
	\cA_s(\De,J)=\frac{2^{d-2\De}2\pi^{\frac{d+2}{2}}\Gamma(\De+2-d)}{\Gamma(\De-\tfrac{d-2}{2})\Gamma(\De+J)\Gamma(\De+2-d-J)}\frac{(-1)^s (\De-1)_s}{(d-\De-1)_s}(-1)^{J+s}.
\ee

An interesting application of~\eqref{eq:twoptfourier} is that it proves sufficiency of the usual unitarity bounds. One can check that when these bounds are satisfied, the combination $(-1)^{J+s}\Pi_{J,s}(p;z_1,z_2)$ corresponds to a positive-definite bilinear form, $(-1)^{J+s}\cA_s(\De,J)$ is positive, and that~\eqref{eq:twoptfourier} is locally integrable. This is sufficient to show that
\be
	\int \frac{d^d p}{(2\pi)^d} f_{\mu_1\ldots \mu_J}(p)|\cO^{\mu_1\ldots \mu_J}(p)\>
\ee
has non-negative norm for any test function $f$. 

Let us look at some simple cases which are relevant for the examples that we discuss below. First of all, if $J=0$, we can only have $s=0$ and
\be
	\cA_0(\De,0)=\frac{2^{d-2\De}2\pi^{\frac{d+2}{2}}}{\Gamma(\De-\frac{d-2}{2})\Gamma(\De)}.
\ee
This is positive for $\De>\frac{d-2}{2}$, in accord with the unitarity bound. As $\De\to\frac{d-2}{2}$, $\cA_0(\De,0)$ goes to 0. Combined with the factor $(-p^2)^{\De-\frac{d}{2}}$ we find that for $\De=\frac{d-2}{2}$ the two-point function is proportional to $\de(p^2)$, as is expected for the theory of free scalars.\footnote{Recall that $\frac{\e}{x^{1-\e}}\to\de(x)$.}

Suppose now $J=1$. We have
\be
	\cA_0(\De,1)&=-\frac{2^{d-2\De}2\pi^{\frac{d+2}{2}}(\De-d+1)}{\Gamma(\De-\tfrac{d-2}{2})\Gamma(\De+1)},\\
	\cA_1(\De,1)&=\frac{2^{d-2\De}2\pi^{\frac{d+2}{2}}(\De-1)}{\Gamma(\De-\tfrac{d-2}{2})\Gamma(\De+1)}.
\ee
This has the positivity properties mentioned above for $\De>d-1$, and for $\De=d-1$ we find
\be\label{eq:A11}
	\cA_0(\De,1)&=0,\\
	\cA_1(\De,1)&=\frac{2^{3-d}\pi^{\frac{d+2}{2}}(d-2)}{\Gamma(\tfrac{d}{2})\Gamma(d)}.
\ee
This is consistent with the fact that spin-1 operators with $\De=d-1$ are conserved currents, i.e.\ they transform in a short multiplet. The condition $\cA_0(\De,1)=0$ simply says that the scalar $s=0$ component vanishes,
\be
	p_\mu \cO^\mu (p)=0.
\ee
In position space this is just the conservation equation
\be
	\ptl_\mu \cO^\mu(x)=0.
\ee
This pattern persists for higher-spin operators: at the unitarity bound $\De=J+d-2$ only the $s=J$ component of the operator survives, i.e.\ only $\cA_J(\De,J)$ is non-zero. This is the CFT analog of the statement that massless particles transform in irreducible representations of the little group $\SO(d-1)$ rather than $\SO(d)$ (here one should think about QFT in $d+1$ dimensions, i.e.\ the flat space limit of AdS${}_{d+1}$/CFT${}_d$ correspondence), see, e.g.~\cite{Kravchuk:2016qvl}.

\subsubsection{Light transform of a general three-point function}
\label{sec:lighttransformgeneralthreept}

We now turn to the calculation of the three-point functions which enter~\eqref{eq:tchannelgeneralblock}. We will first apply the light-transform and then the Fourier transform.

In this section we heavily utilize the embedding formalism~\cite{Costa:2011mg,Costa:2011dw}. Let us briefly review the basic features of this formalism. The space-time points in $\R^{d-1,1}$ are put in one-to-one correspondence with null rays in $\R^{d,2}$. The conformal group $\SO(d,2)$ acts linearly in this space. The points in $\R^{d,2}$ are denoted by $X$ and null rays can be described by $X$ subject to $X^2=0$ and identification $X\sim \l X$ for $\l>0$. If we introduce the components $X^{\pm}, X^\mu$ (where $\mu$ runs over indices of $\R^{d-1,1}$) such that
\be
	X^2=-X^+X^-+X^\mu X_\mu,
\ee
then $x^\mu\in \R^{d-1,1}$ can be embedded as
\be
	(X^+,X^-,X^\mu)=(1,x^2,x^\mu).
\ee
Here we used $X\sim \l X$ to set $X^+=1$.\footnote{The points with $X^+=-1$ correspond to a different Poincare patch of the Lorentzian cylinder. For details see, e.g.,~\cite{Kravchuk:2018htv}.} Local operators can be described by functions $\cO(X)$, defined for $X^2=0$, which are homogeneous
\be
	\cO(\l X)=\l^{-\De}\cO(X).
\ee
Traceless-symmetric spin-$J$ representations are described by adding dependence on a polarization vector $Z$, subject to $Z^2=X\.Z=0$, and
\be\label{eq:Zgauge}
	\cO(X,\l Z+\a X)=\l^J\cO(X,Z).
\ee
In terms of the $\R^{d-1,1}$ polarization vector $z$ and coordinate $x$ we can identify
\be
	(Z^+,Z^-,Z^\mu)=(0,2(x\.z),z^\mu).
\ee
Here, we used the equivalence $Z\sim Z+\a X$ to set $Z^+=0$. 

We will use the embedding formalism to compute the action of the light transform~\eqref{eq:lighttransformdefinition} on correlation functions of local operators. For this, we need its form in embedding space~\cite{Kravchuk:2018htv},
\be\label{eq:EFlighttransform}
	\wL[\cO](X,Z)=\int_{-\oo}^{+\oo}d\a\, \cO(Z-\a X,-X).
\ee
Note that the arguments $X$ and $Z$ are effectively swapped in the right hand side compared to the Minkowski coordinates $x$ and $z$ entering in~\eqref{eq:lighttransformdefinition}.

As shown in~\cite{Costa:2011mg}, a general parity-even three-point function of traceless-symmetric primary operators can be built out of two basic objects $V_{i,jk}$ and $H_{ij}$, defined as
\be
	H_{ij}&=-2(Z_i\.Z_j)(X_i\. X_j)+2(Z_i\.X_j)(Z_j\.X_i)=-4Z_i^{[m}X_i^{n]}Z_{j,[m}X_{j,n]},\nn\\
	V_{i,jk}&=\frac{(Z_i\.X_j)(X_i\.X_k)-(Z_i\.X_k)(X_i\.X_j)}{(X_j\.X_k)}=2\frac{Z_i^{[m}X_i^{n]}X_{j,m}X_{k,n}}{(X_j\.X_k)}.
\ee
Note that due to the condition~\eqref{eq:Zgauge}, $Z_i$ only enters these expressions in the combination $Z_i^{[m}X_i^{n]}$. Moreover, this is also the combination in which $X_i$ enters into those invariants above which contain $Z_i$. 

This fact greatly simplifies the computation of the light transform~\eqref{eq:EFlighttransform} of three-point structures. Indeed, the definition instructs us to replace $X\to Z-\a X$, $Z\to -X$, and integrate over $\a$. The combination $Z_i^{[m}X_i^{n]}$ is invariant under this replacement and thus factors out of the integral. For example, this implies that
\be
	\wL_i [V_{i,jk}F(Z_i,X_i,\cdots)]=V_{i,jk}\wL_i [F(Z_i,X_i,\cdots)],
\ee
where notation $\wL_i$ means that the light transform is applied to point $i$. Similarly, we can factor out all $H_{jk}$ from under $\wL_i$.\footnote{An alternative way to phrase this observation is to say that $Z^{[m}X^{n]}$ is a weight-shifting operator~\cite{Karateev:2017jgd} that commutes with the action of the light-transform. Indeed, it is the only weight-shifting operator in the adjoint representation which shifts $(\De,J)$ by $(-1,1)$, and this shift is invariant under the Weyl reflection associated with $\wL$. The only non-differential weight-shifting operators which enjoy this property are the powers of $Z^{[m}X^{n]}$.}

Therefore, if we start with a three-point tensor structure
\be
	\<\cO_1\cO_2\cO_3\>=\frac{g(V_{1,23},H_{12},H_{13},H_{23})V_{2,31}^{m_2}V_{3,12}^{m_3}}{
		X_{12}^{\frac{\bar\tau_1+\bar\tau_2-\bar\tau_3}{2}}
		X_{13}^{\frac{\bar\tau_1+\bar\tau_3-\bar\tau_2}{2}}
		X_{23}^{\frac{\bar\tau_2+\bar\tau_3-\bar\tau_1}{2}}
	},
\ee
where $g$ is an arbitrary function with appropriate homogeneity, and $\bar\tau_i=\De_i+J_i$, we find that
\be
	\<0|\cO_2\wL[\cO_1]\cO_3|0\>=g(V_{1,23},H_{12},H_{13},H_{23})\<0|\cO'_2\wL[\f_1]\cO'_3|0\>.
\ee
Here we defined the three-point tensor structure
\be\label{eq:intermediate3pt}
	\<\f_1\cO'_2\cO'_3\>=\frac{V_{2,31}^{m_2}V_{3,12}^{m_3}}{
		X_{12}^{\frac{\bar\tau_1+\bar\tau_2-\bar\tau_3}{2}}
		X_{13}^{\frac{\bar\tau_1+\bar\tau_3-\bar\tau_2}{2}}
		X_{23}^{\frac{\bar\tau_2+\bar\tau_3-\bar\tau_1}{2}}
	},
\ee
where new formal operators $\cO'_i$ have spin $J'_i=m_i$ and dimension $\De'_i=\De_i+J_i-m_i$. The scalar $\f_1$ has dimension $\De_\f=\bar\tau_1$. Thus, the light-transform of a general three-point tensor structure is reduced to light-transforms of a special class of three-point tensor structures, where the light-transformed operator is a scalar.

On general grounds, we must have
\be\label{eq:lighttransformansatz}
	\<0|\cO'_2\wL[\f_1]\cO'_3|0\>\propto\frac{
		V_{1,23}^{1-\De_\f}
		V_{2,31}^{J_2'}
		V_{3,12}^{J_3'}
		\, f\p{\frac{H_{12}}{V_{1,23}V_{2,31}},\frac{H_{13}}{V_{1,23}V_{3,12}}}
	}{
		X_{12}^{\frac{\bar\tau'_1+\bar\tau'_2-\bar\tau'_3}{2}}
		X_{13}^{\frac{\bar\tau'_1+\bar\tau'_3-\bar\tau'_2}{2}}
		X_{23}^{\frac{\bar\tau'_2+\bar\tau'_3-\bar\tau'_1}{2}}
	},
\ee
where $f(x,y)=1+O(x,y)$ is a polynomial of degree at most $m_2$ in $x$ and at most $m_3$ in $y$, and we defined $J_1'=1-\De_\f$ and $\De_1'=1$. We did not allow any factors of $H_{23}$ because they contain inner products $(z_2\.z_3)$ and it is easy to see that the light-transform integral cannot produce them.

To further constrain the form of the function $f$ it is useful to step back and discuss some general properties of the light-transform. The light-transform in general acts on continuous-spin operators and yields new continuous spin operators. Here ``continuous-spin'' doesn't necessarily mean $J\notin \Z_{\geq 0}$, but rather that the operator is not polynomial in its polarization vector $z$ (or $Z$ in embedding space notation). In this sense, the $J=0$ operator $\phi_1$ in~\eqref{eq:intermediate3pt} is special in that it is polynomial in $Z_1$.\footnote{In this case it simply means that it is independent of $Z_1$.} We will refer only to the operators which satisfy this requirement as ``integer-spin.'' 

The structure~\eqref{eq:intermediate3pt} is the only three-point tensor structure that is free of $(z_2\.z_3)$ and also consistent with all three operators being of integer spin. Similarly, the structure~\eqref{eq:lighttransformansatz} can be singled out as the only structure which is free of $(z_2\.z_3)$ and corresponds to two integer-spin operators and the \textit{light-transform of an integer-spin operator} $\f_1$.

The fact that $\f_1$ is an integer-spin operator can be expressed as
\be
	D_m \f_1=\p{(\tfrac{d}{2}-2)\frac{\ptl}{\ptl Z^m}-\frac{1}{2}Z_m\frac{\ptl^2}{\ptl Z\.\ptl Z}}\f_1=0,
\ee
where $D_m$ is the Todorov/Thomas operator~\cite{Costa:2011mg}.\footnote{This rather involved form of $D_m$ is required to make sure it is consistent with the fact that $\phi$ is only defined for $Z^2=X^2=Z\.X=0$. Because of this, a single derivative $\ptl/\ptl Z$ isn't good enough.} This should be thought of as a shortening condition for $\phi_1$. It is natural to expect that there exists a differential operator $D_m^L$ which provides a dual shortening condition for $\wL[\f_1]$, i.e.
\be\label{eq:lightshortening}
	D_m\f_1=0\Longrightarrow D^L_m \wL[\f_1]=0.
\ee
It is easy to guess the quantum numbers of $D^L_m\wL[\f_1]$ by assuming that they are just those of $\wL[D_m\f_1]$. A simple exercise shows that it has $\De=2,J=1-\De_\f$, and the index $m$ should be thought of as being in the second row of the Young diagram (the first row is accounted for by $Z$). This allows us to write an ansatz for $D_m^L$ and fix the coefficients by requiring consistency with various embedding space constraints. 
We find that
\be
	W^m D_m^L=(\De_\f-1)(W\.\frac{\ptl}{\ptl X})+(Z\.\frac{\ptl}{\ptl X})(W\.\frac{\ptl}{\ptl Z})
\ee
satisfies all the required properties, including~\eqref{eq:lightshortening}. Here $W$ is a polarization vector for the second-row indices, and satisfies $W^2=W\.X=W\.Z=0$ and $W\sim W+\a Z+\b X$.\footnote{We discuss/review the embedding formalism for general Young diagrams in \cite{AnecOPE}.}

We can therefore constrain the function $f$ by requiring that
\be
	\<0|\cO'_2 (D_m^L \wL[\f_1])\cO'_3|0\>=0.
\ee
By expanding this equation into appropriate conformally-invariant tensor structures, we find the following constraints for $f$,
\be
	&[x(x+2)\ptl_x^2-y(y+2)\ptl_x\ptl_y+((1+x)(\De_\f-J'_2)+(1+y)J'_3+\De'_{23})\ptl_x-J'_2(\De_\f-1)]f(x,y)=0,\nn\\
	&[y(y+2)\ptl_y^2-x(x+2)\ptl_x\ptl_y+((1+y)(\De_\f-J'_3)+(1+x)J'_2-\De'_{23})\ptl_y-J'_3(\De_\f-1)]f(x,y)=0.
\ee 
The solution to these equations with $f(x,y)=1+O(x,y)$ is given by
\be\label{eq:appelF2parameters}
	f(x,y)=F_2(-J_1';-J'_2,-J'_3;\thalf(\tau_1'+\tau_2'-\tau_3'),\thalf(\tau'_1-\tau'_2+\tau'_3);-\thalf x,-\thalf y),
\ee
where as before $J_1'=1-\De_\f$, $\De'_1=1$, and $\tau'_i=\De'_i-J'_i$, while $F_2$ is the Appell $F_2$ hypergeometric function
\be
	F_2(\a;\b,\b';\g,\g';x,y)\equiv \sum_{m=0}^\oo \sum_{n=0}^\oo \frac{(\a)_{m+n}(\b)_m(\b')_n}{m! n! (\g)_m(\g')_n}x^m y^n.
\ee
Note that the coefficients of the Taylor expansion of $F_2$ in either variable are given by ${}_2F_1$ hypergeometric functions in the other variable.

Since we have uniquely fixed the form of the function $f(x,y)$, it only remains to fix the overall coefficient in~\eqref{eq:lighttransformansatz}. This can be done by choosing a degenerate kinematic configuration which simplifies the integrals. We do this in appendix~\ref{app:lighttransformapp}. Here we just quote the result,
\be\label{eq:lighttransformresult}
\<0|\cO'_2\wL[\f_1]\cO'_3|0\>=-2\pi i \frac{e^{i\pi \bar\tau'_2}2^{J_1'}\Gamma(-J_1')}{\Gamma(\tfrac{\tau'_1+\tau_2'-\tau_3'}{2})\Gamma(\tfrac{\tau'_1-\tau_2'+\tau_3'}{2})}&\frac{
	(-V_{1,23})^{J_1'}
	(-V_{2,31})^{J_2'}
	(-V_{3,12})^{J_3'}
}{
	X_{12}^{\frac{\bar\tau'_1+\bar\tau'_2-\bar\tau'_3}{2}}
	X_{13}^{\frac{\bar\tau'_1+\bar\tau'_3-\bar\tau'_2}{2}}
	(-X_{23})^{\frac{\bar\tau'_2+\bar\tau'_3-\bar\tau'_1}{2}}
}\nn\\
&\times f\p{\frac{H_{12}}{V_{1,23}V_{2,31}},\frac{H_{13}}{V_{1,23}V_{3,12}}}\qquad ((3>2)\approx 1).
\ee
which holds for causal relations $(3>2)\approx 1$. In other words, 3 is in the future of 2 and both are spacelike from 1.

Let us apply the results of this section to an example which will be useful below, namely to the three-point function $\<TJJ\>$, where $T$ is the stress-tensor and $J$ is a spin-1 current. We have the general form for the three-point function
\be
\label{eq:myfavoriteequation}
	\<T_1 J_2 J_3\>=\frac{a V_1^2 V_2 V_3+b V_1 H_{12} V_3+c V_{1} H_{13} V_2+ h V_1^2 H_{23} + k H_{12}H_{13}}{
		X_{12}^{\frac{\bar\tau_1+\bar\tau_2-\bar\tau_3}{2}}
		X_{13}^{\frac{\bar\tau_1+\bar\tau_3-\bar\tau_2}{2}}
		X_{23}^{\frac{\bar\tau_2+\bar\tau_3-\bar\tau_1}{2}}
	},
\ee
where $\tau_1=d+2, \tau_2=\tau_3=d$, and we have added subscripts to the operators to indicate at which point they are inserted. Furthermore, we used the standard notation $V_1=V_{1,23},\, V_2=V_{2,31}$ and $V_3=V_{3,12}$. The coefficients $a, b, c, h, k$ are constrained by the conservation conditions for $T$ and $J$, by permutation symmetry in the two $J$s, and by the Ward identity for stress-tensor. These imply
\be\label{eq:conservationandsymmetry}
	b=c,\quad (d+2)h-d b-a=0,
	\quad (d-2)k-2h+2b=0,
\ee
and
\be\label{eq:TJJward}
	C_J=\frac{(k-b)S_d}{d}, \qquad S_d=\vol S^{d-1}=\frac{2\pi^{d/2}}{\Gamma(\frac{d}{2})},
\ee
where $C_J$ is defined as
\be
	\<J_2J_3\>=C_J\frac{H_{23}}{X_{23}^{\bar\tau_3}}.
\ee

In this section, however, it is more convenient to treat the structures that are multiplied by $a,b,c,h,k$ independently. Let us focus on the structure with coefficient $a$ in (\ref{eq:myfavoriteequation}). Using the results above we find that
\be
	\<J_2 \wL[T_1] J_3\>=V_1^2 \wL_1\left[\frac{V_2 V_3}{X_{12}^{\frac{d+2}{2}}
		X_{13}^{\frac{d+2}{2}}
		X_{23}^{\frac{d-2}{2}}}\right],
\ee
where the light transform is applied to the correlation function with
\be
	&J_1'=-d-1, \quad \De_1'=1,\\
	&J_2'=J_3'=1, \quad \De_2'=\De_3'=d-1.
\ee
We can now use equations~\eqref{eq:lighttransformresult} and~\eqref{eq:appelF2parameters} to write (for $(3>2)\approx 1$),
\be
	\wL_1\left[\frac{V_2 V_3}{X_{12}^{\frac{d+2}{2}}
		X_{13}^{\frac{d+2}{2}}
		X_{23}^{\frac{d-2}{2}}}\right]=
	2\pi i \frac{2^{-d-1}\Gamma(d+1)}{\Gamma(\tfrac{d+2}{2})^2}&\frac{
		V_{1}^{-d-1}
		V_{2}
		V_{3}
	}{
	X_{12}^{\frac{-d}{2}}
	X_{13}^{\frac{-d}{2}}
	(-X_{23})^{\frac{3d}{2}}
} f\p{\frac{H_{12}}{V_{1}V_{2}},\frac{H_{13}}{V_{1}V_{3}}},
\ee
where
\be
	f(x,y)=&F_2(-J_1';-1,-1;\thalf(d+2),\thalf(d+2);-\thalf x,-\thalf y)\nn\\
	=&1+\frac{d+1}{d+2}(x+y)+\frac{d+1}{d+2}x y.
\ee
Therefore, in this case
\be
	\<J_2 \wL[T_1] J_3\>=2\pi i \frac{2^{-d-1}\Gamma(d+1)}{\Gamma(\tfrac{d+2}{2})^2}&\frac{
		V_{1}^{-d+1}V_{2}V_{3}
		+\tfrac{d+1}{d+2}\p{V_{1}^{-d}V_{3}H_{12}
		+V_{1}^{-d}V_{2}H_{13}
		+V_{1}^{-d-1}H_{12}H_{13}}
	}{
		X_{12}^{\frac{-d}{2}}
		X_{13}^{\frac{-d}{2}}
		(-X_{23})^{\frac{3d}{2}}
	}.
\ee
Calculation of the light transforms for other structures, corresponding to coefficients $b,c,h,k$, is completely analogous. The complete result is
\be\label{eq:completeTJJlighttransform}
	\<J_2 \wL[T_1] J_3\>=2\pi i \frac{2^{-d-1}\Gamma(d+1)}{\Gamma(\tfrac{d+2}{2})^2}&V_{1}^{-d-1}\frac{a' V_1^2 V_2 V_3+b' V_1 H_{12} V_3+c' V_{1} H_{13} V_2+ h' V_1^2 H_{23} + k' H_{12}H_{13}}{
	X_{12}^{\frac{-d}{2}}
	X_{13}^{\frac{-d}{2}}
	(-X_{23})^{\frac{3d}{2}}
},
\ee
where
\be
	&a'=a,\quad b'=-\frac{d}{d+2}b+\frac{d+1}{d+2}a,\quad c'=-\frac{d}{d+2}c+\frac{d+1}{d+2}a,\nn\\
 	&k'=k+\frac{d+1}{d+2}(a-b-c).
\ee
Note that the algorithm for computing the light transform is much simpler than in the case of the shadow transform~\cite{Karateev:2018oml}.

\subsubsection{Fourier transform of three-point functions}
\label{sec:three-pt-fourier}

Above we have described how to compute the light transform 
\be\label{eq:lighttransformbeforefourier}
	\<0|\cO_2(x_2,z_2) \wL[\cO_1](x_1,z_1) \cO_3(x_3,z_3)|0\>
\ee
for a general three-point structure. We now need to set $x_1=\oo$ and Fourier-transform $\cO_2$ and $\cO_3$. Since after setting $x_1=\oo$ the three-point function becomes translation-invariant in $x_2$ and $x_3$, it suffices to only Fourier-transform $\cO_3$. Therefore, we want to compute the Fourier transforms
\be\label{eq:fouriertarget}
\<\cO_2(p,z_2)|\wL[\cO_1](\oo,z_1)|\cO_3(p,z_2)\>=\int d^dx e^{ipx} \<0|\cO_2(0,z_2)\wL[\cO_1](\oo,z_1)\cO_3(x,z_3)|0\>.
\ee

The configuration in the integrand corresponds to
\be\label{eq:fourierframe}
	&X_1=(0,1,\vec 0),\quad X_2=(1,0,\vec 0),\quad X_3=(1,x^2,x),\nn\\
	&Z_1=(0,0,z_1),\quad Z_2=(0,0,z_2),\quad Z_3=(0,2(x\.z_3),z_3).
\ee
Under this substitution we have
\be\label{eq:fourierframeinvariants}
	&V_{1,23}=-x^{-2}(x\.z_1),\quad V_{2,31}=(x\.z_2),\quad V_{3,12}=(x\.z_3),\nn\\
	&H_{12}=(z_1\.z_2),\quad H_{23}=x^2(z_2\.I(x)\.z_3),\quad H_{31}=(z_1\.z_3).
\ee
Using these identities, the three-point function under the integral in~\eqref{eq:fouriertarget} can be reduced to a linear combination of terms of the form
\be\label{eq:fourierframemonomial}
	(z_1\.z_2)^{n_{12}}(z_2\.z_3)^{n_{23}}(z_3\.z_1)^{n_{31}}(-x\.z_2)^{m_2}(-x\.z_3)^{m_3}(-x\.z_1)^{1-\De_1-n_{12}-n_{31}}(-x^2)^{\l/2}
\ee
where
\be\label{eq:mnconstraints}
	\l&=(1-J_1)-\De_2-\De_3-m_2-m_3-(1-\De_1-n_{12}-n_{31}),\nn\\
	J_2&=n_{23}+n_{12}+m_2,\nn\\
	J_3&=n_{31}+n_{23}+m_3,
\ee
and $m_2,m_3,n_{12},n_{23},n_{31}$ are non-negative integers. 

In the simplest case when $J_2=J_3=0$, there is only one structure
\be
	\frac{(-x\.z_1)^{1-\De_1}}{(-x^2)^{\tfrac{\De_2+\De_3-(1-J_1)+(1-\De_1)}{2}}}.
\ee
It is straightforward to compute the Fourier transform
\be
	&\int d^dx e^{ipx}\frac{(-x\.z_1)^{1-\De_1}}{(-x^2)^{\tfrac{\De_2+\De_3-(1-J_1)+(1-\De_1)}{2}}}\nn\\
	&=\hat\cF_{\De_2+\De_3-(1-J_1),1-\De_1}(-p\.z_1)^{1-\De_1}(-p^2)^{\tfrac{\De_2+\De_3-(1-J_1)-(1-\De_1)-d}{2}}\theta(p),
\ee
where the $i\e$ prescription $x^0\to x^0+i\e$ has to be used, and the coefficient $\hat\cF_{\De,J}$ is given by
\be\label{eq:fouriercoefficient}
	\hat \cF_{\De,J}=2\pi \frac{e^{-i\pi\De/2}2^{d-\De}\pi^{\frac{d}{2}}}{\Gamma(\frac{\De+J}{2})\Gamma(\frac{\De+2-d-J}{2})}.
\ee

We can reuse this result for general $J_2$ and $J_3$. Note that exactly the same calculation as above works for structures with $m_2=m_3=0$. To obtain the result for non-zero $m_2$ and $m_3$ we can introduce the following auxiliary basis,
\be\label{eq:fourierframeharmonic}
&\{z_1^{J_1}z_2^{J_2}z_3^{J_3}|x^{-\De}\}_{n_{12}n_{23}n_{31}}\nn\\
&\equiv(-x^2)^{\frac{-\De-m_1-m_2-m_3}{2}}(z_1\.z_2)^{n_{12}}(z_2\.z_3)^{n_{23}}(z_3\.z_1)^{n_{31}}(z_2\.D_{z_1})^{m_2}(z_3\.D_{z_1})^{m_3}(-x\.z_1)^{m_1+m_2+m_3},
\ee
where $m_2$ and $m_3$ are given by~\eqref{eq:mnconstraints}, $m_1=J_1-n_{12}-n_{31}$, and $D_z$ is the Thomas/Todorov operator~\cite{Costa:2011mg}. Acting with, for example, $(z_2\.D_{z_1})$ on $(-x\.z_1)^\a$ produces terms of two types. One contains $(x\.z_2)$, which is the desired term. If we had only this term, then~\eqref{eq:fourierframeharmonic} would be proportional to~\eqref{eq:fourierframemonomial}. However, there is a second term, proportional to $(z_1\.z_2)$. Nevertheless, it is clear that this term leads to contributions to~\eqref{eq:fourierframeharmonic} which have fewer powers of $(x\.z_i)$ than~\eqref{eq:fourierframemonomial}. This means that the relationship between the structures~\eqref{eq:fourierframeharmonic} and~\eqref{eq:fourierframemonomial} is given by a triangular matrix, and thus can be straightforwardly inverted. In particular~\eqref{eq:fourierframemonomial} and~\eqref{eq:fourierframeharmonic} span the same space of structures.

The advantage of using~\eqref{eq:fourierframeharmonic} is that, obviously,
\be\label{eq:fouriermaster}
	\int d^dx e^{ipx} \{z_1^{J_1}z_2^{J_2}z_3^{J_3}|x^{-\De}\}_{n_{12}n_{23}n_{31}}=\hat\cF_{\De,J_1+J_2+J_3-2n_{12}-2n_{23}-2n_{31}}\{z_1^{J_1}z_2^{J_2}z_3^{J_3}|p^{\De-d}\}_{n_{12}n_{23}n_{31}}\theta(p).
\ee
Therefore, the Fourier transform~\eqref{eq:fouriertarget} can be computed by expanding the integrand in the basis~\eqref{eq:fourierframeharmonic} and applying~\eqref{eq:fouriermaster}.

This method works well in practice if $J_2$ and $J_3$ are some concrete integers which are not very large. For example, let us use it to compute the example we studied above, namely
\be
	\<J(p,z_2)|\wL[T](\oo,z_1)|J(p,z_3)\>.
\ee
which corresponds to $J_2=J_3=1$. We have to look at five structures,
\be
	&\<J(0,z_2)|\wL[T](\oo,z_1)|J(x,z_3)\>=\nn\\
	&\qquad\tl a\{z_1^{1-d}z_2 z_3|x^{-2d+1}\}_{000}+\tl b
	\{z_1^{1-d}z_2 z_3|x^{-2d+1}\}_{100}+\tl c
	\{z_1^{1-d}z_2 z_3|x^{-2d+1}\}_{010}\nn\\
	&\qquad+\tl h\{z_1^{1-d}z_2 z_3|x^{-2d+1}\}_{001}+
	\tl k\{z_1^{1-d}z_2 z_3|x^{-2d+1}\}_{110}.
\ee
For example,
\be
	\{z_1^{1-d}z_2 z_3|x^{-2d+1}\}_{000}=\frac{(d-3)_4(-x\.z_1)^{-1-d}}{4(-x^2)^{\frac{d+2}{2}}}\Big(
		&\frac{d-2}{d-1}(x\.z_1)^2(x\.z_2)(x\.z_3)-x^2(x\.z_1)(z_1\.z_2)(x\.z_3)\nn\\
		&-x^2(x\.z_1)(x\.z_2)(z_1\.z_3)+x^4 (z_1\.z_2)(z_1\.z_3)\nn\\
		&+\frac{1}{d-1}x^2 (x\.z_1)^2(z_2\.z_3).
	\Big).
\ee
After computing the other structures, it is straightforward to plug~\eqref{eq:fourierframeinvariants} into~\eqref{eq:completeTJJlighttransform} to find the coefficients $\tl a,\tl b,\tl c,\tl h,\tl k$ and then use~\eqref{eq:fouriermaster}. These intermediate steps get somewhat messy and we do not reproduce them explicitly here.

To write down the final result, it is convenient to apply a Lorentz transformation and a dilatation so that $p=(1,\vec 0)$. We can then choose $z_i=(1,n_i)$, where $n_i$ are unit vectors. In this frame the Fourier transform becomes
\be\label{eq:boundaryJEJ}
	&\<J_2(p,z_2)|\wL[T](\oo,z_1)|J_3(p,z_3)\>\nn\\
	&=C_JS_d^2\frac{2^{1-2d}\pi^{2-d}(d-2)}{d-1}\p{(n_2\.n_3)+a_2\p{(n_1\.n_2)(n_1\.n_3)-\frac{(n_2\.n_3)}{d-1}}},
\ee
where $C_J$ and $a_2$ define $b$ by
\be
	b=\frac{C_J(d-2)d(a_2+d(d-1))}{S_d(d-1)^3}.
\ee
The other constants $a,c,h,k$ are determined by $b$ and $C_J$ through equations~\eqref{eq:conservationandsymmetry} and~\eqref{eq:TJJward}.

Notice that this method of computing Fourier transforms quickly gets out of hand if, say, $J_3$ is large, or if we want to keep it as a free parameter. The latter is important if we want to compute all the conformal blocks for a particular event shape. In section~\ref{sec:scalarfourier} we describe the calculation of Fourier transforms relevant for scalar event shapes at generic $J_3$. This can be used as a seed for calculation of Fourier transforms for more complicated event shapes at generic $J_3$, although we will not pursue this direction.

\subsection{Holographic multi-point event shapes}
\label{sec:boundarymultipoint}

In section~\ref{sec:computinginbulk} we have computed the following $n$-point even shapes in the bulk theory,
\be
	\<\cE(n_1)\cdots \cE(n_k)\>_\f,\quad \<\cE(n_1)\cdots \cE(n_k)\>_{\e\.J},,\quad \<\cE(n_1)\cdots \cE(n_k)\>_{\e\.T}.
\ee
From the boundary $t$-channel point of view, this calculation corresponds to keeping only the ``comb'' $t$-channel $k$-point blocks which exchange, respectively, $\f,J$, or $T$ in all intermediate channels. We discussed the computation of such $t$-channel $k$-point blocks in section~\ref{sec:tchannelblocks}, here we would like to see how they reproduce the bulk calculations. 

In this section we write all event shapes in the configuration $p=(1,\vec 0)$, $z_i=(1,n_i)$. In the simplest scalar case we have
\be
	\<\f(p)|\cE(n_1)\cdots \cE(n_k)|\f(p)\>=&\<\f(p)|\cE(n_1)|\f(p)\>\frac{1}{\cA(\De)}\<\f(p)|\cE(n_2)|\f(p)\>\cdots\nn\\
	&\times\frac{1}{\cA(\De)}\<\f(p)|\cE(n_k)|\f(p)\>
\ee
while 
\be
	\<\f(p)|\f(p)\>=\cA(\De),
\ee
where we again use notation where the momentum conserving delta-functions $(2\pi)^d\de^d(0)$ are implicitly removed. Combining these expressions together, we find that
\be\label{eq:boundaryscalar}
	\<\cE(n_1)\cdots \cE(n_k)\>_\f=&\frac{\<\f(p)|\cE(n_1)\cdots \cE(n_k)|\f(p)\>}{\<\f(p)|\f(p)\>}\nn\\
	=&\p{\frac{\<\f(p)|\cE(n_1)|\f(p)\>}{\cA(\De)}}^k=\p{\frac{1}{\vol S^{d-2}}}^{k}.
\ee
The last equality follows from the Ward identity
\be
	\int_{S^{d-2}}d^{d-2}n\<\f(p)|\cE(n)|\f(p)\>=p^0 \<\f(p)|\f(p)\>,
\ee
together with the fact that because of Lorentz invariance, $\<\f(p)|\cE(n)|\f(p)\>$ is independent of $n$. Of course, we can also explicitly compute $\<\f(p)|\cE(n)|\f(p)\>$ using the algorithm described in the previous subsection, with the same result. Clearly, given our choice of $p$,~\eqref{eq:boundaryscalar} is equivalent to~\eqref{eq:ansGR}.

This straightforwardly generalizes to the event shapes in $\e\.J$ and $\e\.T$ states. When spinning operators are exchanged, according to~\eqref{eq:tchannelgeneralblock}, we need to glue the three-point functions using $\SO(d-1)$ projectors while summing over different $\SO(d-1)$ components. However, when we are working with $J$ or $T$, the three-point functions only have a single $\SO(d-1)$ component --- of spin-1 or spin-2, respectively --- because of the shortening conditions. Thus, the projectors act trivially. For example, in the case of an $\e\.J$ event shape, we have
\be
	\<\e\.J(p)|\cE(n_1)\cdots \cE(n_k)|\e\.J(p)\>=&\<\e\.J(p)|\cE(n_1)|J_i(p)\>\frac{1}{C_J\cA_1(\De,1)}\<J_i(p)|\cE(n_2)|J_j(p)\>\cdots\nn\\
	&\times\frac{1}{C_J\cA_1(\De,1)}\<J_l(p)|\cE(n_k)|\e\.J(p)\>,
\ee
while 
\be
	\<\e\.J(p)|\e\.J(p)\>=C_J\cA_1(\De,1)\e^\dagger\.\e.
\ee
Notice that we only sum over spatial indices above. This is because for $p=(1,\vec 0)$, $|J_0(p)\>=0$.
We thus match the bulk result~\eqref{eq:Jeventshapebulk} if
\be
	\frac{\<J_i(p)|\cE(n)|J_j(p)\>}{C_J\cA_1(\De,1)}=\frac{H_{ij}(n)}{\vol S^{d-2}}.
\ee
Using~\eqref{eq:boundaryJEJ},~\eqref{eq:A11}, and recalling that $\cE=2\wL[T]$, we can see that this is indeed true if the parameters $a_2$ in~\eqref{eq:boundaryJEJ} and~\eqref{eq:transfermatrixA} are identified. Effectively, we are saying that if one-point event shapes match, so do the higher-point event shapes (in the setting where only on single-trace operator is exchanged). In the case of one-point event shapes this matching is well-known. 

Similarly, for the stress-tensor we need to check
\be
	\frac{\<T_{ij}(p)|\cE(n)|T_{kl}(p)\>}{C_T\cA_2(\De,2)}=\frac{H_{ij,kl}(n)}{\vol S^{d-2}}.
\ee
This is also well-known to be true, given appropriate identifications of OPE coefficients. This result is also easily reproduced using our algorithm.

Let us finally consider an example of a non-minimal coupling of scalars to gravity analogous to the one considered in the end of section~\ref{sec:higherdergravity}. That is, we consider the contribution of a scalar primary $\phi$ to 
\be\label{eq:T2pt}
	\<T(p,z_4)|\cE(n_1)\cE(n_2)|T(p,z_3)\>.
\ee
This contribution is given by
\be\label{eq:massivescalarcontribution}
	\frac{\<T(p,z_4)|\cE(n_1)|\phi(p)\>\<\phi(p)|\cE(n_2)|T(p,z_3)\>}{\<\phi(p)|\phi(p)\>}.
\ee
Conservation and tracelessness of $T$ imply that
\be
	\<T(p,z_4)|\cE(n_1)|\phi(p)\>\propto n_4^a n_4^b\p{n_1^a n_1^b-\frac{\de^{ab}n_1\.n_1}{d-1}}=(n_1\.n_4)^2-\frac{1}{d-1}.
\ee
Explicit calculation shows that if
\be
	\<T_1 T_2 \f\> = \l_{TT\f} \frac{V_{1,23}^2V_{2,31}^2+\ldots}{
		X_{12}^{\frac{\bar\tau_1+\bar\tau_2-\bar\tau_3}{2}}
		X_{13}^{\frac{\bar\tau_1+\bar\tau_3-\bar\tau_2}{2}}
		X_{23}^{\frac{\bar\tau_2+\bar\tau_3-\bar\tau_1}{2}}
	},
\ee
where $\ldots$ contain contributions from $H_{12}^2$ and $V_{1,23}V_{2,31}H_{12}$ which are fixed by conservation of $T$,
then 
\be
	\<T(p,z_4)|\cE(n_1)|\phi(p)\>=\l_{TT\f}\frac{2^{1-d-\De}\pi^{2+\frac{d}{2}}e^{\frac{i\pi}{2}(d-\De)}(d-1)\G(d+1)}{(d-2)\G(d-\tfrac{\De}{2})\G(2+\tfrac{\De}{2})^2\G(\tfrac{d+\De}{2})}\p{(n_1\.n_4)^2-\frac{1}{d-1}}.
\ee
We see that this is non-zero unless $\De=2d+2n$, i.e.\ unless $\phi$ has the dimension of a double-trace $[TT]_{0,n}$. This means that the contribution of a generic single-trace $\phi$ to~\eqref{eq:T2pt} is non-zero and proportional to
\be
	\p{(n_1\.n_4)^2-\frac{1}{d-1}}\p{(n_2\.n_3)^2-\frac{1}{d-1}}.
\ee
Computing the commutator $[\cE(n_1),\cE(n_2)]$ is equivalent to antisymmetrizing in $n_1$ and $n_2$, which clearly gives a non-zero result when applied to the above expression. We therefore see that, similarly to flat space case, a non-minimally coupled scalar ($\l_{TT\f}\neq 0$) leads to a non-zero shock commutator, and must therefore be accompanied by non-minimal couplings to other fields.

\subsection{Structure of the general sum rule}
\label{sec:tsumrule}

In this section we describe the general properties of the sum rule which expresses the commutativity of shocks,
\be\label{eq:generalsumrule}
	\<\cO_4(p,z_4)|[\wL[\cO_1](\oo,z_1),\wL[\cO_2](\oo,z_2)]|\cO_3(p,z_3)\>=0.
\ee
Particularly, we would like to understand some natural components in which this equation can be decomposed, and how various operators in the $t$-channel contribute to these components.

The sum rule is obtained by writing~\eqref{eq:generalsumrule} as
\be\label{eq:commutatorexpansion}
	\<\cO_4|\wL[\cO_1]\wL[\cO_2]|\cO_3\>-\<\cO_4|\wL[\cO_2]\wL[\cO_1]|\cO_3\>=0,
\ee
and expanding both event shapes in $t$-channel conformal blocks. While this is meaningful in any CFT, it is especially interesting to consider this sum rule in a large-$N$ theory. 

Let us assume that $\cO_i$ are single-trace. At leading order in $1/N$, the four-point function $\<\cO_4\cO_1\cO_2\cO_3\>$ is given by the disconnected part. The disconnected part, if at all non-zero, receives contributions from the identity and the double-trace operators only. However, as noted in section~\ref{sec:tchannelblocks}, double-trace operators do not contribute to the event shapes in~\eqref{eq:commutatorexpansion}, and thus to the sum rule. The same is true for the identity operator, since $\wL[\cO_i]|\O\>=0$. The story here is analogous to that of the Lorentzian inversion formula, since each term in~\eqref{eq:commutatorexpansion} can be written as a double commutator. 

Therefore, the leading contribution to the sum rule is given by the single-trace operators.\footnote{Note that it does not make much sense to go to higher $1/N$ orders, since for example in the case $\cO_1=\cO_2=T$ the event shapes are ill-defined beyond the planar order, c.f. section~\ref{sec:convergenceinperturbationtheory}.} As we have seen in the examples above and in section~\ref{sec:computinginbulk}, there exist special, minimal couplings of single trace operators, with which they do not contribute to the sum rule. The sum rule is therefore satisfied if all single trace operators have minimal three-point functions. However, if some single trace operator has a non-minimal coupling, its contribution must be canceled by non-minimal couplings of some other operators.

In the rest of this section we will study the symmetries of the sum rule, and the constraints that these symmetries impose on the potential cancellation of non-minimal couplings.

\subsubsection{Tensor structures}

First, let us discuss the symmetries of equation~\eqref{eq:generalsumrule}. It contains the momentum $p$, which we are free to set to any value. We can choose $p=(1,\vec 0)$. After this, the only symmetry remaining is the $\SO(d-1)$ of spatial rotations transverse to $p$. It is therefore convenient to decompose the spin degrees of freedom of all four operators under this subgroup. 

For the integer-spin operators $\cO_4$ and $\cO_3$ the decomposition is simple and is described, for example, in~\cite{RepresentationsAndSpecialFunctions,Kravchuk:2017dzd}. In the simplest case of a $\SO(1,d-1)$ traceless-symmetric tensor of spin $J$, upon reduction to $\SO(d-1)$ we get traceless-symmetric tensors of spins $s=0,\ldots, J$. If the operator is conserved, then only $s=J$ survives. In general, let us denote by $\rho_i$ the $\SO(1,d-1)$ irreps of these operators, and by $\l_i\in \rho_i$ their $\SO(d-1)$ components.

Decomposition of continuous-spin operators $\wL[\cO_1]$ and $\wL[\cO_2]$ is a bit more non-trivial. Let us explain how it works in the case when the original $\cO_1$ and $\cO_2$ are traceless-symmetric, so that $\wL[\cO_i]$ are as well. In this case, all that we know about $\wL[\cO_1](\oo,z_1)$ as a function of $z_1$ is that it is homogeneous. This homogeneity allows us to completely encode this function by its values for $z_1=(1,n_1)$, and these values are completely unconstrained. We therefore conclude that $\wL[\cO_1](\oo,z_1)$ is equivalent to a scalar function on $S^{d-2}$ parametrized by $n_1$. As is well-known, under the action of $\SO(d-1)$, the space of such functions decomposes into all possible traceless-symmetric representations
\be
	\{\text{functions on }S^{d-2}\}\simeq \bigoplus_{j=0}^\oo j,
\ee
where $j$ denotes a traceless-symmetric irrep of $\SO(d-1)$ of spin $j$. Alternatively, we can say that we are allowed to smear $\wL[\cO_1](\oo,z_1)$ with a spherical harmonic of $n_1$, and such smeared operators transform nicely under $\SO(d-1)$. Furthermore, any smearing function can be decomposed into spherical harmonics.

The left-hand side of~\eqref{eq:generalsumrule} is an $\SO(d-1)$ invariant of the four operators, i.e.~an element of
\be
	\p{\bigoplus_{j_1,j_2=0}^\oo\bigoplus_{\l_3\in\r_3\atop \l_4\in \r_4} j_1\otimes j_2\otimes \l_3\otimes \l_4}^{\SO(d-1)}.
\ee
Such invariants can be conveniently labeled by 
\be
	\{j_1,j_2|\l|\l_3,\l_4\}_s,
\ee
where $\l\in j_1\otimes j_2$ and $\l^*\in \l_3\otimes \l_4$, and $s$ stands for $s$-channel. This invariant is obtained by restricting to a particular term in the direct sums above, and by selecting a particular irrep in $j_1\otimes j_2$.\footnote{It may be the case that $\l^*$ appears in $\l_3\otimes \l_4$ with multiplicity. In this case, we need to add an extra label.}${}^,$\footnote{In the case $\cO_1=\cO_2$, the sum rule is explicitly antisymmetric in $n_1$ and $n_2$. This is reflected in a restriction $j_1\leq j_2$ and a selection rule on $\l$ for $j_1=j_2$. Also the definition of the invariant for $j_1<j_2$ must be altered slightly.} The left hand side of the sum rule can then be expanded
\be
	\<\cO_4|[\wL[\cO_1],\wL[\cO_2]]|\cO_3\>=\sum_{j_1,j_2=0}^\oo\sum_{\l_3\in\r_3\atop \l_4\in \r_4}\sum_{\l\in j_1\otimes j_2\atop \l^*\in \l_3\otimes \l_4} c_{j_1,j_2|\l|\l_3,\l_4}\{j_1,j_2|\l|\l_3,\l_4\}_s.
\ee
The sum rule can be written in components as
\be
	c_{j_1,j_2|\l|\l_3,\l_4}=0.
\ee
Note that these are scalar equations, i.e.\ they contain no cross-ratios.

We would now like to understand which $t$-channel operators these components receive contributions from. First, note that the $t$-channel computes not the commutator, but the individual event shapes (note the arguments in the second event shape),
\be
	\<\cO_4|\wL[\cO_1](n_1)\wL[\cO_2](n_2)|\cO_3\>&=\sum E^{12}_{j_1,j_2|\l|\l_3,\l_4}\{j_1,j_2|\l|\l_3,\l_4\}_s\\ \<\cO_4|\wL[\cO_2](n_1)\wL[\cO_1](n_2)|\cO_3\>&=\sum E^{21}_{j_1,j_2|\l|\l_3,\l_4}\{j_1,j_2|\l|\l_3,\l_4\}_s.
\ee
The coefficients in the sum rule are given by\footnote{There might be an additional relative coefficient between the two terms which depends on the convention for Clebsch-Gordan coefficients and normalization of the invariants. Note that in case $j_1=j_2$ the operation of permuting $j_1$ and $j_2$ has a definite eigenvalue $\pm 1$ depending on $\l$. In the case $\cO_1=\cO_2$ the two terms either cancel or add up, depending on the sign. This corresponds to the selection rule on $\l$ mentioned above.}
\be
	c_{j_1,j_2|\l|\l_3,\l_4}=E^{12}_{j_1,j_2|\l|\l_3,\l_4}-E^{21}_{j_2,j_1|\l|\l_3,\l_4},
\ee
so it is sufficient to understand how $t$-channel operators contribute to $E^{12}$ and $E^{21}$.

The structures defined above are natural from the point of view of computing the commutator, but for the $t$-channel expansion the more natural structures are
\be\label{eq:tchannelstruct}
	\{\l_4,j_1|\l|j_2,\l_3\}_t,
\ee
which are obtained similarly to $\{\cdots\}_s$ structures, but with
\be
	\l\in\otimes j_1\otimes \l_4,\quad \l^*\in j_2\otimes \l_3.
\ee
The usefulness of these structures comes from the fact that $\l$ here is exactly the same as the one summed over in~\eqref{eq:tchannelgeneralblock}. In other words, an operator $\cO$ only contributes to~\eqref{eq:tchannelstruct} with $\l\in \rho_\cO$, where $\rho_\cO$ is the $\SO(d-1,1)$ irrep of $\cO$. Since there are finitely many choices for $\l_3$ and $\l_4$ (for a given event shape), this implies that given an $\cO$, there is a selection rule on possible $j_1$ and $j_2$.

For example, let us consider the sum rule for
\be\label{eq:TTTTsuperconvergence}
	\<T|[\cE,\cE]|T\>=0.
\ee
In this case, we have only one choice for $\l_3$ and $\l_4$ because of the conservation of $T$, i.e.\ $\l_3=\l_4=2$ --- the spin-two traceless-symmetric irrep. Let us for simplicity consider contributions to the sum rule of traceless-symmetric operators. For an operator of spin $J$, the allowed contributions are $\l=0,\ldots J$. The condition $\l\in \l_4\otimes j_1$ then implies $j_1\in\{\l-2,\l,\l+2\}$, and similarly for $j_2$. We then conclude that an operator of spin $J$ contributes only to the structures (either $\{\ldots\}_t$ or $\{\ldots\}_s$) with $j_1, j_2\in \{0,\ldots, J+2\}$. 

This is already non-trivial, since it tells us that contributions to the sum rule from operators of bounded spin live in a finite-dimensional space. This also implies, for example, that in the sum rule, a generic contribution of a spin-$6$ operator cannot be canceled by a spin-$0$ operator. It is less obvious whether a spin-$J$ operator can be completely canceled by spin-$J$ operators. In principle, we can have lots of spin-$J$ operators all contributing to the same finitely many components of the sum rule, so it might seem that there are enough free parameters to cancel out all components. However, in a unitary theory, due to the reality properties of OPE coefficients, the contributions of operators have fixed signs, and it might be that it is impossible to satisfy the sum rule by non-minimal couplings of operators of a single spin $J$. It might even be true that no finite set of spins is sufficient. We leave the investigation of this question to future work.

\subsubsection{$\<T|\cE\cE|T\>$ example}

Here, let us consider two simple contributions to~\eqref{eq:TTTTsuperconvergence}, from the exchange of a stress-tensor itself and from a massive scalar. We will only consider the structures to which the scalar contributes non-trivially. According to the above discussion, before taking the commutator the scalar only contributes to
\be
	\{2,2|0|2,2\}_t,
\ee
which after taking the commutator becomes a combination of
\be
	\{2,2|\l|2,2\}_t
\ee
with all allowed $\l$, i.e. $\l=0,2,4,(3,1),(2,2),(1,1)$, where $(\ell_1,\ell_2)$ denotes a Young diagram with two rows $\ell_1\geq \ell_2$. As we explained above, for the commutator it is more natural to look at the $\{2,2|\l|2,2\}_s$ structures. From the point of view of the $\{2,2|\l|2,2\}_s$ structures, the commutator only contributes to $\l$ which are in the antisymmetric product $2\otimes 2$, i.e.\ to $\l=(3,1)$ and $\l=(1,1)$. This means that we will only get 2 equations involving scalar contributions.

In fact, we can compute that under taking the commutator
\be
	\{2,2|0|2,2\}_t\to&
	\frac{d^2-d-4}{(d+1)(d-2)}\{2,2|0|2,2\}_t
	-\frac{d-1}{(d+3)(d-3)}\{2,2|2|2,2\}_t
	-\frac{1}{36}\{2,2|4|2,2\}_t\nn\\
	&-\frac{1}{12}\{2,2|(2,2)|2,2\}_t
	+\frac{1}{d+1}\{2,2|(1,1)|2,2\}_t
	+\frac{1}{8}\{2,2|(3,1)|2,2\}_t\nn\\
	&=\frac{1}{4}\{2,2|(3,1)|2,2\}_s+\frac{2}{d+1}\{2,2|(1,1)|2,2\}_s,
\ee
where the explicit expressions for structures $\{\cdots\}_t$ and $\{\cdots\}_s$ are given in appendix~\ref{app:structs}.

On the other hand, stress-tensor contribution to $j_1=j_2=2$ is only through
\be
	\{2,2|2|2,2\}_t
\ee
because $T$ has only a spin-2 $\SO(d-1)$ component. Under taking the commutator it goes to
\be
	\{2,2|2|2,2\}_t \to&
	-\frac{2(d-3)(d+3)}{(d+1)(d-1)(d-2)}\{2,2|0|2,2\}_t
	+\frac{d^2-2d+9}{2(d+3)(d-3)}\{2,2|2|2,2\}_t \nn\\
	&
	-\frac{d-3}{18(d-1)}\{2,2|4|2,2\}_t+\frac{d+3}{12(d-1)}\{2,2|(2,2)|2,2\}_t\nn\\
	&
	+\frac{(d-3)(d+3)}{2(d-1)(d+1)}\{2,2|(1,1)|2,2\}_t
	-\frac{1}{2(d-1)}\{2,2|(3,1)|2,2\}_t\nn\\
	&=-\frac{1}{d-1}\{2,2|(3,1)|2,2\}_s+\frac{(d-3)(d+3)}{(d-1)(d+1)}\{2,2|(1,1)|2,2\}_s.
\ee

Scalar exchange of dimension $\De$ contributes, according to the result of section~\ref{sec:boundarymultipoint},
\be
	\<T(n_4)|\cE(n_1)\cE(n_2)|T(n_3)\>&\ni |\l_{TT\f}|^2 q(\De) \p{(n_1\.n_4)^2-\frac{1}{d-1}}\p{(n_2\.n_3)^2-\frac{1}{d-1}}\nn\\
	&=|\l_{TT\f}|^2 q(\De)\{2,2|0|2,2\}_t\, ,
\ee
where
\be
	q(\De)=\frac{2^{1-3d}\pi^{3+\frac{d}{2}}(d-1)^2\G(d+1)^2}{(d-2)^2\G(d-\tfrac{\De}{2})^2\G(2+\tfrac{\De}{2})^4\G(\tfrac{d+\De}{2})^2}{\Gamma(\De)\Gamma(\De+1-\tfrac{d}{2})}\geq 0
\ee
is a non-negative function. The stress-tensor exchange contribution to $j_1=j_2=2$ is given by
\be
	\<T(n_4)|\cE(n_1)\cE(n_2)|T(n_3)\>&\ni\frac{C_T2^{-1-3d}\pi^{3-\frac{d}{2}}((4+d)t_2+4t_4)^2\Gamma(d+3)}{(d-1)d(d+1)^2(d+4)\Gamma(3+\tfrac{d}{2})\Gamma(\tfrac{d}{2})^2}\{2,2|2|2,2\}_t.
\ee
We therefore get two sum rules in which scalars participate, corresponding to $\{2,2|(3,1)|2,2\}_s$ and $\{2,2|(1,1)|2,2\}_s$,
\be
	-\frac{C_T2^{-1-3d}\pi^{3-\frac{d}{2}}((4+d)t_2+4t_4)^2\Gamma(d+3)}{(d-1)^2d(d+1)^2(d+4)\Gamma(3+\tfrac{d}{2})\Gamma(\tfrac{d}{2})^2}+\frac{1}{4}\sum_\f |\l_{TT\f}|^2 q(\De_\f)+\text{non-scalar}&=0,\nn\\
	\frac{C_T2^{-1-3d}(d^2-9)\pi^{3-\frac{d}{2}}((4+d)t_2+4t_4)^2\Gamma(d+3)}{(d-1)^2d(d+1)^3(d+4)\Gamma(3+\tfrac{d}{2})\Gamma(\tfrac{d}{2})^2}+\frac{2}{d+2}\sum_\f |\l_{TT\f}|^2 q(\De_\f)+\text{non-scalar}&=0.
\ee
For example, in $d=4$ this reduces to
\be
	-\frac{C_T\pi(t_4+2t_2)^2}{15\.2^{13}}+\frac{1}{4}\sum_\f |\l_{TT\f}|^2 q(\De_\f)+\text{non-scalar}&=0,\nn\\
	\frac{7C_T\pi(t_4+2t_2)^2}{75\.2^{13}}+\frac{1}{3}\sum_\f |\l_{TT\f}|^2 q(\De_\f)+\text{non-scalar}&=0,
\ee
which we quoted in \eqref{eq:superconvergenceexamples} in the introduction.
Here ``non-scalar'' represents contributions of higher-spin operators, starting from massive spin-2. We see explicitly that there can be no cancellation between massive scalars. Furthermore, there is a component of the contribution of scalars which cannot be canceled by the stress-tensor exchange. (The reverse is obvious since the stress-tensor contributes also to components other than $j_1=j_2=2$.) One can also take appropriate linear combinations of these equations to obtain separate sum rules for $(t_4+2t_2)^2$ and scalar contributions.

\subsection{General $t$-channel blocks for scalar event shapes}
\label{sec:scalarblocks}

In this section we derive a closed form expression for all $t$-channel conformal blocks appearing in 
\be
	\<\f_4|\wL[\f_1]\wL[\f_2]|\f_3\>,
\ee
where $\f_i$ are all scalars. The only essential difference from the algorithm of section~\ref{sec:tchannelblocks} is that we perform Fourier transform in a slightly different way, and we keep the intermediate spin as a free parameter.

\subsubsection{Fourier transform for scalar event shapes}
\label{sec:scalarfourier}

We start with the three-point tensor structure 
\be\label{eq:threeptnorm}
\<\f_1\f_2\cO_3\>=\frac{V_{3,12}^{J_3}}{
	X_{12}^{\frac{\De_1+\De_2-\De_3-J_3}{2}}
	X_{13}^{\frac{\De_1+\De_3-\De_2+J_3}{2}}
	X_{23}^{\frac{\De_2+\De_3-\De_1+J_3}{2}}
}.
\ee
Using results of section~\ref{sec:tchannelblocks} we find for $1\approx (3>2)$
\be
&\<0|\f_2\wL[\f_1]\cO_3|0\>\nn\\&=-2\pi i \frac{e^{i\pi \De_2}2^{1-\De_1}\Gamma(\De_1-1)}{\Gamma(\tfrac{\De_1+\De_2-\De_3+J_3}{2})\Gamma(\tfrac{\De_1-\De_2+\De_3-J_3}{2})}\frac{
	(-V_{1,32})^{1-\De_1}
	(-V_{3,12})^{J_3}
}{
	X_{12}^{\frac{2-\De_1+\De_2-\De_3-J_3}{2}}
	X_{13}^{\frac{2-\De_1+\De_3-\De_2+J_3}{2}}
	(-X_{23})^{\frac{\De_2+\De_3-2+\De_1+J_3}{2}}
}\nn\\
&\qquad\times {}_2F_1\p{\De_1-1,-J_3;\thalf(\De_1+\De_2-\De_3+J_3);-\half\frac{H_{13}}{V_{1,23}V_{3,12}}}.
\ee
We now want to find the Fourier transform~\eqref{eq:fouriertarget}, so we have to specialize to configuration~\eqref{eq:fourierframe}. Using~\eqref{eq:fourierframeinvariants} we find
\be\label{eq:scalarblockfourierframe}
&\<0|\f_2(0)\wL[\f_1](\oo,z_1)\cO_3(x,z_3)|0\>\nn\\&=-2\pi i \frac{e^{i\pi \De_2}2^{1-\De_1}\Gamma(\De_1-1)}{\Gamma(\tfrac{\De_1+\De_2-\De_3+J_3}{2})\Gamma(\tfrac{\De_1-\De_2+\De_3-J_3}{2})}\frac{
	(-x\.z_1)^{1-\De_1}
	(-x\.z_3)^{J_3}
}{
	(-x^2)^{\frac{\De_2+\De_3-\De_1+J_3}{2}}
}\nn\\
&\qquad\times {}_2F_1\p{\De_1-1,-J_3;\thalf(\De_1-\De_2+\De_3-J_3);\half\frac{x^2(z_1\.z_3)}{(x\.z_1)(x\.z_3)}}.
\ee

In~\eqref{eq:scalarblockfourierframe} we have a rather non-trivial function of $x$, and it is not obvious whether it should have a simple Fourier transform. Let us define functions
\be\label{eq:zzharmonicbasis}
[z_1^{m_1} z_2^{m_2}|x]\equiv (-z_1\.x)^{m_1}(-z_2\.x)^{m_2}{}_2F_1(-m_1,-m_2;1-\nu-m_1-m_2;\tfrac{x^2 (z_1\.z_2)}{2(x\.z_1)(x\.z_2)}).
\ee
These functions are homogeneous in $x$ and satisfy
\be
\ptl_x^2[z_1^{m_1} z_2^{m_2}|x]=0.
\ee
This means that
\be
	[z_1^{m_1} z_2^{m_2}|x]=\Q^{m_1,m_2}_{\mu_1\ldots \mu_{m_1+m_2}}(z_1,z_2)(x^{\mu_1}\ldots x^{\mu_{m_1+m_2}}-\text{traces})
\ee
for some function $\Q$. Therefore, for the purposes of computing the Fourier transform, we can treat these functions as $(z\. x)^{m_1+m_2}$, where $z$ is a null vector. In other words,
\be\label{eq:zzharmonicfourier}
\int d^dx e^{ipx} [z_1^{m_1} z_2^{m_2}|x](-x^2)^{-\tfrac{\De+m_1+m_2}{2}}=\hat\cF_{\De,m_1+m_2} [z_1^{m_1} z_2^{m_2}|p](-p^2)^{\tfrac{\De-m_1-m_2-d}{2}}\theta(p).
\ee
We can find the decomposition
\be
&(-x\. z_1)^{1-\De_1}(-x\. z_3)^{J_3}{}_2F_1(\De_1-1,-J_3;\thalf(\De_1-\De_2+\De_3-J_3);\tfrac{x^2 (z_1\.z_3)}{2(x\.z_1)(x\.z_3)})\nn\\
&=\sum_{k=0}^{J_3} \a_k(-z_1\.z_3)^k (-x^2)^k[z_1^{1-\De_1-k}z_3^{J_3-k}|x].
\ee
where
\be
\a_k = 2^k\frac{
	(\De_1-1)_k(-J_3)_k(\tfrac{2-d+\De_1+\De_2-\De_3-J_3}{2})_k
}{
	k!(-\tfrac{d}{2}+\De_1-J_3+k)_k(\tfrac{\De_1-\De_2+\De_3-J_3}{2})_k
}.
\ee
This yields the following Fourier transform
\be
&\<\f_2(p)|\wL[\f_1](\oo,z_1)|\cO(p,z_3)\>=-2\pi i 
\frac{e^{i\pi \De_2}2^{1-\De_1}\Gamma(\De_1-1)}{\Gamma(\tfrac{\De_1+\De_2-\De_3+J_3}{2})\Gamma(\tfrac{\De_1-\De_2+\De_3-J_3}{2})}\nn\\
&\qquad\times\sum_{k=0}^{J_3} \a_k\hat\cF_{\De_2+\De_3-1,1-\De_1+J_3-2k}(-z_1\.z_3)^k [z_1^{1-\De_1-k}z_3^{J_3-k}|p] (-p^2)^{\frac{\De_2+\De_3+\De_1-J_3+2k-2-d}{2}}\theta(p).
\ee
Surprisingly, this sum reassembles into another hypergeometric function,
\be\label{eq:seedfourier}
&\<\f_2(p)|\wL[\f_1](\oo,z_1)|\cO(p,z_3)\>\nn\\
&=-2\pi i 
\frac{e^{i\pi \De_2}2^{1-\De_1}\Gamma(\De_1-1)}{\Gamma(\tfrac{\De_1+\De_2-\De_3+J_3}{2})\Gamma(\tfrac{\De_1-\De_2+\De_3-J_3}{2})}\hat\cF_{\De_2+\De_3-1,1-\De_1+J_3}(-p^2)^{\frac{\De_2+\De_3+\De_1-J_3-2-d}{2}}\theta(p)\nn\\
&\quad\times (-p\.z_1)^{1-\De_1}(-p\.z_3)^{J_3}{}_3F_2(\De_1-1,-J_3,\De_3-1;\tfrac{\De_1+\De_2+\De_3-d-J_3}{2},\tfrac{\De_1-\De_2+\De_3-J_3}{2};\tfrac{p^2 (z_1\.z_3)}{2(p\.z_1)(p\.z_3)}).
\ee
It would be interesting to understand whether one can arrive at this expression in a more direct way, which generalizes to more complicated three-point functions.

Note that one can in principle use this result as a ``seed'' to compute more complicated objects, such as
\be
\<T(p,z_2)|\wL[T](\oo,z_1)|\cO_3(p,z_3)\>
\ee
by using weight-shifting operators~\cite{Costa:2011dw,Karateev:2017jgd}. We can always choose the weight-shifting operator acting on the point at infinity to be powers of $Z_1^{[m}X_1^{n]}$, which evaluates to something $x$-independent. Weight-shifting operators acting on points $2$ and $3$ can be rewritten as differential operators in momentum space, since any weight-shifting operator is polynomial in both coordinates and derivatives~\cite{Karateev:2017jgd}. This suggests that an expression in terms of linear combinations of ${}_3F_2$ functions can always be found for this type of objects. 

\subsubsection{Decomposition into $\SO(d-1)$ components}

The last step in the computation of $t$-channel event-shape conformal blocks is to compute the $\SO(d-1)$-invariant contraction
\be
\<\cO_4(p)|\wL[\cO_1](\oo,z_1)|\cO^\a(p)\>^{(a)}\Pi_{\a\b,\l}(p)\<\cO^\b(p)|\wL[\cO_2](\oo,z_2)|\cO_3(p)\>^{(b)}.
\ee
For this it is convenient to decompose the spin degrees of freedom of $\cO^\a(p)$ in each three-point function into irreducible components under the $\SO(d-1)$. 

Take the scalar structure
\be\label{eq:tchannelscalarsod-1}
f(z_3)=\<\f_2(p)|\wL[\f_1](\oo,z_1)|\cO_3(p,z_3)\>.
\ee
For the moment we are considering it just as a function of $z_3$. We can write
\be
f(z_3)=f_{\mu_1\ldots \mu_{J_3}}z_3^{\mu_1}\cdots z_3^{\mu_{J_3}}=\sum_{s=0}^{J_3} f_{\mu_1\ldots \mu_{J_3}}\Pi_{J_3,s}^{\mu_1\ldots\mu_{J_3}}{}_{\nu_1\ldots \nu_{J_3}}(p)z_3^{\nu_1}\cdots z_3^{\nu_{J_3}}
\ee
The indices of the traceless-symmetric $f_{\mu_1}\ldots f_{\mu_J}$ have to be provided by $p$ and $z_1$.\footnote{The contribution of $h^{\mu\nu}$ is fixed by tracelessness condition and in any case vanishes after contracting with traceless symmetric projector.} It may appear that there are several choices of how many indices to fill with $p$, but in fact all possibilities are exhausted by using no $p$ at all. The reason is that there is only one way to obtain a given $\SO(d-1)$ irrep from a given $\SO(d-1,1)$, and in this case we are trying to extract spin-$s$ irrep from $\wL[\f_1]$. We thus find that 
\be
&\<\f_2(p)|\wL[\f_1](\oo,z_1)|\cO_3(p,z_3)\>\nn\\
&=\sum_{s=0}^{J_3}\<\f_2|\wL[\f_1]|\cO_3^{(s)}\>(-p\.z_1)^{1-\De_1-J_3}\Pi_{J_3,s}(p;z_1,z_3)(-p^2)^{\tfrac{\De_2+\De_3-2+\De_1+J_3-d}{2}}\theta(p)
\ee
for some numbers $\<\f_2|\wL[\f_1]|\cO_3^{(s)}\>$.

Another way of arriving at this conclusion is to specialize to kinematics $p=(1,\vec 0)$ and $z_i=(1,n_i)$. In this kinematics the three-point function~\eqref{eq:tchannelscalarsod-1} is necessarily a function of $(n_1\.n_2)$, where $n_i$ live on the unit sphere. The question of decomposing into $\SO(d-1)$ representations is then equivalent to decomposition of this function into spherical harmonics of $n_2$, which are proportional to $\Pi_{J_3,s}(n_1\.n_2)$ since the latter is essentially a Gegenbauer polynomial.

To perform the decomposition explicitly, we rewrite the result~\eqref{eq:seedfourier} in the special kinematics,
\be
&\<\f_2(p)|\wL[\f_1](\oo,z_1)|\cO(p,z_3)\>\nn\\
&=-2\pi 
\frac{e^{i\pi \De_2}2^{1-\De_1}\Gamma(\De_1-1)}{\Gamma(\tfrac{\De_1+\De_2-\De_3+J_3}{2})\Gamma(\tfrac{\De_1-\De_2+\De_3-J_3}{2})}\hat\cF_{\De_2+\De_3-1,1-\De_1+J_3}\nn\\
&\qquad\times {}_3F_2(\De_1-1,-J_3,\De_3-1;\tfrac{\De_1+\De_2+\De_3-d-J_3}{2},\tfrac{\De_1-\De_2+\De_3-J_3}{2};\tfrac{1-\eta}{2}), 
\ee
where $\eta=(n_1\.n_2)$. The hypergeometric function truncates to a polynomial in $\eta$ thanks to the argument $-J_3$, making the decomposition straightforward for each given $J_3$. Interestingly, we can find a closed-form expression for the coefficients,
\be
{}_3F_2(\De_1-1,-J_3,\De_3-1;\tfrac{\De_1+\De_2+\De_3-d-J_3}{2},\tfrac{\De_1-\De_2+\De_3-J_3}{2};\tfrac{1-\eta}{2})=\sum_{s=0}^{J_3} \gamma_s(\De_1,\De_2,\De_3,J_3) \Pi_{J_3,s}(\eta),
\ee
where
\be
\gamma_s(\De_1,\De_2,\De_3,J_3)&=\frac{(-1)^{J_3-s}2^{3-d-J_3-2s}\sqrt{\pi}\Gamma(J_3+s+d-2)(\De_1-1)_s(\De_3-1)_s}
{\Gamma(J_3+\tfrac{d-2}{2})\Gamma(s+\tfrac{d-1}{2})(\thalf(\De_1-\De_2+\De_3-J_3))_s(\thalf(\De_1+\De_2+\De_3-d-J_3))_s}\nn\\
&\quad\times {}_4F_3\p{{{s+\tfrac{d}{2}-1,-J_3+s,\De_1+s-1,\De_3+s-1}\atop{
			2s+d-2,\tfrac{\De_1-\De_2+\De_3-J_3+2s}{2},\tfrac{\De_1+\De_2+\De_3-d-J_3+2s}{2}
	}};1}.
\ee
We thus conclude that
\be\label{eq:scalarscalarspinLFourier}
&\<\f_2|\wL[\f_1]|\cO_3^{(s)}\>\nn\\
&=-2\pi i
\frac{e^{i\pi \De_2}2^{1-\De_1}\Gamma(\De_1-1)}{\Gamma(\tfrac{\De_1+\De_2-\De_3+J_3}{2})\Gamma(\tfrac{\De_1-\De_2+\De_3-J_3}{2})}\hat\cF_{\De_2+\De_3-1,1-\De_1+J_3}\gamma_s(\De_1,\De_2,\De_3,J_3),
\ee
where $\hat\cF$ is given by~\eqref{eq:fouriercoefficient}.

\subsubsection{Complete scalar event shape blocks}

We have found that
\be
&\<\f_2(p)|\wL[\f_1](\oo,z_1)|\cO_3(p,z_3)\>\nn\\
&=\sum_{s=0}^{J_3}\<\f_2|\wL[\f_1]|\cO_3^{(s)}\>(-p\.z_1)^{1-\De_1-J_3}\Pi_{J_3,s}(p;z_1,z_3)(-p^2)^{\tfrac{\De_2+\De_3-2+\De_1+J_3-d}{2}}\theta(p),
\ee
where the coefficients $\<\f_2|\wL[\f_1]|\cO_3^{(s)}\>$ are given in~\eqref{eq:scalarscalarspinLFourier}. By applying Hermitian conjugation we find
\be
&\<\cO_3(p,z_3)|\wL[\f_1](\oo,z_1)|\f_2(p)\>\nn\\
&=\sum_{s=0}^{J_3}\<\cO_3^{(s)}|\wL[\f_1]|\f_2\>(-p\.z_1)^{1-\De_1-J_3}\Pi_{J_3,s}(p;z_1,z_3)(-p^2)^{\tfrac{\De_2+\De_3-2+\De_1+J_3-d}{2}}\theta(p),
\ee
where
\be\label{eq:threeptreducedconj}
\<\cO_3^{(s)}|\wL[\f_1]|\f_2\>\equiv \<\f_2|\wL[\f_1]|\cO_3^{(s)}\>^*.
\ee
Using~\eqref{eq:tchannelgeneralblock} we then find the scalar event shape t-channel conformal block
\be\label{eq:tchannelscalarfinal}
G^t_{\De,J}(p;z_1,z_2)
&=\sum_{s=0}^J \<\f_4|\wL[\f_1]|\cO^{(s)}\>\cA_s(\De,J)^{-1}\<\cO^{(s)}|\wL[\f_2]|\f_3\>\Pi_{J,s}(p;z_1,z_2)\nn\\
&\qquad\times(-p^2)^{\tfrac{\De_3+\De_4-4+\De_1+\De_2+2J-d}{2}}(-p\.z_1)^{1-\De_1-J}(-p\.z_2)^{1-\De_2-J}.
\ee
To find the above expression we used the identity $\Pi_{J,s}\Pi_{J,s'}=\delta_{ss'}\Pi_{J,s}$.

As we discussed above, in order for the $t$-channel expansion to converge, we should smear the event shape over the polarizations of detectors
\be
\int D^{d-2} z_1 D^{d-2}z_2 f(z_1,z_2)\<\f_4(p)|\wL[\f_1](\oo,z_1)\wL[\f_2](\oo,z_2)|\f_3(p)\>.
\ee
It is convenient to use the following smearing functions
\be
f_s(z_1,z_2)\propto(-p\.z_1)^{\De_1+1-d}(-p\.z_2)^{\De_2+1-d}C_s^{(\tfrac{d-3}{2})}(\eta),
\ee
so that the result simply picks out the coefficient of the Gegenbauer polynomial $C_s^{(\tfrac{d-3}{2})}(\eta)$ in the event shape. We can define the corresponding blocks $G^t_{\De,J}(s)$ by the identity, for $p=(1,\vec 0)$,
\be\label{eq:tchannelscalarexpansion}
&\<\f_4(p)|\wL[\f_1](\oo,z_1)\wL[\f_2](\oo,z_2)|\f_3(p)\>=\sum_s\sum_{\cO}\l_{14\cO}\l_{23\cO}^*G^t_{\De_\cO,J_\cO}(s)C_s^{(\tfrac{d-3}{2})}(\eta).
\ee
Using~\eqref{eq:tchannelscalarfinal} and~\eqref{eq:sod-1projector} we find
\be
\label{eq:tchannelscalarblockresult}
G^t_{\De,J}(s)=\frac{2^{-J}J!(d+J-2)_J(d-2)_J}{(\tfrac{d-1}{2})_J}\frac{(-1)^{s+J}(d+2s-3)}{(J-s)!(d-3)_{J+s+1}}\frac{\<\f_4|\wL[\f_1]|\cO^{(s)}\>\<\cO^{(s)}|\wL[\f_2]|\f_3\>}{\cA_s(\De,J)}.
\ee
Here, the reduced three-point functions $\<\f_4|\wL[\f_1]|\cO^{(s)}\>$ are given by~\eqref{eq:scalarscalarspinLFourier} and~\eqref{eq:threeptreducedconj} (with appropriate permutation of indices), while the coefficients $\cA_s(\De,J)$ are given by~\eqref{eq:twoptcoeffs}. In interpreting this result, it is also important to note our normalizations of two- and three-point tensor structures~\eqref{eq:twoptnorm} and~\eqref{eq:threeptnorm}.

\subsubsection{Example: scalar event shape in (generalized) free scalar CFT}
\label{sec:tchanexpample}

In this section we will consider an example of the $t$-channel decomposition of the scalar event shape
\be\label{eq:scalareventshape}
	\<\O|\f_4\wL[\f_1]\wL[\f_2]\f_3|\O\>
\ee
for the four-point function of scalars of dimension $\De=2$ in 4d,
\be\label{eq:example4pt}
	\<\f_1\f_2\f_3\f_4\>=\frac{1}{x_{14}^2 x_{23}^2 x_{13}^2 x_{24}^2}+\text{disconnected}.
\ee
This four-point function is given by a box Wick contraction, and the disconnected part is any sum of products of two-point functions, such as $\<\f_1\f_2\>\<\f_3\f_4\>$. The scalars $\f_i$ are not necessarily identical. For example, we can think about $\f_i$ as being double-trace operators $\f_1=\r^2,\f_2=\s^2,\f_3=\f_4=\r\s$, where $\r$ and $\s$ are fundamental free scalars. In this case, we are looking at a state with one $\r$ and one $\s$ particle, and $\wL[\f_1]$ detects $\r$ while $\wL[\f_2]$ detects $\s$.

First, we observe that the disconnected part does not contribute to~\eqref{eq:scalareventshape}, since $\wL[\f_i]$ annihilates the vacuum state. We then note that
\be
	\<\f_1\f_2\f_3\f_4\>=\<\f_1\f'_3\f'_4\>\<\f_2\f'_3\f'_4\>,
\ee
where $\f'_i$ are fictitious scalars of scaling dimension $\De/2=1$. This allows us to compute~\eqref{eq:scalareventshape} by reusing the results of section~\ref{sec:lighttransformgeneralthreept}. We find
\be
\<0|\f'_4\wL[\f_1]\f'_3|0\>&=-2\pi i \frac{e^{i\pi \De/2}2^{1-\De}\Gamma(\De-1)}{\Gamma(\tfrac{\De}{2})^2}&\frac{
	(-V_{1,43})^{1-\De}
}{
	(x_{14}^2)^{\frac{2-\De}{2}}
	(x_{13}^2)^{\frac{2-\De}{2}}
	(-x_{43}^2)^{\frac{2\De-2}{2}}
}\qquad ((3>4)\approx 1)\nn\\
&=i\pi\frac{
	(-V_{1,43})^{-1}
}{
	-x_{43}^2,
}
\ee
and similarly for $\<0|\f'_4\wL[\f_2]\f'_3|0\>$. Sending $1$ and $2$ to infinity, $4$ to $0$, and $3$ to $x>0$, we find
\be
\<0|\f'_4\wL[\f_1]\f'_3|0\>=i\pi
	(-(x\.z_1))^{-1},
\ee
and multiplying by $\<0|\f'_4\wL[\f_2]\f'_3|0\>$ we get
\be
\<0|\f_4\wL[\f_1]\wL[\f_2]\f_3|0\>=-\pi^2 (-(x\.z_1))^{-1}(-(x\.z_2))^{-1}.
\ee
Now, we need to compute the Fourier transform of the above expression with the $i\e$-prescription $x^0\to x^0+i\e$. For this, it is convenient to use Lorentz invariance to set $z_1=(1,1,0,0)$ and $z_2=(1,-1,0,0)$. Introducing the lightcone coordinates $x^\pm = x^0\pm x^1$, we find
\be
	\int d^4 x e^{ipx}(-(x\.z_1))^{-1}(-(x\.z_2))^{-1}=\half(2\pi)^2\de^2(\vec p)f(p^+)f(p^-),
\ee
where $\vec p=(p^2,p^3)$ and
\be
	f(p)=\int dx e^{-\frac{i}{2}px} (x+i\e)^{1-\De}=2^{3-\De}\pi \frac{e^{\frac{i\pi}{2}(1-\De)}}{\G(\De-1)}p^{\De-2}\theta(p)=-2\pi i \theta(p).
\ee
We thus find
\be
	\<\f_4(p)|\wL[\f_1]\wL[\f_2]|\f_3(p)\>=8\pi^6\theta(p)\de^2(\vec p).
\ee
It is easy to find the covariant form
\be
	\<\f_4(p)|\wL[\f_1]\wL[\f_2]|\f_3(p)\>=8\pi^6\theta(p)\de^2(\vec p)(-p^2)(-z_1\.p)^{-1}(-z_2\.p)^{-1},
\ee
where
\be
	\de^2(\vec p)=\frac{1}{\pi}\de({\vec{p}}^{\,2})=\frac{1}{\pi}\de\p{p^2-2\frac{(z_1\.p)(z_2\.p)}{(z_1\.z_2)}}.
\ee
Now, setting $p=(1,\vec 0)$ and $z_i=(1,n_i)$, we get
\be\label{eq:eventshapedirectresult}
	\<\f_4(p)|\wL[\f_1]\wL[\f_2]|\f_3(p)\>= 16\pi^5\de((n_1\.n_2)+1).
\ee
The delta-function forces $n_1$ and $n_2$ to point in opposite directions. This corresponds simply to the fact that $\f_3=\r\s$ creates a pair of particles, and by momentum conservation they must fly off in opposite directions, since we have set the spatial component of $p$ to $0$.

We would like to compute this event shape using the $t$-channel OPE. We will first expand the four-point function~\eqref{eq:example4pt} in the $14\rightarrow 23$ channel and then use the resulting expansion to sum the event shape conformal blocks.

The disconnected piece of~\eqref{eq:example4pt} only contains the contributions of the identity and double-trace operators. The double-trace operators do not contribute to the event shape, as seen in the discussion below~\eqref{eq:tchanneldoubletracebyebye}. The identity operator also doesn't contribute, as its contribution is
\be
	\<\f_4(p)|\wL[\f_1]|\O\>\<\O|\wL[\f_2]|\f_3(p)\>=0,
\ee
since light-transforms annihilate the vacuum state~\cite{Kravchuk:2018htv}.

The connected contribution can be rewritten as
\be
	\<\f_1(x_1)\f_2(x_2)\f_3(x_3)\f_4(x_4)\> = \frac{1}{x_{14}^2 x_{23}^2}\<\phi'_2(x_4)\phi'_1(x_1)\phi'_2(x_2)\phi'_1(x_3)\>,
\ee
where $\phi'_1$ and $\phi'_2$ are fictitious canonically normalized free scalars of dimension $\De/2=1$. The prefactor $\frac{1}{x_{14}^2 x_{23}^2}$ simply plays the role of shifting the external dimensions of conformal blocks, and so we have the identity
\be
	\l_{14\cO}\l_{23\cO}=\l_{12\cO}\l_{21\cO}
\ee
where in the right hand side we mean the OPE coefficients which enter into the decomposition of the function $\<\f_2'(x_4)\f_1'(x_1)\f_2'(x_2)\f_1'(x_3)\>$. Since this is a four-point function of free fields, only a single family of higher-spin currents $\cO$ with $\De=J+2$ contribute to its OPE. In our conventions we have~\cite{Dolan:2000ut}
\be
	\l_{12\cO}\l_{21\cO}=(-1)^J\l_{12\cO}^2=(-1)^J\frac{2^J \Gamma(J+1)^2}{\Gamma(2J+1)}.
\ee

Setting $\De_1=\De_2=\De_3=\De_4=2$ we find
\be
	G^t_{J+2,J}(s)=\frac{2^{J+4}\pi^{\frac{9}{2}}\Gamma(J+\tfrac{3}{2})}{\Gamma(J+1)}\de_{J,s}.
\ee
The reason only $s=J$ is allowed is because $\De=J+2$ corresponds to conserved higher-spin currents, which only have one $\SO(d-1)$ component. Using~\eqref{eq:tchannelscalarexpansion} we find
\be
&\<\f_4(p)|\wL[\f_1](\oo,z_1)\wL[\f_2](\oo,z_2)|\f_3(p)\>=\nn\\
&=\sum_s\sum_{\cO}\l_{14\cO}\l_{23\cO}^*G^t_{\De_\cO,J_\cO}(s)C_s^{(\tfrac{d-3}{2})}(\eta)=\sum_{J=0}^\oo(-1)^J\frac{2^J \Gamma(J+1)^2}{\Gamma(2J+1)}\frac{2^{J+4}\pi^{\frac{9}{2}}\Gamma(J+\tfrac{3}{2})}{\Gamma(J+1)}P_J(\eta)\nn\\
&=8\pi^5\sum_{J=0}^\oo (-1)^J (2J+1) P_J(\eta)=16\pi^5\de(\eta+1).
\ee
The last equality follows from the completeness relation for Legendre polynomials and $P_J(-1)=(-1)^J$. This result indeed agrees with~\eqref{eq:eventshapedirectresult}. Note that the convergence here is only in a distributional sense, i.e.\ we have to smear the event shape with some test function in $\eta$ before computing the sum.

A more nontrivial check is to repeat the same calculation in a generalized free theory (GFT). The event shape is the same up to a coefficient,
\be\label{eq:gftexample}
	\<\f_4(p)|\wL[\f_1]\wL[\f_2]|\f_3(p)\>= \frac{16^{3-2\De_f}\pi^5}{\Gamma(\De_f)^4}\de((n_1\.n_2)+1),
\ee
where $\De_f$ is the dimension of the fundamental fields $\r$ and $\s$ ($\De_f=1$ for the free scalar case considered above). We can use the same logic to obtain the relevant OPE coefficients from the known GFT ones~\cite{Heemskerk:2009pn,Fitzpatrick:2011dm}. The main difference is that each Legendre polynomial $P_s(\eta)$ receives contribution from infinitely many operators $[\r\s]_{n,J}$ with $J\geq s$ and $n\geq 0$.\footnote{Note that these operators are not double-traces of $\r\s$ and $\r^2$ or $\s^2$, so they do contribute to the OPE.} The sum is now much more non-trivial and we cannot tackle it analytically. We have focused on the coefficients in front of $P_0(\eta)$ and $P_1(\eta)$, and found numerically that the sum over $n$ converges rather quickly. However, the sum over $J$ appears to behave as
\be
	\sim \sum_J (-1)^J J^{2\De_f-3}\, ,
\ee
and so it diverges for $\De_f>3/2$ and converges for $\De_f<3/2$. In the latter case, convergence can be improved by an Euler transform,\footnote{
	For series $\sum_{n=0}^\oo(-1)^n a_n$ the Euler transform is
	\be
		\sum_{n=0}^\oo(-1)^n a_n=\sum_{n=0}^\oo\frac{(-1)^n b_n}{2^{n+1}},
	\ee
	where $b_n=\sum_{k=0}^n(-1)^k\binom{n}{k}a_k$. It generally tends to improve the rate of convergence of slowly-converging series.
} which allows us to check for a few sample values of $\De_f$ with $1<\De_f<3/2$ that the $t$-channel sum agrees to high precision with
\be
	\frac{16^{3-2\De_f}\pi^5}{2\Gamma(\De_f)^4}P_0(\eta)-3\frac{16^{3-2\De_f}\pi^5}{2\Gamma(\De_f)^4}P_1(\eta)+\ldots,
\ee
which is the expansion of~\eqref{eq:gftexample}.\footnote{In fact, Euler transform makes the sum over $J$ convergent for all values of $\De_f$.}

Based on intuition from $\nu$-space described in~\cite{AnecOPE}, we expect that the divergence for $\De_f\geq 3/2$ is due to the behavior of our test functions near the collinear limit $\eta=1$. Recall that if
\be
	\<\f_4(p)|\wL[\f_1]\wL[\f_2]|\f_3(p)\>=\sum_s a_s P_s(\eta),
\ee
then the coefficients $a_s$ are given by smearing the event shape with the test functions
\be
	f_s(\eta)=\thalf (2s+1)P_s(\eta).
\ee
To moderate the contribution of $\eta=1$, we can smear with linear combinations of functions $f_s$ which vanish at $\eta=1$ as $(\eta-1)^k$ for sufficiently high $k$. Suppose that $\sum_s \a_s f_s(\eta)$ is one such combination for a fixed $k$. We are then led to consider the expansion
\be
\sum_{\cO}\l_{14\cO}\l_{23\cO}^*\sum_s \a_s G^t_{\De_\cO,J_\cO}(s)C_s^{(\tfrac{d-3}{2})}(\eta).
\ee
Numerically, we find that the sum over primary operators with different spin behaves as
\be
	\sum_J (-1)^J J^{\w_k}
\ee
where $w_k$ is monotonically non-increasing function of $k$. Specifically, for any choice of $\De_f$, we find that $w_k$ starts at $w_0=2\De_f-3$ as described above, and then decreases monotonically with $k$ until it saturates at some value $w_*<0$.\footnote{Typically $w_k\approx w_{k-1}-2$, but we have not studied this in sufficient detail either numerically or analytically.} This means that for any value of $\De_f$ the expansion converges for test functions which vanish sufficiently quickly at $\eta=1$.

\section{Discussion}

\subsection{Bounds on heavy contributions to non-minimal couplings}
\label{sec:bounds}

A quantitative understanding of the superconvergence sum rule requires some extra analysis which we postpone for  future work. Here, we sketch its qualitative implications. For simplicity, we consider a toy model for a gravitational scattering amplitude, but the argument for the CFT correlator is essentially the same. In our discussion, the rough correspondence between amplitudes and CFT correlators is
\begin{center}
\begin{tabular}{l|l}
amplitude & four-point function \\
\hline 
$t$ & $(z \bar z)^{-1/2}$ \\
$s$ & $\p{\De-\frac d 2}^2$ \\
$J$ & $J$ \\
$a_J^\pm(s)$ & $C^\pm(\De,J)$ \\
$\mathrm{Disc}$ & $\mathrm{dDisc}$ \\
Froissart-Gribov & Lorentzian inversion 
\end{tabular}
\end{center}

The basic idea was explained in \cite{Caron-Huot:2017vep} and goes as follows. Let us imagine a theory with a large gap $\Delta_{\mathrm{gap}} \gg 1$ in the spectrum of particles (or operators). We would like to bound the contribution of heavy, or stringy, modes to the superconvergence sum rule and therefore determine the maximal allowed value of non-minimal three-point graviton couplings. To do so, recall that given a polynomially bounded ``amplitude'' $\cA(s,t)$ with ``partial wave" expansion
\be
\cA(s,t) &= \sum_J a_J(s) \nu^J ,~~~ \nu=\frac{t-u}{2} ,
\ee
we can write a ``Froissart-Gribov inversion formula'' which takes the following form
\be
\label{eq:FGtoy}
a^\pm_J(s) &= \int_0^\oo \frac{d\nu}{\nu} \,\nu^{-J} \Disc_t \cA(s,\nu) \pm \int_{-\oo}^{0} \frac{d\nu}{-\nu}(-\nu)^{-J} \Disc_u \cA(s,\nu) .
\ee
where $\Disc_u = -\Disc_t$. Convergence properties of the integral (\ref{eq:FGtoy}) depend on the behavior of the amplitude at large $t$ and fixed $s$. In a consistent theory, the integral converges for $J>1$. In
particular, we can evaluate (\ref{eq:FGtoy}) at $J=2$, for which the integral (\ref{eq:FGtoy}) must reproduce the graviton pole at $s=0$,
\be \label{eq:graviton pole}
a_{J=2}^+(s) \sim {1 \over c_T} \frac{1}{s}\, .
\ee
More generally, away from the graviton pole, the integral over the discontinuity should correctly reproduce the Pomeron pole
\be
\label{eq:aJestimate}
a_J^+(s) &\sim \frac{C_{\f\f P}^2(s)}{J-J(s)},
\ee
where we expect $a_J(s)$ is suppressed by $1/c_T$. This suppression does not follow from our knowledge of the three-point coefficients that control the residue of the pole. Indeed the value of the residue in~\eqref{eq:aJestimate} solely reflects the asymptotic behavior of the discontinuity in (\ref{eq:FGtoy}), $\Disc_t \cA(s,\nu) \sim C_{\f\f P}^2(s) \nu^{J(s)}$ at large $\nu$. In particular we can imagine an isolated ``outlier state'' at some intermediate scale $\nu^*$ that contributes to $\Disc_t \cA(s,\nu) \sim C_{\f\f P^*}^2(s) \delta(\nu - \nu^*)$  with a large coefficient $C_{\f\f P^*}^2(s)$. Through (\ref{eq:FGtoy}), such an outlier state would invalidate the estimate $a_J^+(s) \sim {1 \over c_T}$ away from the pole. In what follows we assume that there are no outliers. The same assumption was made in \cite{Caron-Huot:2017vep}.

In a large-gap theory we expect $J(s) = 2 + {s \over \Delta_{\mathrm{gap}}^2}$. Correctly reproducing the ${1 \over c_T}$ residue of the graviton pole \eqref{eq:graviton pole} from (\ref{eq:aJestimate}) thus requires that $C_{\f\f P}^2(s) \sim \frac{1}{c_T} {1 \over \Delta_{\mathrm{gap}}^2}$. Note also that by plugging $\Disc_t \cA(s,\nu) \sim C_{\f\f P}^2(s) \nu^{J(s)}$ for $\nu>\De_{\text{gap}}^2$ into (\ref{eq:FGtoy}) (and thus assuming no outliers) we get
\be
\label{eq:aJestimateB}
a_J^+(s) &\sim \frac{1}{c_T} {1 \over \Delta_{\mathrm{gap}}^2} \frac{1}{J-J(s)} (\Delta_{\mathrm{gap}}^2)^{J(s) - J}.
\ee

The superconvergence sum rule is the statement that $a_{J=3}^{-}(s) = 0$. Note that only the square of the non-minimal three-point coupling contributes to $a_{J=3}^{-}(s)$, see e.g.\ (\ref{eq:gravitoncommutatorflat}). We can write it as follows
\be
\label{eq:sumrule}
\alpha_{\mathrm{GB}}^2(s) + \int_{\Delta_{\mathrm{gap}}^2}^\oo \frac{d\nu}{\nu} \,\nu^{-3} \Disc_t \cA(s,\nu) - \int_{-\oo}^{-\Delta_{\mathrm{gap}}^2} \frac{d\nu}{-\nu}(-\nu)^{-3} \Disc_u \cA(s,\nu) = 0 ,
\ee
where we separated the contribution from the graviton pole $\alpha_{\mathrm{GB}}(s)$ from the rest  (``GB" stands for Gauss-Bonnet --- a particular type of nonminimal coupling that contributes to $\a_\mathrm{GB}$).

Next we would like to bound the contribution of heavy states in (\ref{eq:sumrule}). The estimate goes as follows
\be
|\alpha_{\mathrm{GB}}^2(s)| &= \left| \int_{\Delta_{\mathrm{gap}}^2}^\oo \frac{d\nu}{\nu} \,\nu^{-3} \Disc_t \cA(s,\nu) - \int_{-\oo}^{-\Delta_{\mathrm{gap}}^2} \frac{d\nu}{-\nu}(-\nu)^{-3} \Disc_u \cA(s,\nu) \right| \nn \\
&\leq a_{J=3}^+(s)  \sim  {1 \over \Delta_{\mathrm{gap}}^4} {1 \over c_T} \qquad(s\ll \De_\text{gap}^2)
\ee
where we used the positivity of $\Disc_t \cA(s,\nu)$ and $\Disc_u \cA(s,\nu)$, which follow from unitarity in an appropriate kinematical region.
We see that what follows is the same qualitative conclusion as was obtained in \cite{Camanho:2014apa}. This time, however, we have a precise sum rule that must be satisfied. 
We will show in \cite{AnecOPE} that the superconvergence sum rule in CFT can be written as 
\be C^-(\De=\frac d 2 + i\nu, J=3) = 0,\ee where $C^\pm(\De,J)$ is the quantity computed by the Lorentzian inversion formula, so the argument in the CFT case proceeds analogously to the one here.

\subsection{Conclusions and future directions}
\label{sec:conclusions}

In this work, we found connections between commutativity of coincident shocks, superconvergence sum rules, and boundedness in the Regge and lightcone limits. These connections hold both in flat space and in AdS.

In flat space, we defined ``shock amplitudes" as amplitudes with special external wavefunctions. We showed that boundedness of amplitudes in the Regge limit is a sufficient condition for commutativity of coincident shocks. Furthermore, when coincident shocks commute, one obtains superconvergence sum rules that constrain the matter content and three-point couplings of the theory. It was argued in \cite{Camanho:2014apa} that causal theories of gravity should have Regge intercept $J_0\leq 2$. Assuming this, it follows that coincident gravitational shocks commute. The associated superconvergence sum rules relate non-minimal gravitational couplings to three-point couplings of stringy states.

In AdS, commutativity of coincident shocks is dual to the question of commutativity of certain null-integrated operators (e.g.\ ANEC operators) in a CFT. This question can be studied on its own using CFT techniques. In particular, we show using CFT methods that ANEC operators on the same null plane commute. (This result holds both nonperturbatively and in the planar limit, but it can be violated at fixed loop order in bulk perturbation theory.) This establishes commutativity of coincident gravitational shocks in AdS. We conjecture that coincident gravitational shocks commute in UV-complete gravitational theories in flat space, AdS, dS, and possibly beyond, dubbing this a ``stringy equivalence principle."

The CFT version of superconvergence sum rules can be obtained by inserting complete sets of states between the null-integrated operators. In large-$N$ theories, such sum rules relate non-minimal bulk couplings to the massive single-trace spectrum. However, the resulting sum rules are completely general, independent from holography, and interesting on their own.

Let us discuss some open questions and future directions.

\subsubsection{Constraints on UV-complete gravitational theories}

Higher derivative gravitational couplings are inconsistent with commutativity of coincident shocks, unless their effects are cancelled in the superconvergence sum rule. This cancellation can occur in different ways. In weakly-coupled (tree-level) gravity theories, the cancellation must involve massive (stringy) states. More generally, the cancellation could involve loop effects. An important problem is to compute independent bounds on non-graviton contributions to the superconvergence sum rules. This would give quantitative bounds on the size of non-minimal gravitational couplings. A toy version of this argument was presented in the previous section. However, it will be important to make it precise.

Another interesting question is the extent to which the low-energy matter content of the Standard Model is consistent with commutativity of coincident gravitational shocks.  If one could compute the Standard Model contribution to superconvergence sum rules (including loop effects) and find that it is nonzero, that would establish the necessity of additional massive states, and possibly hint at their properties.

To incorporate loop effects (e.g.\ loops in the Standard Model or $1/N$ effects in CFT), it may be necessary to use eikonal techniques to re-sum gravitational exchanges. The reason is that $n$-graviton exchange on its own leads to Regge growth with spin $J_0=1+n$, which would invalidate superconvergence sum rules.

\subsubsection{Bootstrapping amplitudes and four-point functions}
\label{sec:bootstrappingstuff}

In the context of flat-space amplitudes, superconvergence sum rules have been used to bootstrap the Veneziano amplitude \cite{SCHMID1968348}. This result relies on assuming linear Regge trajectories. The assumption of linear trajectories has two nice effects. Firstly, it removes the necessity of bootstrapping masses --- one can focus only on three-point couplings. Secondly, because Regge trajectories behave as $J(s) = \mathrm{const.} + \a' s$, one can make $J(s)$ arbitrarily negative by making $s$ arbitrarily negative. When $J(s) \leq -k,$ for integer $k$, one obtains a new superconvergence sum rule obtained by inserting $t^k$ into a dispersion relation.

It would be interesting to perform an analogous exercise in CFT, with the goal of using superconvergence sum rules to bootstrap a planar four-point function in AdS with finite $\a'$. In CFT, we do not have linear Regge trajectories. However, perhaps one could take the known single-trace spectrum computed from integrability \cite{Gromov:2009tv,Gromov:2013pga,Gromov:2014caa} as input and try to bootstrap the three-point couplings. In $\cN=4$ SYM theory, the analog of $J(s) = \mathrm{const.} + \a' s$ is \cite{Brower:2006ea}
\be
J(\nu) &= 2 - \frac{\De(4-\De)}{2\sqrt \l} + \dots=2 - \frac{\nu^2 + 4}{2\sqrt \l} + \dots,
\ee
where $\l$ is the 't Hooft coupling, $\De=2 + i \nu$, and ``$\dots$" represents subleading corrections in $1/\l$. Here, $-\nu^2/(2\sqrt \l)$ is analogous to $\a' s$. In the flat space limit, $\l\to \oo$ with $\nu^2/\sqrt \l$ held fixed, it is possible to make $J(\nu)$ arbitrarily negative by making $\nu^2/(2\sqrt \l)$ large \cite{Cornalba:2007fs}.\footnote{We thank David Meltzer for discussions on this point.} It would be interesting to know whether this is true more generally (e.g.\ at finite $\l$): can $J(\nu)$ always be made arbitrarily negative by going to large $\nu$?

If so, one should be able to obtain additional superconvergence sum rules beyond the ones we have discussed at sufficiently large $\nu^2$. We expect such sum rules should come from commutativity of the operators $L^n(\vec y)$ defined in \cite{Casini:2017roe} and other descendants of $\wL[T](x,z)$, such as those studied in \cite{Cordova:2018ygx}. Note that these operators only need to commute when transformed to $\nu$ space and placed at sufficiently large $\nu$. To transform to $\nu$-space, one must smear the operators in the transverse positions $\vec y$ against an appropriate $d{-}2$-dimensional three-point structure, see e.g.\ \cite{AnecOPE}.

It would be interesting to understand how our flat space analysis is embedded into the corresponding limit of the AdS/CFT duality. Indeed,  in theories with sub-AdS locality there is no problem in localizing both the shocks and the probes in the region of spacetime much less than $L_{AdS}$ \cite{Maldacena:2015iua}. Therefore the flat space analysis should apply in this limit. Shockwaves that are well-localized in the AdS interior were analyzed for example in \cite{Afkhami-Jeddi:2017rmx}. The same idea should apply to high energy scattering in nontrivial backgrounds, e.g.\ in the vicinity of a black hole horizon \cite{Shenker:2013pqa}. Presumably, the commutativity of shocks in this case is related to the consistency conditions on the spinning scramblon couplings \cite{Kitaev:2017awl}. More generally, the statement that gravitational shocks commute locally at every point in AdS should constrain spinning couplings to the ``modulon'' \cite{Faulkner:2018faa}, a mode that saturates the modular bound on chaos and captures local high energy gravitational scattering in the bulk. It seems plausible that saturation of the modular chaos bound together with the corresponding superconvergence conditions uniquely select Einstein gravity in AdS as the dual theory.\footnote{We thank Tom Faulkner for discussions on this point.} 

\subsubsection{Generalizations}

Although we have focused mostly on superconvergence sum rules coming from commutativity of ANEC operators, one additionally gets sum rules from studying $[\wL[\cO_1],\wL[\cO_2]]$ for any pair of operators $\cO_1,\cO_2$ with sufficiently large $J_1+J_2$. A further generalization could come from studying commutativity of more general continuous-spin light-ray operators defined in \cite{Kravchuk:2018htv}. Can one show that such general light-ray operators commute on the same null plane as well, $[\mathbb{O}_1,\mathbb{O}_2]=0$? One way to obtain such light-ray operators is from OPEs of more traditional null integrals, $\wL[\cO_1]\wL[\cO_2]\sim \sum_k \mathbb{O}_k$ \cite{AnecOPE}. In this case, commutativity of $\mathbb{O}_k$ would follow from commutativity of null-integrated operators. However, the general construction of continuous-spin light-ray operators is more complicated.

In addition to introducing continuous-spin light-ray operators in the CFT context, it is interesting to ask whether they can be introduced in the amplitudes context as well. In CFT, null integrated operators $\wL[\cO_{i,J}]$ can be analytically continued in spin to obtain more general light-ray operators $\mathbb{O}^\pm_{i,J}$, where $i$ labels Regge trajectories. It is natural to guess that the shock amplitudes defined in section~\ref{sec:shockwaveamplitudes} can be analytically continued in the spin of the shocks. This would provide a vast generalization of the amplitudes usually considered. It would be interesting to investigate such analytically continued amplitudes in string theory.

\subsubsection{Further applications}

In the main text we discussed a set of extra consistency conditions (superconvergence relations) on the CFT spectrum which follow from commutativity of average null energy operators. Together with ANEC, commutativity also implies that products of multiple average null energy operators are positive-semidefinite. Therefore, we can further require that
\be
\label{eq:positivity}
\langle \Psi | \wL[T] \cdots \wL[T] | \Psi \rangle \geq 0 .
\ee
Using the results of section 5, this leads to extra conditions on the OPE data of the theory. In numerical bootstrap calculations, boundedness of the Regge limit is manifest, and therefore superconvergence relations are automatically true. On the other hand, from the point of view of four-point functions, the positivity conditions (\ref{eq:positivity}) are extra nontrivial conditions. For example, positivity of two-point energy correlators follows from unitarity of six-point functions, and thus is not manifest from the conformal block expansion of a four-point function. It would be interesting to include these positivity constraints in the numerical bootstrap. In particular, it will be interesting to see how they affect the stress-tensor bootstrap results \cite{Dymarsky:2017yzx}.

It will be interesting to apply the $t$-channel OPE formulas to QCD-like theories, say ${\cal N}=4$ SYM. An appealing feature of the $t$-channel OPE is that in the planar limit the contribution of double-trace operators is suppressed by an extra power of ${1 \over N^2}$. Therefore, to leading order only the single trace data is needed. 
We hope that one day these will be computed at finite $\lambda$ using integrability techniques, see e.g.\ \cite{Beisert:2010jr,Gromov:2013pga,Basso:2015zoa}. It will be also interesting to use the $t$-channel OPE to reproduce the known weak coupling results \cite{Belitsky:2013ofa,Henn:2019gkr}, and try to extend the $t$-channel analysis to actual QCD, where the state of the art is the NLO analytic result \cite{Dixon:2018qgp}. Note that in the case of ${\cal N}=4$ SYM, due to superconformal symmetry energy-energy correlation can be computed using the four-point function of scalars to which (\ref{eq:tchannelscalarblockresult}) can be directly applied. 
	
\section*{Acknowledgements}	

We thank Alex Belin, Xi\'an Camanho, Cliff Cheung, Simon Caron-Huot, Thomas Dumitrescu, Tom Faulkner, Nima Lashkari, Raghu Mahajan, Juan Maldacena, David Meltzer, Shiraz Minwalla, Eric Perlmutter, Douglas Stanford, and Eduardo Teste for discussions. DSD thanks the KITP and the organizers of the ``Chaos and Order" workshop for hospitality during some of this work. DSD is supported by Simons Foundation grant 488657 (Simons Collaboration on the Nonperturbative Bootstrap), a Sloan Research Fellowship, and a DOE Early Career Award under grant No.\ DE-SC0019085. PK is supported by DOE grant No.\ DE-SC0009988. This research was supported in part by the National Science Foundation under Grant No.\ NSF PHY-1748958, as well as by the U.S.\ Department of Energy, Office of Science, Office of High Energy Physics, under Award Number DE-SC0011632.

\pagebreak
	
\appendix

\section{More on superconvergence in flat space}
\label{app:moresuperflat}

\subsection{Commutativity and the Regge Limit}

It is also very easy to see the connection between the Regge limit and commutativity at the level of the four-point function. 
As the simplest example consider $\phi^3$ theory where we write different couplings in different channels. The amplitude takes the form
\be
\label{eq:scalarampl}
{\cal A}(s,t) = {\alpha_s \over s} + {\alpha_t \over t} + {\alpha_u \over u} ,
\ee
where $s+t+u = 0$.

Let us consider the Regge limit of (\ref{eq:scalarampl}), $t \to \infty$ and $s$ fixed. The amplitude takes the form
\be
{\cal A}(s,t) = {\alpha_s \over s} + {\alpha_t - \alpha_u \over t} + ... .
\ee
For the superconvergence sum rule to hold, the amplitude should decay faster than ${1 \over t}$. We see that it implies $\alpha_s = 0$ and $\alpha_t = \alpha_u$, the latter condition being the analog of shock commutativity.

A different example is to consider the scattering of a vector particle against the scalar shocks. The amplitude takes the form
\be
\label{eq:scalarvectorampl}
{\cal A}(s,t,\e_3, \e_4) = \left(\alpha_t {p_1 \. p_3 \over p_1 \. p_2} + \alpha_u {p_2 \. p_3 \over p_1 \. p_2} \right) \left( \e_3 \. \e_4 + {\e_3 \. p_1 \e_4 \. p_2 \over p_1 \. p_3} + {\e_3 \. p_2 \e_4 \. p_1 \over p_2 \. p_3}  \right)\, ,
\ee
where the form of the amplitude is fixed by unitarity and we introduced different couplings for the $t$- and $u$-channel residues. The second bracket in (\ref{eq:scalarvectorampl}) is manifestly symmetric under permutations of the shocks $1 \leftrightarrow 2$. On the other hand, we can rewrite the first bracket as
\be
{\cal A}(s,t,\e_3, \e_4) \sim (\alpha_t {u \over s} + \alpha_u {t \over s}) \to (\alpha_u - \alpha_t) t , ~~~ t \to \infty, ~~~ s~\text{fixed} . 
\ee
Again, we explicitly see the relation between the commutativity of shocks and the Regge limit.

\subsection{Scalar-graviton elastic scattering}
\label{sec:scalargravitonscat}

Let us consider a four-point amplitude of scalar-graviton elastic scattering in General Relativity. The scattering amplitude takes the form \cite{Bjerrum-Bohr:2013bxa}
\be
{\cal A} =- {1 \over 2} {p_1 \. p_3 p_2 \. p_3 \over p_1 \. p_2} \left( {1 \over p_1 \. p_3} \e_1 \. p_3 \e_2 \. p_4 + {1 \over p_2 \. p_3} \e_1 \. p_4 \e_2 \. p_3 - \e_1 \. \e_2 \right)^2 ,
\ee
where gravitons have momenta $p_1$ and $p_2$ and scalars have momenta $p_3$ and $p_4$. 

We evaluate this amplitude in the shockwave kinematics (\ref{eq:momentumthreefour}). The result takes the form
\be
\label{eq:grcomptonshock}
{\cal A}_{\phi g^* \to \phi g^*} =  2 (p^u)^3 {\vec q_1 \. \vec q_2 \over p^v ( p^u p^v - 2 \vec q_1 \. \vec q_2)} .
\ee
As expected from the discussion in the main text, at large $p^v$ (or $t$) this goes as ${\cal A}_{\phi g^* \to \phi g^*} \sim {1 \over t^2}$, in particular we can write the superconvergence sum rule. Upon taking the discontinuity at $t=0$ we reproduce (\ref{eq:minimalscalarQ}).

Let us contrast this behavior with the kinematics used in \cite{Camanho:2014apa} which describes a high-energy scattering with physical transverse momenta and polarizations
\be
\label{eq:momentumthreefourCEMZ}
p_1 &= (-{\vec q^2 \over p^v},-p^v,\vec q), ~~~p_2 = ({\vec q^2 \over p^v},p^v,\vec q), \nn\\
p_3 &= ( p^u, {\vec q^2 \over p^u},, -\vec q), ~~~ p_4 = (- p^u, -{\vec q^2 \over p^u}, -\vec q),
\ee
 and polarizations
\be
\label{eq:polarizationsCEMZ}
\e_1 &= (-2 {\vec e_1 \. \vec q \over p^v},0,\vec e_1), ~~~\e_2 = (2{\vec e_2 \. \vec q \over p^v},0,\vec e_2) . 
\ee
The scattering amplitude then evaluates to
\be
\label{eq:grcomptonphys}
{\cal A}_{\phi g \to \phi g} = ( \vec e_1 \. \vec e_2 )^2 {(p^u p^v)^2 \over  16 \vec q^2 } \left(1 - {\vec q^2 \over p^u p^v} \right)^2 \left(1 + {\vec q^2 \over p^u p^v} \right)^2 .
\ee
This time the amplitude grows as ${\cal A}_{\phi g \to \phi g} \sim t^2$ consistent with the usual intuition about the high energy behavior of gravitational scattering. In this case, it is not possible to write the superconvergence sum rule. The difference in the high energy behavior of (\ref{eq:grcomptonshock}) and (\ref{eq:grcomptonphys}) is ${1 \over t^{J_1 + J_2}} = {1 \over t^4}$ as discussed in the main text.

\subsection{Gravitational sum rule in a generic theory}
\label{sec:gravitationalsumrulegeneric}

In this section, we describe the superconvergence sum rule for commutativity of coincident shocks in a generic tree-level theory of gravity. Firstly, a shock three-point amplitude for a massive graviton $\tl g$ with mass $m$ takes the form
\be
{\cal A}_{g_1 g_3 \tilde g}^{\mu \nu} &= \tilde \alpha_2  \left[ \e_1^{\mu} p_3^{\nu} (\e_3 \. p_1) + \e_3^{\mu} p_1^{\nu} (\e_1 \. p_3) - p_1^{\mu} p_3^{\nu} (\e_1 \. \e_3) - \e_1^{\mu} \e_3^{\nu} (p_1 \. p_3)\right] \nn\\
&\quad \qquad\x\left[(\e_1 \. \e_3) (p_1 \. p_3) - (\e_3 \. p_1) (\e_1 \. p_3) \right] \nn\\
&\quad- \tilde \alpha_4 p_1^{\mu} p_3^{\nu} \left[  (\e_1 \. p_3)  (\e_3 \. p_1) -  (\e_1 \. \e_3)  (p_1 \. p_3)  \right]^2,
\ee
where the $\mu\nu$ indices should be contracted with the polarization tensor of the massive graviton. The massive graviton can appear as an intermediate state in the propagation of a graviton through two shocks.
Its contribution to the shock commutator is given by
\be
\label{eq:massivespintwo}
\Delta {\cal Q}_{J=2} = {2 \pi \over p^u} \left( {\cal A}_{g_1 g_3 \tilde g}^{\mu \nu} \Pi^{\tilde g}_{\mu \nu, \rho \sigma} {\cal A}_{g_2 g_4 \tilde g}^{\rho \sigma} - {\cal A}_{g_2 g_3 \tilde g}^{\mu \nu} \Pi^{\tilde g}_{\mu \nu, \alpha \beta} {\cal A}_{g_1 g_4 \tilde g}^{\alpha \beta} \right) ,
\ee
where
\be
 \Pi^{\tilde g}_{\mu \nu, \alpha \beta } &= {1 \over 2} \left( P_{\mu \alpha} P_{\nu \beta} + P_{\mu \beta} P_{\nu \alpha} \right) - {1 \over D-1} P_{\mu \nu} P_{\alpha \beta} , \nn \\
 P_{\alpha \beta} &= \eta_{\alpha \beta} + {p_{\alpha} p_\beta \over m^2}.
\ee
The explicit result for (\ref{eq:massivespintwo}) is
\be
\Delta {\cal Q}_{J=2} 
&\sim 
{1 \over 2} \tilde \alpha_{2}^2 \vec{e}_{3}\cdot \vec{e}_{4} \vec{q}_{1}\cdot \vec{q}_{2} \left(\vec{e}_{3}\cdot \vec{q}_{1} \vec{e}_{4}\cdot \vec{q}_{2} - \vec{e}_{3}\cdot \vec{q}_{2} \vec{e}_{4}\cdot \vec{q}_{1} \right) 
\nn\\
&\quad
-\tilde \alpha_{2} \tilde \alpha_{4} \vec{q}_{1}\cdot \vec{q}_{2} \left(\vec{e}_{3}\cdot \vec{q}_{1} \vec{e}_{4}\cdot \vec{q}_{2} - \vec{e}_{3}\cdot \vec{q}_{2} \vec{e}_{4}\cdot \vec{q}_{1} \right)  \left(\vec{e}_{3}\cdot \vec{q}_{1} \vec{e}_{4}\cdot \vec{q}_{1}+\vec{e}_{3}\cdot \vec{q}_{2} \vec{e}_{4}\cdot \vec{q}_{2} \right)
\nn\\
&\quad
+ \left( \tilde \alpha_{4}^2 [ (\vec{q}_{1}\cdot \vec{q}_{2})^2 - {1 \over 2} m^2 \vec{q}_{1}\cdot \vec{q}_{2} ] + {{D-2 \over 4} m^4 \tilde \alpha_4^2 - (D-4) m^2 \tilde \alpha_2 \tilde \alpha_4 + (D-10)\tilde \alpha_2^2 \over 4 (D-1)}\right) \nn \\
&\quad \left[ (\vec{e}_{3}\cdot \vec{q}_{1})^2 (\vec{e}_{4}\cdot \vec{q}_{2})^2 - (\vec{e}_{3}\cdot \vec{q}_{2})^2 (\vec{e}_{4}\cdot \vec{q}_{1})^2 \right] .
\ee
From this form of the commutator, it is immediately obvious that massive spin-2 particles cannot cancel the contributions of higher-derivative gravitational interactions.

For higher spin particles, the shock three-point amplitude has the same structure, with extra polarization indices contracted with the momentum ${1 \over m^{J-2}}\e_{\mu_1 \mu_2 \mu_3 ... \mu_s} p_3^{\mu_3} ... p_3^{\mu_J}$. An extra structure takes the form
\be
{\cal A}_{gg \tilde g} =&{1 \over m^{J-4}} \tilde \alpha_0^J \e_{\mu \nu \rho \sigma} \cdot p_3 ... p_3 \left[ \e_1^{\mu} p_3^{\nu} (\e_3 . p_1) + \e_3^{\mu} p_1^{\nu} (\e_1 . p_3) - p_1^{\mu} p_3^{\nu} (\e_1 . \e_3) - \e_1^{\mu} \e_3^{\nu} (p_1 . p_3)\right] \nn \\
&\left[ \e_1^{\rho} p_3^{\sigma} (\e_3 . p_1) + \e_3^{\rho} p_1^{\sigma} (\e_1 . p_3) - p_1^{\rho} p_3^{\sigma} (\e_1 . \e_3) - \e_1^{\rho} \e_3^{\sigma} (p_1 . p_3)\right] ,
\ee
where we added the proper powers of mass to have the same dimensionality for different couplings that will contribute to the superconvergence sum rules.

To compute the blocks we simply need to square these and sum over intermediate states in our kinematics. A convenient way to do that is to first contract a spin-$J$ symmetric traceless projection operator with null vectors. The result takes the form
\be
\Pi_J(p,z_1,z_2) &= z_1^{\mu_1} ... z_1^{\mu_J} \Pi_{\mu_1 ... \mu_J ; \nu_1 ... \nu_J} z_2^{\nu_1} ... z_2^{\nu_J} \nn \\
&={\Gamma(J+1) \Gamma({D-3 \over 2}) \over 2^J \Gamma({D-3 \over 2}+J)} (- p^2)^{-J} (z_1 \. p)^J (z_2 \. p)^J C_J^{({D-3 \over 2})}\left(\eta \right) , ~~~ \eta = 1 - {p^2 z_1 \. z_2 \over (z_1 \. p) (z_2 \. p)}  .
\ee
The commutator is then given by
\be
\label{eq:massivespinJ}
\Delta {\cal Q}_J &={2 \pi \over p^u} \left( {\cal A}_{g_1 g_3 \tilde g}^{\mu_1 \mu_2 \mu_3 \mu_4} \Pi_{\mu_1 \mu_2 \mu_3 \mu_4 ; \nu_1 \nu_2 \nu_3 \nu_4} (-p_1-p_3; p_3, p_4) {\cal A}_{g_2 g_4 \tilde g}^{\nu_1 \nu_2 \nu_3 \nu_4} \right. \nn \\
&\quad\left. - {\cal A}_{g_2 g_3 \tilde g}^{\mu_1 \mu_2 \mu_3 \mu_4} \Pi_{\mu_1 \mu_2 \mu_3 \mu_4 ; \nu_1 \nu_2 \nu_3 \nu_4}(-p_2 -p_3 ; p_3, p_4) {\cal A}_{g_1 g_4 \tilde g}^{\nu_1 \nu_2 \nu_3 \nu_4} \right) ,
\ee
where we defined
\be
\Pi_{\mu_1 \mu_2 \mu_3 \mu_4 ; \nu_1 \nu_2 \nu_3 \nu_4} = D^{(1)}_{\mu_1}  D^{(1)}_{\mu_2}  D^{(1)}_{\mu_3}  D^{(1)}_{\mu_4}   D^{(2)}_{\nu_1}  D^{(2)}_{\nu_2}  D^{(2)}_{\nu_3}  D^{(2)}_{\nu_4} \Pi_J(p,z_1,z_2) 
\ee
with $D_\mu = \left( {D\over 2} - 1 + z \cdot \pa_z \right) \pa_{\mu} - {1\over 2} z_\mu \pa^2$ being the standard Thomas-Todorov operator. In both channels, the argument of the Gegenbauer polynomial becomes $\eta = 1+{4 \vec q_1 \cdot \vec q_2 \over m^2}$. The final superconvergence sum rule holds for any $\vec q_1 \cdot \vec q_2$. 

The result for the commutator takes the form
\be
\Delta {\cal Q}_J&=\vec{e}_{3}\cdot \vec{e}_{4} \vec{q}_{1}\cdot \vec{q}_{2} \left(\vec{e}_{3}\cdot \vec{q}_{1} \vec{e}_{4}\cdot \vec{q}_{2} - \vec{e}_{3}\cdot \vec{q}_{2} \vec{e}_{4}\cdot \vec{q}_{1} \right) \Delta {\cal Q}_{J}^{(1)} (\eta)
\nn\\
&
+ \vec{q}_{1}\cdot \vec{q}_{2} \left(\vec{e}_{3}\cdot \vec{q}_{1} \vec{e}_{4}\cdot \vec{q}_{2} - \vec{e}_{3}\cdot \vec{q}_{2} \vec{e}_{4}\cdot \vec{q}_{1} \right)  \left(\vec{e}_{3}\cdot \vec{q}_{1} \vec{e}_{4}\cdot \vec{q}_{1}+\vec{e}_{3}\cdot \vec{q}_{2} \vec{e}_{4}\cdot \vec{q}_{2} \right)  \Delta {\cal Q}_{J}^{(2)} (\eta)
\nn\\
&
+\left[ (\vec{e}_{3}\cdot \vec{q}_{1})^2 (\vec{e}_{4}\cdot \vec{q}_{2})^2 - (\vec{e}_{3}\cdot \vec{q}_{2})^2 (\vec{e}_{4}\cdot \vec{q}_{1})^2 \right]  \Delta {\cal Q}_{J}^{(3)} (\eta) ,
\ee
where $ \Delta {\cal Q}_{J}^{(i)}$ are computable polynomials of $\eta$ of maximal power $\eta^J$ and quadratic functions of the couplings $(\tilde \alpha_0^J, \tilde \alpha_2^J, \tilde \alpha_4^J)$. We get three superconvergence sum rules, each should be satisfied identically for any $\eta$.

Let us go back to the non-commutativity introduced by the Gauss-Bonnet coupling $\alpha_2$. Adding a spin four particle with $ \tilde \alpha_2^J =\tilde \alpha_4^J = 0$ leads to 
\be
\Delta {\cal Q}_{4}^{(1)}(\eta) &\sim - (\tilde \alpha_0^4)^2 , \nn \\
\Delta {\cal Q}_{4}^{(3)}(\eta) &\sim - (\tilde \alpha_0^4)^2 , \nn \\
\Delta {\cal Q}_{4}^{(2)}(\eta) &=0 ,
\ee
where we omitted irrelevant positive-definite $D$-dependent coefficients. We see that by adding to the Gauss-Bonnet theory a single spin four particle and a non-minimally coupled scalar we can satisfy the superconvergence relations in flat space.  With one spin four particle in the spectrum, the theory would still have pathological Regge behavior which should become visible in a slightly different kinematics, see e.g.\ \cite{Camanho:2014apa}.

\section{Noncommutativity of light-transformed scalars}	
\label{app:lighttransformscalarsubtleties}

Let us consider a simple model that illustrates some of the subtleties involved in computations of light transforms at coincident points. Imagine four free complex scalar fields of different masses in AdS. The dual theory is a version of generalized free field theory with scalar operators $\phi_{k}$ with dimensions $\Delta_k$.
We consider the following correlator\footnote{See fig. 3 in \cite{Belitsky:2013bja}, where this model of detector cross-talk is discussed. }

\be
{\cal O}_1 = \phi_a \phi_b , ~ {\cal O}_2 =\phi^\dagger_b \phi^\dagger_d,~  {\cal O}_3 =\phi^\dagger_c \phi_d , ~ {\cal O}_4 = \phi^\dagger_a \phi_c   ,
\ee
where $\cO_{1,2}$ model detectors and $\cO_{3,4}$ model sinks. The four-point function takes the form
\be
\la {\cal O}_4 {\cal O}_1 {\cal O}_2 {\cal O}_3  \ra = {1 \over x_{12}^{2 \Delta_b} x_{34}^{2 \Delta_c}} {1 \over x_{14}^{2 \Delta_a} x_{23}^{2 \Delta_d}} .
\ee
An immediate observation about this four-point function is that 
\be \label{eq:trivComm}
\la [{\cal O}_4, {\cal O}_2] [{\cal O}_1, {\cal O}_3] \ra &= 0  . 
\ee
Therefore, doing the light transform and taking the coincident limit for the $\wL[\cO_2] \wL[\cO_1] $ ordering of light ray operators always exist and is trivially equal to zero. 

For the other ordering of operators we get
\be \label{eq:dDiscEx}
\la [{\cal O}_4, {\cal O}_1] [{\cal O}_2, {\cal O}_3] \ra &= {4 \sin \pi \Delta_a \sin \pi \Delta_d \over x_{12}^{2 \Delta_b} x_{34}^{2 \Delta_c}} {{\rm sgn} [x_{14}^0] \theta(-x_{14}^2) {\rm sgn} [ x_{23}^0 ] \theta(-x_{23}^2) \over (-x_{14}^2)^{\Delta_a} (-x_{23}^2)^{\Delta_d}}  . 
\ee

To do the light transforms, it is convenient to specialize to simple kinematics in the lightcone coordinates $(u,v,\vec y)$ for which we get
\be \label{eq:frameWightmanEx}
&\<\Omega|\cO_4(1,-1,\vec 0) \cO_1(- \delta u,v_1,-\vec y) \cO_2(\delta u,v_2,\vec y) \cO_3(-1,1,\vec 0)|\Omega\> \\ \nn
&={1 \over (2 \delta u (v_1 - v_2 - i \e) + 4 \vec y^2)^{\Delta_b} 4^{\Delta_c} } {1 \over ( (1 + \delta u)(v_1 + 1+ i \e) + \vec y^2)^{\Delta_a} ( (1+ \delta u)(1 - v_2 + i \e) + \vec y^2)^{\Delta_d} } , \nn  
\ee
where we explicitly wrote the $i \e$ prescription dictated by the ordering of operators and introduced a small separation in the $u$-direction between the detector operators.

Let us first analyze the integral (\ref{eq:integraloveralpha}) which guarantees both the existence and commutativity of the coincident limit. To do that we set $\delta u = 0$ in (\ref{eq:frameWightmanEx}) and perform the integral over $v_1$ and $v_2$
\be
\int_{- \infty}^{\infty} dv_1 d v_2 {1 \over (4 \vec y^2)^{\Delta_b} 4^{\Delta_c} } {1 \over ( v_1 + 1 + \vec y^2 + i \e)^{\Delta_a} ( 1 - v_2  + \vec y^2 + i \e)^{\Delta_d} } .
\ee
This integral is zero for $\Delta_a, \Delta_d >1$ and diverges otherwise. Let us check that it agrees with the sufficient conditions for the existence of the integral derived in the main text. For the local operators at hand we have $J_1 = J_2 = 0$ and $\Delta_1 = \Delta_a+ \Delta_b$ and $\Delta_2 = \Delta_b + \Delta_d$. The Euclidean OPE, light-cone OPE, and the Regge limit are all controlled by the leading operator that appears in the OPE of $\cO_1$ and $\cO_2$, which has dimension $\Delta_a + \Delta_d$ and spin $0$. The strongest constraint comes from the light-cone OPE (\ref{eq:lightconecondition}) which for the case at hand becomes
\be
- | \Delta_a - \Delta_d | > 2 - (\Delta_a + \Delta_d) .
\ee
This can be rewritten as 
\be
4(\Delta_a -1) (\Delta_d-1) > 0 ,
\ee
which indeed coincides with the direct analysis above. Note also that in this case the behavior of the double discontinuity (\ref{eq:dDiscEx}) is no different from the behavior of the original Wightman function. Using the analysis from the main text we again conclude that the sufficient condition for the existence of the coincident limit of the light transform is $\Delta_a, \Delta_d >1$.

Let us now do the light transform first, while keeping $\delta u \neq 0$. Again it is clear from the position of poles that the result depends on the ordering of the operators $\cO_1$ and $\cO_2$, one of them trivially producing zero in agreement with (\ref{eq:trivComm}) . For the nontrivial ordering we get
\be
&\int_{-\infty}^{\infty} d v_1 d v_2 \<\Omega|\cO_4(1,-1,\vec 0) \cO_1(- \delta u,v_1,-\vec y) \cO_2(\delta u,v_2,\vec y) \cO_3(-1,1,\vec 0)|\Omega\> \\ \nn
&=(2 \pi i)^2 e^{- i \pi (\Delta_a + \Delta_d)} {\Gamma(\Delta_a + \Delta_b + \Delta_d -2) \over \Gamma(\Delta_a) \Gamma(\Delta_b) \Gamma(\Delta_d)  4^{\Delta_c}} (2 \delta u)^{\Delta_a + \Delta_d - 2} (4 \vec y^2)^{2- \Delta_a - \Delta_b - \Delta_d} . \nn
\ee
Next, we take the $\delta u \to 0$ limit for the nontrivial ordering of light ray operators $\wL[\cO_1] \wL[\cO_2] $. We see that the result is zero for $\Delta_a + \Delta_d >2$, finite and non-trivial for $\Delta_a + \Delta_d = 2$, and divergent for $\Delta_a +\Delta_d < 2$. Thus, we observe that the coincident limit exists beyond the range found by the sufficient conditions described in the text.

Apart from variations of the correlator above where subtleties related to the coincident limit of the light-ray operators can be demonstrated, one can also consider exchange Witten diagrams in AdS where similar subtleties occur. This can be easily seen using the Mellin space approach used to efficiently compute light transforms in \cite{Belitsky:2013xxa}. 
	
\section{Fourier transform of two-point functions}
\label{app:twoptforier}

In this appendix we compute the Fourier transform of the two-point function which in Euclidean signature the takes form
\be
\<\cO_J(x_1,z_1)\cO_J(x_2,z_2)\>=\frac{(z_1\.I(x_{12})\.z_2)^J}{x_{12}^{2\De}}.
\ee
Continuation to Lorentzian signature for 
\be\label{eq:2ptwightman}
	\<\O|\cO_J(0,z_1)\cO_J(x,z_2)|\O\>
\ee
amounts to using the Euclidean expression for spacelike $x$ and the $i\e$ prescription $x^0\to x^0+i\e$ for analytic continuation to
timelike $x$.

The basic strategy is the same as for three-point functions. We first decompose into harmonic functions~\eqref{eq:zzharmonicbasis}
\be
(z_1\.I(x)\.z_2)^J=\sum_{k=0}^J\b_k (z_1\.z_2)^k (-x^2)^{k-J} [z_1^{J-k}z_2^{J-k}|x]\, ,
\ee
with
\be
\b_k=\frac{2^{J-k}J!(\tfrac{d-2}{2}+J-k)_{J-k+1}}{k!(J-k)!(\tfrac{d-2}{2}+J)_{J-k+1}}.
\ee
Therefore, for $x>0$ the Wightman two-point function~\eqref{eq:2ptwightman} takes form
\be
	\<\O|\cO_J(0,z_1)\cO_J(x,z_2)|\O\>=\sum_{k=0}^J e^{i\pi\De}\b_k (z_1\.z_2)^k (-x^2)^{k-J-\De} [z_1^{J-k}z_2^{J-k}|x].
\ee
Using~\eqref{eq:zzharmonicfourier} we find that the Fourier transform is given by
\be\label{eq:2ptfourierpre}
&\<\O|\cO_J(0,z_1)\cO_J(p,z_2)|\O\>\nn\\
&=\sum_k e^{i\pi\De}\b_k \hat\cF_{2\De,2J-2k} (-p^2)^{\De-J+k-d} (z_1\.z_2)^k [z_1^{J-k}z_2^{J-k}|p] \theta(p)\nn\\
&=e^{i\pi\De}\hat\cF_{2\De,2J}2^J(-z_1\.p)^J(-z_2\.p)^J{}_2F_1\p{-J,\De-1;\De-J-\tfrac{d-2}{2};\frac{1}{2}\frac{p^2 (z_1\.z_2)}{(z_1\.p)(z_2\.p)}}(-p^2)^{\De-d/2-J}\theta(p),
\ee
where $\hat\cF$ is given by~\eqref{eq:fouriercoefficient}.

We would now like to understand its decomposition into $\SO(d-1)$ projectors. For this, we set $p=(1,0,\ldots)$, $z_i=(1,n_i)$, and decompose into $(d-1)$-dimensional Gegenbauer polynomials of $\eta=(n_1\.n_2)$. In particular, we have
\be
&{}_2F_1\p{-J,\De-1;\De-J-\tfrac{d-2}{2};\frac{1-\eta}{2}}\nn\\
&=
\sum_{s=0}^J \frac{
	2^{-2J}J!(d+J-2)_J(d-2)_J(d-\De-1)_J
}{
	(\tfrac{d-1}{2})_J(\tfrac{d-2\De}{2})_J
}\frac{
	(-1)^s(d+2s-3)
}{
	(J-s)!(d-3)_{J+s+1}
}
\frac{
	(\De-1)_s
}{
	(d-\De-1)_s
}C_s^{(\frac{d-3}{2})}(\eta).
\ee
Plugging this into~\eqref{eq:2ptfourierpre} and by using~\eqref{eq:sod-1projector}, we reproduce the claimed result~\eqref{eq:twoptfourier}.

\section{Details on the light-transform of three-point structures}
\label{app:lighttransformapp}

In this section we fix the overall normalization in our light-transform ansatz~\eqref{eq:lighttransformansatz}
\be\label{eq:lighttransformansatzapp}
\<0|\cO'_2\wL[\f_1]\cO'_3|0\>\propto\frac{
	V_{1,23}^{1-\De_\f}
	V_{2,31}^{J_2'}
	V_{3,12}^{J_3'}
	\, f\p{\frac{H_{12}}{V_{1,23}V_{2,31}},\frac{H_{13}}{V_{1,23}V_{3,12}}}
}{
	X_{12}^{\frac{\bar\tau'_1+\bar\tau'_2-\bar\tau'_3}{2}}
	X_{13}^{\frac{\bar\tau'_1+\bar\tau'_3-\bar\tau'_2}{2}}
	X_{23}^{\frac{\bar\tau'_2+\bar\tau'_3-\bar\tau'_1}{2}}
}.
\ee
We have fixed the form of the function $f$ in~\eqref{eq:appelF2parameters}. To fix the coefficient, we will compute this light-transform in special kinematics. In particular, we take $x_2=0$, $x_3=\oo$, and $x_1=x$ in the absolute past of $0$. Since light transform annihilates the vacuum state, we have
\be
	\<0|\cO'_2\wL[\f_1]\cO'_3|0\>=\<0|[\cO'_2,\wL[\f_1]]\cO'_3|0\>,
\ee
and the integral in $\wL$ in the right-hand side is only over positions of $\f_1$ which are in the past of $\cO'_2$. The rest of the integral over $\f_1$ is spacelike from $\cO_2'$ and the commutator vanishes.

Therefore, we first need to find $\<0|[\cO'_2, \f_1] \cO'_3|0\>$ in the configuration where $(2>1)\approx 3$, starting from the Euclidean expression~\eqref{eq:intermediate3pt}. We obtain
\be
	\<0|\cO'_2 \f_1 \cO'_3|0\>=\frac{e^{-i\pi\frac{\bar\tau_1+\bar\tau_2-\bar\tau_3}{2}}V_{2,31}^{J_2'}V_{3,12}^{J_3'}}{
		(-X_{12})^{\frac{\bar\tau_1+\bar\tau_2-\bar\tau_3}{2}}
		X_{13}^{\frac{\bar\tau_1+\bar\tau_3-\bar\tau_2}{2}}
		X_{23}^{\frac{\bar\tau_2+\bar\tau_3-\bar\tau_1}{2}}
	}.
\ee
We then specialize to our kinematics, which gives (using also $\bar\tau_2=\bar\tau_2'$ and $\bar\tau_3=\bar\tau_3'$)
\be
	\<0|\cO'_2 \f_1 \cO'_3|0\>=\frac{e^{-i\pi\frac{\bar\tau_1+\bar\tau_2'-\bar\tau_3'}{2}}(-(x\.z_2))^{J_2'}(x^{-2}(x\.z_3))^{J_3'}}{
	(-x^2)^{\frac{\bar\tau_1+\bar\tau_2'-\bar\tau_3'}{2}}
}.
\ee
We can further set $z_2=z_3=z$ to get
\be
\<0|\cO'_2 \f_1 \cO'_3|0\>=\frac{e^{-i\pi\frac{\bar\tau_1+\bar\tau_2-\bar\tau_3}{2}}(-(x\.z))^{J_2'+J_3'}}{
	(-x^2)^{\frac{\bar\tau_1+\bar\tau_2'-\bar\tau_3'+2J_3'}{2}}}.
\ee
The opposite ordering comes with the opposite phase, and so
\be
\<0|[\cO'_2, \f_1]\cO'_3|0\>=-\frac{2i\sin\p{\pi\frac{\bar\tau_1+\bar\tau_2-\bar\tau_3}{2}}(-(x\.z))^{J_2'+J_3'}}{
	(-x^2)^{\frac{\bar\tau_1+\bar\tau_2'-\bar\tau_3'+2J_3'}{2}}}.
\ee
We can evaluate the light transform $\wL[\f_1]$ at $z_1=z$,
\be
	\<0|[\cO'_2, \wL[\f_1]] \cO'_3|0\>=-\int_{-\oo}^{2(z\.x)/x^2} d\a (-\a)^{-\bar\tau_1}\frac{2i\sin\p{\pi\frac{\bar\tau_1+\bar\tau_2-\bar\tau_3}{2}}(-(x\.z))^{J_2'+J_3'}}{
		(-x^2+2(x\.z)/\a)^{\frac{\bar\tau_1+\bar\tau_2'-\bar\tau_3'+2J_3'}{2}}},
\ee
where the upper bound of integration is due to the vanishing of the commutator. The integral is easy to perform, and by comparing to the ansatz~\eqref{eq:lighttransformansatzapp} we can fix the overall coefficient. After some simple manipulations the result is
\be
\<0|\cO'_2\wL[\f_1]\cO'_3|0\>=-2\pi i \frac{2^{J_1'}\Gamma(-J_1')}{\Gamma(\tfrac{\tau'_1+\tau_2'-\tau_3'}{2})\Gamma(\tfrac{\tau'_1-\tau_2'+\tau_3'}{2})}&\frac{
	(-V_{1,23})^{J_1'}
	(-V_{2,31})^{J_2'}
	(-V_{3,12})^{J_3'}
}{
	(-X_{12})^{\frac{\bar\tau'_1+\bar\tau'_2-\bar\tau'_3}{2}}
	X_{13}^{\frac{\bar\tau'_1+\bar\tau'_3-\bar\tau'_2}{2}}
	X_{23}^{\frac{\bar\tau'_2+\bar\tau'_3-\bar\tau'_1}{2}}
}\nn\\
&\times f\p{\frac{H_{12}}{V_{1,23}V_{2,31}},\frac{H_{13}}{V_{1,23}V_{3,12}}}\qquad ((2>1)\approx 3),
\ee
which holds for causal relations $(2>1)\approx 3$. Note that the explicit $(-)$ signs are inserted so that there are no phase ambiguities in this causal configuration. Other causal configurations may be obtained by analytic continuation using the appropriate $i\e$ prescription. 

In particular, we need to send point $1$ to spatial infinity, which corresponds to $1\approx 2,3$ and we'll also choose $3>2$. During the corresponding analytic continuation, one can check that all $V$ structures go from being negative back to being negative, with trivial monodromy around 0.\footnote{This really only matters for $V_{1,23}$ and it is a general result that the value of this structure cannot wind around $0$ with the $i\e$ prescriptions corresponding to the Wightman ordering in question~\cite{Kravchuk:2018htv}.} Then, the only phases come from the distances $x_{ij}^2$ in the denominator, and it is easy to see that after the analytic continuation we find
\be
\<0|\cO'_2\wL[\f_1]\cO'_3|0\>=-2\pi i \frac{e^{i\pi \bar\tau'_2}2^{J_1'}\Gamma(-J_1')}{\Gamma(\tfrac{\tau'_1+\tau_2'-\tau_3'}{2})\Gamma(\tfrac{\tau'_1-\tau_2'+\tau_3'}{2})}&\frac{
	(-V_{1,23})^{J_1'}
	(-V_{2,31})^{J_2'}
	(-V_{3,12})^{J_3'}
}{
	X_{12}^{\frac{\bar\tau'_1+\bar\tau'_2-\bar\tau'_3}{2}}
	X_{13}^{\frac{\bar\tau'_1+\bar\tau'_3-\bar\tau'_2}{2}}
	(-X_{23})^{\frac{\bar\tau'_2+\bar\tau'_3-\bar\tau'_1}{2}}
}\nn\\
&\times f\p{\frac{H_{12}}{V_{1,23}V_{2,31}},\frac{H_{13}}{V_{1,23}V_{3,12}}}\qquad ((3>2)\approx 1).
\ee
This can also be derived by observing that if $(2>1)\approx 3$ then $(3>2^-)\approx 1$, which is the desired ordering if we replace $2\to 2^+$. Using the fact that $\<0|\cO'_2\tsym^{-1}=e^{-i\pi\bar\tau'_2}\<0|\cO'_2$, where $\cT$ is defined in~\cite{Kravchuk:2018htv}, we can arrive at the same result.

\section{Structures for the sum rule}
\label{app:structs}

In this appendix, we describe the tensor structures $\{2,2|\l|2,2\}_t$ and $\{2,2|\l|2,2\}_s$ used in the main text. We start with $\{2,2|\l|2,2\}_t$. In principle, these structures are harmonic polynomials of $n_i$ with appropriate homogeneity degree 2. To describe them, it is convenient to restrict to complex $n_i$ subject to $n_i^2=0$. For example, the structure $\{2,2|0|2,2\}_t$ is given in main text as
\be
	\p{(n_1\.n_4)^2-\frac{1}{d-1}}\p{(n_2\.n_3)^2-\frac{1}{d-1}}.
\ee
We can first restore homogeneity as
\be
	\p{(n_1\.n_4)^2-\frac{n_1^2n_4^2}{d-1}}\p{(n_2\.n_3)^2-\frac{n_2^2n_3^2}{d-1}},
\ee
and then set $n_i^2\to 0$ to get
\be
	(n_1\.n_4)^2(n_2\.n_3)^2,
\ee
which is a more economical encoding of the original structure.

Using this convention, we have
\be
	\{2,2|0|2,2\}_t&=n_{14}^2n_{23}^2,\nn\\
	\{2,2|2|2,2\}_t&=4n_{14}n_{23}\p{\frac{n_{12}n_{34}+n_{13}n_{24}}{2}-\frac{n_{14}n_{23}}{d-1}},\nn\\
	\{2,2|4|2,2\}_t&=6 \left(\frac{8 n_{14}^2 n_{23}^2}{(d+1)(d+3)}-\frac{8 n_{14} \left(n_{13} n_{24}+n_{12} n_{34}\right) n_{23}}{d+3}+n_{13}^2 n_{24}^2+n_{12}^2 n_{34}^2+4 n_{12} n_{13} n_{24} n_{34}\right)\, ,\nn\\
	\{2,2|(1,1)|2,2\}_t&=\frac{8 n_{14}^2 n_{23}^2}{d^2-5 d+6}-\frac{8 n_{14} \left(n_{13} n_{24}+n_{12} n_{34}\right) n_{23}}{d-3}+4 \left(n_{13} n_{24}-n_{12} n_{34}\right){}^2\, , \nn\\
	\{2,2|(2,2)|2,2\}_t&=2 n_{14} n_{23} \left(n_{12} n_{34}-n_{13} n_{24}\right),\nn\\
	\{2,2|(3,1)|2,2\}_t&=-\frac{4 \left(n_{13} n_{24}-n_{12} n_{34}\right) \left((d+1) \left(n_{13} n_{24}+n_{12} n_{34}\right)-4 n_{14} n_{23}\right)}{d+1},
\ee
where $n_{ij}=(n_i\.n_j)$.

The structures $\{2,2|0|2,2\}_s$ are obtained from $\{2,2|0|2,2\}_t$ by exchanging $2\leftrightarrow 4$.

\bibliographystyle{JHEP}
\bibliography{refs}

\end{document}